\documentclass[11pt]{article}
\usepackage[utf8]{inputenc} % allow utf-8 input
\usepackage[T1]{fontenc} 
\usepackage{amsthm}
\usepackage{mathtools}

\usepackage{graphicx} % support the \includegraphics command and options
\usepackage{array} % for better arrays (eg matrices) in maths

\usepackage{amsmath, amssymb, amsfonts, verbatim}
\usepackage{hyphenat,epsfig,subcaption,multirow}
\usepackage{nicefrac}
\usepackage{paralist}
\usepackage{enumitem}

\usepackage[font=footnotesize,labelfont=bf]{caption}

\usepackage[usenames,dvipsnames]{xcolor}
\usepackage[ruled]{algorithm2e}

\DeclareFontFamily{U}{mathx}{\hyphenchar\font45}
\DeclareFontShape{U}{mathx}{m}{n}{
      <5> <6> <7> <8> <9> <10>
      <10.95> <12> <14.4> <17.28> <20.74> <24.88>
      mathx10
      }{}
\DeclareSymbolFont{mathx}{U}{mathx}{m}{n}
\DeclareMathSymbol{\bigtimes}{1}{mathx}{"91}

\usepackage{tcolorbox}
\tcbuselibrary{skins,breakable}
\tcbset{enhanced jigsaw}

\usepackage[normalem]{ulem}
\usepackage[compact]{titlesec}

\definecolor{DarkRed}{rgb}{0.5,0.1,0.1}
\definecolor{DarkBlue}{rgb}{0.1,0.1,0.5}
\definecolor{RURed}{rgb}{0.8,0.1,0.1}

\usepackage{nameref}
\definecolor{ForestGreen}{rgb}{0.1333,0.5451,0.1333}
\definecolor{Red}{rgb}{0.9,0,0}
\usepackage[linktocpage=true,
	pagebackref=true,colorlinks,
	linkcolor=DarkRed,citecolor=ForestGreen,
	bookmarks,bookmarksopen,bookmarksnumbered]
	{hyperref}
\usepackage[noabbrev,nameinlink]{cleveref}
\crefname{property}{property}{Property}
\creflabelformat{property}{(#1)#2#3}
\crefname{equation}{eq}{Eq}
\creflabelformat{equation}{(#1)#2#3}

\usepackage{bm}
\usepackage{url}
\usepackage{xspace}
\usepackage[mathscr]{euscript}

\usepackage{tikz}
\usetikzlibrary{arrows}
\usetikzlibrary{arrows.meta}
\usetikzlibrary{shapes}
\usetikzlibrary{backgrounds}
\usetikzlibrary{positioning}
\usetikzlibrary{decorations.markings}
\usetikzlibrary{patterns}
\usetikzlibrary{calc}
\usetikzlibrary{fit}
\usetikzlibrary{snakes}
\tikzset{vertex/.style={circle, black, fill=Yellow, line width=1pt, draw, minimum width=8pt, minimum height=8pt, inner sep=0pt}}
	
\usepackage{mdframed}

\usepackage[noend]{algpseudocode}
\makeatletter
\def\BState{\State\hskip-\ALG@thistlm}
\makeatother

\usepackage{cite}
\usepackage{enumitem}
\usepackage{todonotes}

\usepackage[margin=1in]{geometry}

\newtheorem{theorem}{Theorem}
\newtheorem{lemma}{Lemma}[section]
\newtheorem{proposition}[lemma]{Proposition}

\newtheorem{claim}[lemma]{Claim}
\newtheorem{fact}[lemma]{Fact}

\newtheorem{definition}[lemma]{Definition}

\newtheorem{problem}{Problem}
\newtheorem{example}{Example}

\newtheorem*{claim*}{Claim}
\newtheorem*{theorem*}{Theorem}
\newtheorem*{proposition*}{Proposition}
\newtheorem*{lemma*}{Lemma}
\newtheorem*{problem*}{Problem}

\crefname{lemma}{Lemma}{Lemmas}
\crefname{proposition}{Proposition}{Propositions}
\crefname{claim}{Claim}{Claims}
\crefname{example}{Example}{Examples}

\newtheorem{mdresult}{Result}
\newenvironment{result}{\begin{mdframed}[backgroundcolor=lightgray!40,topline=false,rightline=false,leftline=false,bottomline=false,innertopmargin=2pt, innerleftmargin=10pt]\begin{mdresult}}{\end{mdresult}\end{mdframed}}

\newtheorem{remark}[lemma]{Remark}
\newtheorem*{remark*}{Remark}
\newtheorem{observation}[lemma]{Observation}

\newtheoremstyle{restate}{}{}{\itshape}{}{\bfseries}{~(restated).}{.5em}{\thmnote{#3}}
\theoremstyle{restate}

\theoremstyle{definition}
\newtheorem{mdalg}{algorithm}

\newtheorem{mddist}{Distribution}

\allowdisplaybreaks

\DeclareMathOperator*{\argmin}{arg\,min}

\renewcommand{\qed}{\nobreak \ifvmode \relax \else
      \ifdim\lastskip<1.5em \hskip-\lastskip
      \hskip1.5em plus0em minus0.5em \fi \nobreak
      \vrule height0.75em width0.5em depth0.25em\fi}

\newcommand{\myqed}[1]{\let\qed\relax #1 \ensuremath{\square}}

\newcommand*\samethanks[1][\value{footnote}]{\footnotemark[#1]}

\let\originalleft\left
\let\originalright\right
\renewcommand{\left}{\mathopen{}\mathclose\bgroup\originalleft}
\renewcommand{\right}{\aftergroup\egroup\originalright}

\newcommand{\tvd}[2]{\ensuremath{\norm{#1 - #2}_{\mathrm{TVD}}}}
\newcommand{\Ot}{\ensuremath{\widetilde{O}}}

\newcommand{\eps}{\ensuremath{\varepsilon}}

\newcommand{\bracket}[1]{\left[#1\right]}
\newcommand{\paren}[1]{\ensuremath{\left(#1\right)}\xspace}
\newcommand{\card}[1]{\left\vert{#1}\right\vert}

\newcommand{\IF}{\ensuremath{\mathbb{F}}}

\newcommand{\norm}[1]{\ensuremath{\|#1\|}}

\newcommand{\prob}[1]{\Pr\bracket{#1}} % change paren --> bracket -vik
\newcommand{\expect}[1]{\Exp\bracket{#1}}
\newcommand{\expectr}[2]{\Exp_{#1}\bracket{#2}}

\newcommand{\eexp}[1]{\exp\paren{#1}}
\newcommand{\set}[1]{\ensuremath{\left\{ #1 \right\}}}

\newcommand{\polylog}{\textnormal{polylog}\xspace}

\newcommand{\OPT}{\ensuremath{\mbox{\sc opt}}\xspace}

\newcommand{\alg}{\ensuremath{\mathcal{A}}\xspace}
\newcommand{\algline}{\ensuremath{\text{Process}^{*}}\xspace}

\DeclareMathOperator*{\Exp}{\ensuremath{{\mathbb{E}}}}
\DeclareMathOperator*{\Prob}{\ensuremath{\textnormal{Pr}}}
\renewcommand{\Pr}{\Prob}

\newenvironment{tbox}{\begin{tcolorbox}[
		enlarge top by=5pt,
		enlarge bottom by=5pt,
		 breakable,
		 boxsep=2pt,
                  left=5pt,
                  right=7pt,
                  top=10pt,
                  arc=0pt,
                  boxrule=1pt,toprule=1pt,
                  colback=white
                  ]%%
	}
{\end{tcolorbox}}

\newcommand{\II}{\ensuremath{\mathbb{I}}}
\newcommand{\HH}{\ensuremath{\mathbb{H}}}

\newcommand{\mireal}[1][]{
  \ifx\relax#1\relax%
    \II(\mione \,; \mitwo)%
  \else%
    \II(\mione \,; \mitwo\mid #1)%
  \fi
}

\renewcommand{\OPT}{\textnormal{\textsf{OPT}}\xspace}

\newcommand{\frust}[2]{\ensuremath{\textnormal{\textsf{frust}}(#1,\, #2)}\xspace}
\newcommand{\frest}[3]{\ensuremath{\textnormal{\textsf{frust}}_{#3}(#1,\, #2)}\xspace}
\newcommand{\frustind}[1]{\ensuremath{\textnormal{\textsf{frust}}(#1)}\xspace}
\newcommand{\plab}{\text{`$+$'}\xspace}
\newcommand{\mlab}{\text{`$-$'}\xspace}
\newcommand{\Ep}{\ensuremath{E^+}}
\newcommand{\Em}{\ensuremath{E^-}}
\newcommand{\Gp}{\ensuremath{G^+}\xspace}
\newcommand{\Gm}{\ensuremath{G^-}\xspace}
\newcommand{\Np}[1]{\ensuremath{N^+}\paren{#1}}
\newcommand{\Nm}[1]{\ensuremath{N^-}\paren{#1}}
\newcommand{\EQUALITY}{\ensuremath{\mathsf{EQUALITY}}}
\newcommand{\INDEX}{\ensuremath{\mathsf{INDEX}}}
\newcommand{\grph}[3]{\ensuremath{#1 = \paren{#2, #3}}}
\newcommand{\dedge}{\ensuremath{\Ep \cup \Em}}
\newcommand{\merging}{\textsc{merging}\xspace}
\newcommand{\Merging}{\textsc{Merging}\xspace}
\newcommand{\minimerging}{\textsc{mini-merging}\xspace}
\newcommand{\Minimerging}{\textsc{Mini-merging}\xspace}
\newcommand{\switching}{\textsc{switching}\xspace}
\newcommand{\Switching}{\textsc{Switching}\xspace}

\newcommand{\disag}[2]{\ensuremath{\mathsf{Dis}\paren{#1,#2}}}
\newcommand{\sdisag}[3]{\ensuremath{\mathsf{Dis}_{#3}\paren{#1, #2}}}

\newcommand{\Vg}{\ensuremath{V_g}}
\newcommand{\Vb}{\ensuremath{V_b}}

\newcommand{\nsens}{\emph{noise-sensitive}\xspace}
\newcommand{\nres}{\emph{noise-resistant}\xspace}
\newcommand{\Nsens}{\emph{Noise-sensitive}\xspace}
\newcommand{\Nres}{\emph{Noise-resistant}\xspace}

\newcommand{\eve}[1]{\ensuremath{\mathcal{E}_{#1}}\xspace}
\newcommand{\evecon}{\eve{S_L, S_R}}
\newcommand{\eveconl}{\eve{S'_L, S'_R}}
\newcommand{\evemerge}{\eve{\text{\merging}(\Vg)}}
\newcommand{\eveswitch}{\eve{\text{\switching}(\Vg)}}
\newcommand{\everes}{\eve{\text{\nres}(\Vb)}}

\newcommand{\evemini}{\eve{\text{\Minimerging}(\Vg)}}
\newcommand{\eveminif}{\eve{\hat{S}_L, \hat{S}_R}}

\newcommand{\evesep}{\eve{\text{sep}}\xspace}

\newcommand{\Eebg}{\ensuremath{\card{E^{(err)}_{b,g}}}\xspace}
\newcommand{\Ebg}{\ensuremath{\card{E_{b,g}}}\xspace}
\newcommand{\Egg}{\ensuremath{\card{E_{g,g}}}\xspace}
\newcommand{\Ebb}{\ensuremath{\card{E_{b,b}}}\xspace}

\newcommand{\flp}{\emph{flip-edge}\xspace}
\newcommand{\flps}{\emph{flip-edges}\xspace}
\newcommand{\bpart}[1]{\ensuremath{\mathcal{P}\paren{#1}}\xspace}
\newcommand{\flpgrph}[2]{\textnormal{\ensuremath{\textsf{flip-graph}_{#1}\paren{#2}}\xspace}}
\newcommand{\vst}{\ensuremath{v^*}\xspace}
\newcommand{\sst}{\ensuremath{s^*}\xspace}

\newcommand{\lod}{\textsc{LowDegreeSampler}\xspace}
\newcommand{\hid}{\textsc{HighDegreeSampler}\xspace}
\newcommand{\dhid}{\textsc{DerandomizedHighDegreeSampler}\xspace}
\newcommand{\btest}{\textsc{BalanceTester}\xspace}
\newcommand{\hidbs}{\hid\textsc{-based Sketch}\xspace}

\newcommand{\indep}[1]{$#1$-wise independent hash function\xspace}
\newcommand{\indepce}[1]{$#1$-wise independence\xspace}
\newcommand{\slod}{\ensuremath{S_{\text{low}}}\xspace}
\newcommand{\hlod}{\ensuremath{h_{\text{low}}}\xspace}
\newcommand{\shid}[1]{\ensuremath{#1_{\text{hi}}}\xspace}
\newcommand{\hhid}[1]{\ensuremath{h_{\text{hi}}^{(#1)}}\xspace}
\newcommand{\cands}{\ensuremath{$\set{2^i \mid i \in \bracket{2 \log{\card{V}}}}$}\xspace}

\newcommand{\mwt}{\ensuremath{\widetilde{m}}\xspace}
\newcommand{\kwt}{\ensuremath{\widetilde{k}}\xspace}
\newcommand{\nbr}[1]{\ensuremath{N\paren{#1}}\xspace}
\newcommand{\edgs}[1]{\ensuremath{E\paren{#1}}\xspace}

\newcommand{\lmid}{\ensuremath{\Bigg|}}

\newcommand{\Otilde}{\ensuremath{\tilde{O}}}

\newcommand{\GG}{\text{Giotis-Guruswami}\xspace}

\newcommand{\ist}{\ensuremath{i^*}}
\newcommand{\jst}{\ensuremath{j^*}}
\newcommand{\dc}{\ensuremath{d}}
\newcommand{\Pstar}{\ensuremath{P^{*}}\xspace}

\title{Evaluating Stability in Massive Social Networks: \\ Efficient Streaming Algorithms for Structural Balance}
 
\author{Vikrant Ashvinkumar \footnote{Department of Computer Science, Rutgers University. \texttt{\{va264,sa1497,cd751,jg1555,wc497\}@rutgers.edu}. \\ Assadi 
is supported in part by a Alfred P. Sloan Research Fellowship, and Assadi and Wang are supported in part by a NSF CAREER Grant CCF-2047061, a gift from Google Research, and a Rutgers Research Council Fulcrum Award. Ashvinkumar, Deng and Gao have been supported by NSF Grant CCF-2118953, CRCNS-2207440, CCF-2208663, and CNS-2137245.} 
\and Sepehr Assadi\samethanks 
\and Chengyuan Deng\samethanks 
\and Jie Gao\samethanks 
\and Chen Wang\samethanks}

\date{}

\begin{document}

\maketitle

\pagenumbering{roman}

% !TeX root = main.tex 
%!TEX root = main.tex
\begin{abstract}

Structural balance theory studies stability in networks.  Given a $n$-vertex complete graph $G=(V,E)$ whose edges are labeled positive or negative, the graph is considered \emph{balanced} if every triangle either consists of three positive edges (three mutual ``friends''), or one positive edge and two negative edges (two ``friends'' with a common ``enemy''). From a computational perspective, structural balance turns out to be a special case of correlation clustering with the number of clusters at most two. The two main algorithmic problems of interest are: $(i)$ detecting whether a given graph is balanced, or $(ii)$ finding a partition that approximates the \emph{frustration index}, i.e., the minimum number of edge flips that turn the graph balanced. 

We study these problems in the streaming model where edges are given one by one and focus on \emph{memory efficiency}. We provide randomized single-pass algorithms for: $(i)$ determining whether an input graph is balanced with $O(\log{n})$ memory, and $(ii)$ finding a partition that induces a $(1 + \eps)$-approximation to the frustration index with $O(n \cdot \polylog(n))$ memory. We further provide several new lower bounds, complementing different aspects of our algorithms such as the need for randomization or approximation.

To obtain our main results, we develop a method using pseudorandom generators (PRGs) to sample edges between independently-chosen \emph{vertices} in graph streaming. Furthermore, our algorithm that approximates the frustration index improves the running time of the state-of-the-art correlation clustering with two clusters (Giotis-Guruswami algorithm [SODA 2006]) from $n^{O(1/\eps^2)}$ to $O(n^2\log^3{n}/\eps^2 + n\log n \cdot (1/\eps)^{O(1/\eps^4)})$ time for $(1+\eps)$-approximation. These results may be of independent interest.

\end{abstract}

\clearpage

\setcounter{tocdepth}{3}
\tableofcontents

\clearpage

\pagenumbering{arabic}
\setcounter{page}{1}

% !TeX root = main.tex 
%!TEX root = main.tex
\section{Introduction}
\label{sec:intro}

Structural balance theory~\cite{Heider1946-bt, Heider1982-rg,Cartwright1956-fc, davis1967clustering} arises in the study of social relationships with  positive and negative relations.
Positive links describe friendship/agreement and negative links describe antagonism/disagreement. 
The theory dates back to work by Heider~\cite{Heider1946-bt}.
It describes the stability of relations among three individuals -- only two kinds of triangles are stable: the ones where all three ties are positive, indicating mutual friendship; and the ones with two negative ties and one positive tie, describing the folklore that ``the enemy of your enemy is your friend.''
A network that is far from being balanced (i.e., with many unstable triangles) accumulates stress which may lead to major re-arrangements of edges.

The structural balance theory appears in many application areas such as international relations~\cite{Moore1978-by,Antal2006-jy}, biological networks~\cite{Iacono2010-iz,DasGupta2007-ig}, portfolio analysis in financial networks~\cite{Harary2002-vf}, Ising model~\cite{Barahona1982-ma} in statistical physics, and online social media and opinion formation, dynamics, and evolution~\cite{altafini2012dynamics, tang2016survey, Teixeira2017-vk, wang22coevolution} which bears itself on, for example, Facebook~\cite{Sibona2014-ni} and Twitter~\cite{Xu:2013:SBT:2441776.2441875,kivran2011impact}.

One fundamental question is to understand whether a network is close to being balanced or not.
In a \emph{complete signed} graph where each pair of vertices is labeled positive or negative---the focus of our study---, the \emph{Cartwright-Harary Theorem}~\cite{Harary1953-na,Cartwright1956-fc} states that every balanced network  must have the nodes partitioned into (at most) two `camps', inside each of which the edges are all positive and between them the edges are negative.
For graphs that are not balanced, a natural measure to characterize its distance from total balance is the \emph{frustration index}, defined as the minimum number of edges whose negation of signs results in balance~\cite{alberson58symbolic,Zaslavsky1987-gi, Harary2007-yy,Aref2020-vf}. 
These questions can be distilled into the following algorithmic problems (see \Cref{prob:structural-balance} and \Cref{prob:min-frust-partition} for the formal definitions):
\begin{itemize}
\item \textbf{Structural Balance Testing}: Given a  complete signed graph $G$, decide whether or not it is balanced, namely, does not contain any imbalanced triangle.
\item \textbf{Frustration-minimizing Partition}: Given a complete signed graph $G$, find a partition of vertices into two camps such that the minimum number of sign flips on the edges is required for the resulting graph to be balanced.
\end{itemize}

The problems we consider are closely related to (min-disagreement) \emph{correlation clustering} where the goal is to partition the graph into clusters, so as to minimize the total number of negative edges inside clusters and positive edges across the clusters~\cite{CharikarGW03,BansalBC04,AilonCN05,ChawlaMSY15,giotis2005correlation,CohenAddadLN22}, except that structural balance enforces the number of clusters to be two.
Structural balance testing is straightforward to solve in $O(n^2)$ time on $n$-vertex graphs: place an arbitrary vertex and all its positive neighbors on one side of the bi-partition $L$, and its negative neighbors on the other side $R$, and then verify.
On the other hand, the classical work by Giotis and Guruswami~\cite{giotis2005correlation} on correlation clustering 
with two clusters implies the NP-hardness of the frustration-minimizing partition. It further provides a PTAS with running time $n^{O(1/\eps^2)}  + n \cdot (1/\eps)^{O(1/\eps^4)}$ for $(1+\eps)$-approximation of this problem. To our knowledge, this is the state of the art for structural balance testing and frustration-minimizing partition. 

Prior work primarily focused on \emph{running time} of algorithms and assume \emph{unrestricted access} to the entire graph.
In many modern applications of large-scale networks, however, there are many other considerations to take into account.
For instance, the algorithms may only have \textbf{limited access} to and a \textbf{memory} much smaller than the input.
One of the most popular models capturing the above scenario is the \textbf{graph streaming model}.
In this model, the edges arrive one after another and the algorithm needs to process this stream `on-the-fly' with limited memory.
The algorithm is allowed to make one or multiple passes over the stream, and the memory is usually substianlly smaller than the (worst-case) input size, i.e. $\Theta(n^2)$.
We focus on the single-pass setting and ask the following motivating question: 
\textit{How well can we solve structural balance testing and frustration-minimizing partition problems in the single-pass graph streaming setting?}

The golden spots for streaming algorithms are $(i)$ the $\polylog(n)$ memory regime, which is polynomial to the memory that is necessary to represent a single edge, and $(ii)$ the $\Ot(n):=O(n\cdot \polylog(n))$ memory regime --- often referred to as the \emph{semi-streaming} model.
We study structural balance in the graph streaming model in these parameter regimes.

\subsection{Our Contributions}

\paragraph{Structural balance testing.}
One way of testing structural balance of $G$ is to check whether the $2$-lift (see \Cref{def:two-lift}) of $G$ associated with its edge signing is connected.
Computing the $2$-lift can be done on-the-fly, and testing connectivity in the streaming model can be done with $O(n\log{n})$ space.
This is already a quadratic improvement over the trivial algorithm with $O(n^2)$ space, and it is also known that graph connectivity requires $\Omega(n\log{n})$ space in the streaming model. 
Is this also the limit for structural balance testing?
Our first result is a simple algebraic algorithm, which stems from testing for complete bipartiteness of negative edges, that uses \emph{exponentially improved} space complexity.
Further, we provide a companion algorithm that, while more complicated to describe, is combinatorial and introduces a technique that partially derandomizes a vertex sketch we use where hash functions with limited-independence alone do not suffice.

\begin{result}[Informal statement of \Cref{thm:structural-balance-ub}]
\label{rst:structural-balance}
    There is a single-pass randomized algorithm that given a complete signed graph $G$ in a graph stream, tests whether $G$ is balanced with probability at least $\frac{99}{100}$ using $O\paren{\log {n}}$ space.
\end{result}

\Cref{rst:structural-balance} shows that structural balance testing can be done with high space efficiency -- note that $\Theta(\log{n})$ space is necessary to simply write down a single edge.
As we elaborate later, our algorithms in~\Cref{rst:structural-balance} and \Cref{app:S-sampler-alt} are \emph{sketching-based} algorithms that are randomized and crucially use the fact that each edge of the graph appears precisely once in the stream.
We further complement our algorithms with two new lower bounds  (\Cref{prop:structural-balance-dlb,prop:structural-balance-rlb}) that show the necessity of both conditions for obtaining any $o(n)$-space algorithm for this problem.
Additionally, we show an $\Omega(n \log n)$ space lower bound (\Cref{prop:general-graph-test-lb}) for input graphs that are not necessarily complete via a reduction from bipartiteness testing; the generalized problem and the $o(n)$ space regime are simply incompatible.

\paragraph{Frustration-minimizing partition.}
Outputting the solution to frustration-minimizing partition already requires $\Omega(n)$ space, thus, as is standard, we focus on semi-streaming algorithms for this problem.
One can use \emph{cut sparsifiers} (see~\Cref{sec:cut-sparsifiers}) and a variant of an argument by~\cite{AhnCGMW15} (for correlation clustering) to obtain an algorithm with $\Ot(n/\eps^2)$ space that can return the `frustration value' of any bi-partition of vertices to within a $(1\pm \eps)$ factor (for completeness, we provide this argument in~\Cref{subsec:eval-frus}).
By enumerating over all bi-partitions, then, we can find the frustration-minimizing one up to a $(1+\eps)$-approximation.
The problem with this approach is that the resulting algorithm takes exponential time (to enumerate all bi-partitions), which is quite prohibitive, especially in large graphs\footnote{This scenario is not uncommon
 in the streaming model.
 For instance, the Tournament Feedback Arc Set problem admits an ``easy'' exponential-time $(1+\eps)$-approximation semi-streaming algorithm~\cite{ChakrabartiG0V20} that was improved very recently by~\cite{BawejaJW22} to a PTAS albeit with $O(\log{n})$ passes over the stream; see~\cite{BateniEM17} for another example.}.

To bypass this challenge, we present a `streamified' version of the correlation clustering algorithm with two clusters of\cite{giotis2005correlation} (henceforth, the Giotis-Guruswami algorithm); namely, an approach that allows us to collect just enough information from the stream to \emph{weakly} simulate (meaning not entirely faithfully) the Giotis-Guruswami algorithm in polynomial time at the end of the stream. 

\begin{result}[Informal statement of \Cref{thm:frus-algo-main}]
\label{rst:min-frust-partition}
    There is a single-pass randomized algorithm that given a complete signed graph in a graph stream, with high probability\footnote{
    Here, and throughout, with high probability means with probability at least $1-1/n$.} finds a partition of vertices with frustration value at most $(1+\eps)$ factor of the frustration index using $\Ot(n/\eps^2)$ space and polynomial (in $n$ but not $\eps$) time.
\end{result}

\Cref{rst:min-frust-partition} gives an efficient semi-streaming algorithm for $(1 + \eps)$-approximation of the frustration-minimizing partition problem, which is NP-hard to solve exactly. 
We further prove (\Cref{prop:frust-min-lb}) that even if we allow the streaming algorithm to use exponential (or more) time, solving this problem exactly requires $\Omega(n^2)$ space (the same as storing the entire input).
This fully rules out any non-trivial streaming algorithms for solving this problem exactly.

There is still one missing piece in obtaining ``truly efficient'' semi-streaming algorithms for our problem. As stated earlier, the Giotis-Guruswami algorithm that~\Cref{rst:min-frust-partition} builds on has running time (roughly) $n^{O(1/\eps^2)}$ even ignoring any streaming aspects\footnote{A simple modification of the Giotis-Guruswami algorithm and a slightly more careful analysis actually reduces the running time of their algorithm to
roughly $O(n^{100}+n \cdot (1/\eps)^{O(1/\eps^4)})$ (see the analysis in \Cref{sec:frustration-index}). 
While this is faster than the original $n^{O(1/\eps^2)}$ bound, it is still quite far from being practical.}, which makes this algorithm quite impractical. 
Our final contribution remedies this state of affairs. By building on our weak simulation of the Giotis-Guruswami algorithm in~\Cref{rst:min-frust-partition}, 
we design an improved offline (non-streaming) algorithm for frustration-minimizing partition with nearly-linear running time for any fixed $\eps > 0$ (note that
the input is of size $\Theta(n^2)$ in this problem over a complete signed graph). 

\begin{result}[Informal statement of \Cref{thm:offline-eptas}]
\label{rst:EPTAS-alg}
    There is a randomized (classical) algorithm that given a complete signed graph, with high probability, finds a partition of vertices with frustration value at most $(1+\eps)$ factor of the frustration index in $\Ot(n^2/\eps^2 + n \cdot (1/\eps)^{O(1/\eps^4)})$ time. 
\end{result}

\Cref{rst:EPTAS-alg} presents the first improvement after two decades over the \GG algorithm for frustration-minimizing partition and correlation clustering with two fixed clusters \textit{outside of graph streaming models}, which can be of independent interest.
Moreover, the algorithm in~\Cref{rst:EPTAS-alg} can also be used in~\Cref{rst:min-frust-partition}, improving the running time to nearly-linear for any constant $\eps > 0$ in the semi-streaming model.

Our algorithmic results combined with our new lower bounds (in~\Cref{sec:lbs}) collectively complete the picture of streaming and efficient algorithms for structural balance testing and frustration-minimizing partition problems.

\subsection{Our Techniques}
\label{subsec:technique}

\paragraph{Structural balance testing.}
The lower bounds in~\Cref{prop:structural-balance-dlb,prop:structural-balance-rlb} show that any $o(n)$-space algorithm for testing structural balance has to be randomized and, more importantly, uses the fact that it sees the sign between every pair of vertices \emph{exactly once}.
This motivates us to consider \emph{sketching-based} algorithms that are based on collecting aggregate statistics of large subsets of edges in the graph when seeing them in the stream (without having to store explicit information).
We proceed by proving the following structural result (stated informally): 
$(i)$ In any balanced graph, for any set $S$ of \emph{odd} size, there is an \emph{even} number 
of pairs of vertices in $S$ with a negative sign (easy direction); 
$(ii)$ Conversely, in any unbalanced graph, a constant fraction of odd-size sets $S$ (of certain cardinality) have an \emph{odd} number of negatively-signed pairs inside (hard direction). \par

This result naturally suggests the following algorithm (stated with some oversimplification): sample an odd-size set $S$ of vertices uniformly at random
at the beginning of the stream and count the number of negative edges in the stream with both endpoints in $S$. In the balanced case, 
this number will be even, while in the unbalanced case, it has a non-trivial chance of being odd.  \par

There is a serious challenge in making this strategy work: determining $S$ and therefrom which edges should be counted requires $\Omega(n)$ space to \emph{store the random bits}.
We provide two solutions that surmount this obstacle, the first being a simple algebraic algorithm and the second being a combinatorial algorithm (up to the use of limited-independence and pseudorandom generators (PRGs)).
While the combinatorial algorithm is weaker by logarithmic factors, it goes through without any appeal to low-degree polynomials and may hence provide a solution to sampling vertices with random bits even when the problem does not reduce to evaluating such a polynomial.

\begin{enumerate}[leftmargin=0.5cm]
    \item \textit{Algebraic Solution (\Cref{sec:stuctural-balance-ub}):} 
       
        Observe that the function for the parity of $\mlab$ edges in a sampled set $S$ can be viewed as a \emph{degree-2 polynomial} with one variable per vertex.
        As such, we turn from sampling vertices to using PRGs for low-degree polynomials (see \Cref{def:prg-poly}) based on sums of small bias generators so that $O(\log(n))$ truly random bits and $O(\log(n))$ extra space suffice to generate pseudorandom bits whose $0$-set on the polynomial is in proportion with that of truly random bits.
    \item \textit{Combinatorial Solution (\Cref{app:S-sampler-alt}):}
    
        Using Nisan's PRG to derandomize sketching-based algorithms was first addressed by~\cite{indyk2006stable}, and later on in several works in graph streams~\cite{AhnGM12b,KapralovLMMS14,KapralovMMMNST20}.
        These ideas target algorithms for dynamic streams (with edge deletions) that assign independent random bits to \emph{edges} of the graph: the truly random bits are read-once when the order of the stream is such that all updates to an edge appear next to each other, which is then amenable to being derandomized with Nisan's PRG.
        Given that the output of sketching algorithms are invariant under reordering of inputs, the derandomization goes through generally (pseudorandom bits are generated just-in-time).
        This solution is not immediately applicable as our algorithm samples vertices and not edges; no reordering of the input renders the truly random bits to be read-once.
        
        We introduce new techniques for partially derandomizing the vertex sampling process required of this problem.
        At a high level, we constrain the problem enough that we may then (i) use a limited-independence hash function to find a random cut of the graph, (ii) independently sample vertices on one side of the cut, such that the only events that matter consist solely of edges crossing the cut.
        Given this setup, we can indeed fix an ordering of the edges so that Nisan's PRG may be applied.
        While this solution gives slightly worse space usage for \Cref{rst:structural-balance}, it is more general and the idea can be applied to a broader class of problems.
        To the best of our knowledge, it is a new application of Nisan's PRG to derandomize vertex-samplers with $o(n)$ space in the streaming setting.
\end{enumerate}

\paragraph{Frustration-minimizing partition.}
Our algorithm in~\Cref{rst:min-frust-partition} is obtained via `streamifying' the Giotis-Guruswami algorithm~\cite{giotis2005correlation}.
Their algorithm considers the high versus low frustration index cases separately where the high frustration index scenario means at least a constant fraction of all pairs needs to flip sign.
Among these, the first part has a simple solution in~\cite{giotis2005correlation} which already lends itself to a semi-streaming algorithm immediately.
Our main technical ingredient in~\Cref{rst:min-frust-partition} lies in addressing the low frustration index case.
For simplicity of exposition, in the following, we assume $\eps$ is a constant. 

The Giotis-Guruswami algorithm in the low frustration index case is as follows.
Sample $O(\log{n})$ vertices $S$ and enumerate all $n^{O(1)}$ bi-partitions of $S$.
Then, for every bi-partition $(S_L, S_R)$, perform a \emph{merging} operation followed by a \emph{switching} operation: 
In merging, every vertex $v \in V\setminus S$ is assigned in parallel to the side of $(S_L,S_R)$ that creates less frustration for $v$, i.e., the number of negative edges of $v$ to this side plus its positive edges to the opposing side is minimized.
Let $(L,R)$ be the resulting bi-partition of vertices.
In switching, each vertex $v \in V$ in parallel is re-assigned to the side of $(L,R)$ with the least frustration.
The analysis of~\cite{giotis2005correlation} is as follows: for the bi-partition $(S_L,S_R)$ of $S$ that is consistent with the optimal solution, $(i)$ after merging, most vertices are in the ``correct'' side of $(L,R)$ already, namely, they are consistent with the optimal bi-partition, and $(ii)$ after switching, the vertices already consistent with the optimal solution do not change side, and the rest of vertices are moved to the side that only induce a small additive cost.
The key reasoning behind both steps is that in the low frustration-index case, most vertices have a ``clear'' preference toward which side of the optimal bi-partition they should reside. 

To simulate the Giotis-Guruswami algorithm via a semi-streaming algorithm, we can sample the set $S$ at the beginning of the stream and store all $\Ot(n)$ edges incident on it during the stream.
The merging phase can now be easily implemented as it relies only on the edges connected to $S$.
The switching phase, on the other hand, requires checking \emph{every} edge in the graph and thus cannot be faithfully implemented in $o(n^2)$ space.
We instead develop a way to \emph{approximately} implement this step, by sampling $O(\log{n})$ edges per vertex and using them to estimate the frustration of each vertex in the switching phase.
This allows us to still classify the majority of vertices the same as that of the Giotis-Guruswami algorithm except for the ones with no clear preference between bi-partitions of the optimal solution.
An extra argument here ensures that this step does not increase the cost too much and this new solution is still a $(1+\eps)$-approximation, leading to our semi-streaming algorithm.

The ideas above also allow us to significantly improve the \emph{running time} of the Giotis-Guruswami algorithm even outside of graph streaming models.
The most time-consuming step of that algorithm is the need to enumerate all bi-partitions of the sampled set $S$ of size $O(\log{n})$.
We preface the sampling step of the set $S$ by sampling a smaller set $T$ of size only $O(\log\log{n})$, enumerate only over $(\log{n})^{O(1)}$ bi-partitions of $T$, and use those to find an \emph{approximate} optimal bi-partition of the set $S$.
We then follow a similar strategy as above to simulate the Giotis-Guruswami algorithm, by showing that 
even though the bi-partition of the set $S$ is no longer truly optimal, the extra error occurred by the approximation, does \emph{not} propagate too much in the merging and switching step -- in other words, these operations are ``robust'' enough to handle even an approximately optimal bi-partition of $S$, not a truly optimal one.

\paragraph{Lower Bounds.}
Most of our lower bounds are based on reductions from known problems in communication complexity such as Index or Equality (see~\Cref{sec:communication}).
The  exception is our lower bound in~\Cref{prop:structural-balance-rlb} which proves that, perhaps counter-intuitively, when a stream contains copies of an edge more than once, solving structural balance testing becomes impossible in $o(n)$ space.
We prove this lower bound by presenting a new \emph{3-party} communication problem which is a mixture of the Index and Set Intersection problems.
Roughly speaking, in this new problem, we also provide the information about the hidden index of the Index problem to \emph{all} players, but in a way  that to find this index, they will need to solve an instance of the Set Intersection problem.
We then borrow an idea from~\cite{AssadiCK19,AssadiR20} that allows us to argue that a low-communication protocol cannot even change the distribution of the hidden index by much.
We then combine this with standard information-theoretic arguments to show that the Index problem the players need to solve remains hard as the distribution of its hidden index has not been altered too much.

\subsection{Related Work}

By slightly relaxing structural balance to allow triangles with three negative ties as in~\cite{davis1967clustering}, the frustration-minimizing partitions becomes fully equivalent to the correlation clustering problem (without any constraint on the number of clusters). 
More recently, motivated by similar considerations as in our paper, there has been a flurry of results on this problem in modern models of computation such as sublinear-time, streaming, or massively parallel algorithms~\cite{ChierichettiDK14,AhnCGMW15,Cohen-AddadLMNP21,AssadiW22,BehnezhadCMT22,BehnezhadCMT23}. In particular, for the single-pass streaming setting, the very recent state-of-the-art result by \cite{chakrabarty2023single} obtains a $(3+\eps)$-approximation correlation clustering in $O(n/\eps)$ space and polynomial time. % , and the guarantee is tight for their algorithm. 
Our result shows that we can obtain a better approximation when the number of clusters is two. Furthermore, \cite{BehnezhadCMT23} independently observes a $(1+\eps)$-approximation algorithm for correlation clustering in exponential time using cut sparsifiers, which is similar to our observation in \Cref{lem:eval-frus}.

In a general graph (not necessarily complete), structural balance means that all cycles must contain an even number of negative edges. 
The problem of minimizing frustration index, phrased in the literature as the {Balanced Subgraph} problem~\cite{Huffner2010-wl}, is MaxSNP-hard~\cite{Papadimitriou1988-bh,Yannakakis1981-cz} and even NP-hard to approximate within any constant factor assuming Khot's Unique Games Conjecture~\cite{Khot2002-ue}. On the positive side, this problem admits a polynomial time $O(\sqrt{\log{n}})$-approximation~\cite{Agarwal2005-zr} and
an $O(\log k)$-approximation~\cite{Avidor2007-do} when the frustration index is $k$. Readers can refer to~\cite{Huffner2010-wl} for further details about applications in practice. None of these algorithms consider space constraints or the streaming setting.

\section{Preliminaries}
\label{sec:prelim}

% \subsection{Notation}
% \label{subsec:notation}

\paragraph{Notation.} %Let us first establish some notations that will be used throughout the paper. 
% We introduce the notation we use throughout the paper. Additional notation for section-specific tools can be found in \Cref{app:info-theoretic-facts}. 
Throughout, we use $G=(V,E=\Ep\cup \Em)$ to denote a complete signed graph with $\card{V}=n$ vertices, where $\Ep$ is the set of \plab edges and $\Em$ is the set of \mlab edges. We further use $\Gp=(V,\Ep)$ and $\Gm=(V,\Em)$ to denote the graph restricted to the \plab and \mlab edges. Note that $\Ep$ and $\Em$ are always disjoint and $\card{\Ep \cup \Em} = {n \choose 2}$. For any set $S \subseteq V$ of vertices, 
$G[S]$ denotes the induced subgraph of $G$ on $S$. 

For two sets of vertices $S, \, T \subset V$, we use $E\paren{S,T}$ to denote the set of edges (both \plab and \mlab) with one endpoint in $S$ and the other in $T$. Analogously, we use $E\paren{v,S}$ to denote the set of edges (both \plab and \mlab) with one endpoint $v$ and the other in $S$. Furthermore, $\Ep\paren{S,T}$ (resp. $\Em\paren{S,T}$) refers to the set of \plab (resp. \mlab) edges with one endpoint in $S$ and the other in $T$, and $\Ep\paren{v,S}$ (resp. $\Em\paren{v,S}$) refers to the set of \plab (resp. \mlab) edges with one endpoint $v$ and the other in $S$. In particular, $\Np{S} = \Ep\paren{S,V \setminus S}$ and similarly, $\Np{v} = \Ep\paren{v,V}$ ($\Nm{S}$ and $\Nm{v}$ defined in the same manner).

When the context involves more than one graph (say $G$ and $H$), we add a subscript to make the notation clear as to which graph is being referred to (for example, $E_G\paren{S,T}$ and $E_H\paren{S,T}$).

\subsection{Problem Definition}
\label{subsec:prob-def}

% \chen{Here, or maybe in the intro, mention the (relatively straightforward) lower bound for general graphs.}

In this paper, we consider (strong) structural balance. 
A complete signed graph $G=(V,  \Ep \cup \Em)$ is said to exhibit structural balance property (in short, is balanced) if every triangle has an even number of \mlab edges.
An equivalent characterization of structural balance shown in \cite{Cartwright1956-fc} is that there exists a bi-partition of the vertices into $S$ and $T$, $S\cap T=\emptyset$ and $S\cup T=V$, such that every edge with both endpoints in $S$ (or $T$) is labelled \plab, and every edge connecting $S$ to $T$ is labelled \mlab.
Inspired by this, the notion of frustration index~\cite{alberson58symbolic,Zaslavsky1987-gi, Harary2007-yy} captures how far a graph is from structural balance.

% Let $E\paren{S,T}$ be the set of edges (both \plab and \mlab) with one endpoint in $S \subseteq V$ and the other in $T\subseteq V$. When $S=\{v\}$, we use
% $E\paren{v,T}$ to refer to $E\paren{S,T}$. We use $\Ep\paren{S,T}$ (resp. $\Em\paren{S,T}$) for the set of \plab (resp. \mlab) edges with one endpoint in $S$ and the other in $T$.
% When $T=V\setminus S$, we simplify $\Ep\paren{S,V \setminus S}$ by $\Np{S}$, and similarly, for $\Nm{S}$, $\Np{v}$, and $\Nm{v}$. 

\begin{definition}[Frustration Index]
\label{def:frustind}
    Let $\grph{G}{V}{\dedge}$ be a complete signed graph, and let $(L,R)$ be a bi-partition of $V$.
    Then the \emph{frustration} of $G$ with respect to $L,R$ is
    \begin{align*}
        \frust{G}{L,R} = \card{\Ep\paren{L,R}} + \card{\Em\paren{L}} + \card{\Em\paren{R}}.
    \end{align*}
    When the context is clear, we use the notation $\frust{G}{L}$ instead of $\frust{G}{L,R}$.
    The \emph{frustration index} $\frustind{G}$ is the minimum of $\frust{G}{L,R}$ over all possible bi-partitions $(L,R)$ of $V$.
\end{definition}

Frustration can be equivalently defined using ``disagreement'' notation:
    \begin{align*}
        \frust{G}{L,R} = \frac{1}{2} \sum_{v \in V} \disag{v}{L,R}.
    \end{align*}

% \chen{Here you're assigning $v$ to $T$...}
% Similarly, we count ``Agreement'' from assigning $v$ to $S$ by $\ag{v}{S,T} = \disag{v}{T,S}$.
We use $\disag{v}{S,T}$ to count the disagreement of edges incident to $v$ with respect to $S,T$. For example, when $v \in S$, we have $\disag{v}{S,T} = \card{\Em\paren{v,S}} + \card{\Ep\paren{v,T}}$. When $T=V\setminus S$, we may write $\disag{v}{S}$ instead of $\disag{v}{S, V \setminus S}$.

Now we are ready to formally define our problems.

\begin{problem}[Structural Balance Testing]
\label{prob:structural-balance}
    Given a complete signed graph $G = \paren{V, \Ep \cup \Em}$, decide if there is a partition $\paren{L,R}$ of $V$ such that
    \begin{itemize}
        \item If $(u,v) \in \Ep$, then $u$ and $v$ are both contained in $L$ or both contained in $R$;
        \item If $(u,v) \in \Em$, then either $u$ is in $L$ and $v$ is in $R$, or vice versa.
    \end{itemize}
In other words, the answer is `YES' if and only if the graph is balanced. 
\end{problem}

\begin{problem}[Frustration-minimizing Partition]
\label{prob:min-frust-partition}
Given a complete signed graph $\grph{G}{V}{\dedge}$, find a partition $\paren{L^*,R^*}$ of $V$ such that 
\begin{align*}
\paren{L^*,R^*} = \argmin_{\substack{\paren{L,R}:\\ L \cup R = V\\ L \cap R = \set{}}}  \frust{G}{L,R}.
\end{align*}
In other words, the partition $\paren{L^*,R^*}$ minimizes the frustration. 
\end{problem}

% !TeX root = main.tex 
%!TEX root = main.tex
\section{Structural Balance Testing in Logarithmic Space}
\label{sec:stuctural-balance-ub}

% \textcolor{red}{
% Tentative plan for revising this section:
% \begin{enumerate}
%     \item Keep the formal statement and high-level idea, delete examples, flip graphs, and everything thereafter.
%     \item State that we can construct a 2-deg polynomial over $\mathbb{F}_2$ such that if the graph is balanced, the evaluation is 0. Otherwise becomes 1 with constant probability.
%     \item Cite the low-deg polynomail PRG result and show that with log square space we only lose $\eps$
%     \item Run $O(1)$ copies, present the final algorithm.
% \end{enumerate}
% }
We formally present our main algorithm for structural balance testing (\Cref{prob:structural-balance}) -- a randomized algorithm that uses only $O(\log n)$ bits and returns whether the graph is balanced with high (constant) probability. % \chengyuan{4 -> 2.}

\begin{theorem}[Formalization of \Cref{rst:structural-balance}]
\label{thm:structural-balance-ub}
    There is a randomized single-pass algorithm \btest that solves \Cref{prob:structural-balance} with $O\paren{\log{n}}$ bits of space and the following guarantees:
    \begin{itemize}
        \item If input graph $G$ is balanced, \btest always outputs \textsc{Balanced};
        \item If input graph $G$ is not balanced, \btest outputs \textsc{NotBalanced} with probability at least $\frac{99}{100}$. 
    \end{itemize}
\end{theorem}

A few remarks on \Cref{thm:structural-balance-ub} are in order. First, it is not hard to design a deterministic algorithm that uses $\tilde{O}(n)$ space to test whether the graph is balanced -- such an algorithm can be found in \Cref{app:semi-streaming-testing} as a warm-up. However, by our lower bound in \Cref{prop:structural-balance-dlb}, any single-pass streaming algorithm for structural balance testing with $o(n)$ memory has to be randomized. Furthermore, we remark that the $O\paren{\log{n}}$ memory is asymptotically optimal since it is the number of bits that is necessary to store even a single edge. Finally, since the error is one-sided, we can boost the success probability to $1-\frac{1}{n}$ by running the algorithm $O(\log{n})$ times, which results in $O(\log^2{n})$ overall space complexity.

We now proceed to the design and analysis of our algorithm. In what follows, we assume our graph contains at least $n\geq 3$ vertices, as the structural balance is always satisfied otherwise.

\subsection{A Sample-and-Test Lemma for \mlab Edges}
\label{subsec:structural-lem}
The idea behind our algorithm starts as follows.
If we focus on \mlab edges, our task can be framed as testing if the graph $\Gm=(V, \Em)$ is a \emph{complete bipartite graph}. To this end, observe that if we sample a set $S$ of an odd number of vertices, and count the number of \mlab edges in $G[S]$, the parity will always be even when $\Gm$ is complete bipartite. With a slightly more involved analysis, we can show that if $\Gm$ is not complete bipartite, then by sampling $S$ uniformly at random, the number of \mlab edges in $G[S]$ is odd with constant probability. As such, we can use the parity of such a counter as a signal of the structural balance of the graph. 

We now formalize the above intuition. In particular, in the following lemma, we design the $S$-sampler and state its properties. 

\begin{lemma}[$S$-sampler Lemma]
\label{lem:neg-sample}
    For a given complete signed graph $G=(V,E)$, let $S \subseteq V$ be a subset of vertices sampled with the following \underline{$S$-sampler}:
    \begin{enumerate}
    \item Pick a fixed vertex $v_n$ arbitrarily. % \vik{Reminder to check why we want to specify the order. Seems superfluous if we later say ``(or any fixed order)''} by the lexicographical order (or any fixed order).
    \item\label{line:S-sample-indep}Sample each vertex in $v \in V\setminus \{v_n\}$ independently with probability $\frac{1}{2}$ to get $S'$.
    \item If $\card{S'}$ is odd, let $S=S'$; otherwise, let $S=S'\cup \{v_n\}$. 
    \end{enumerate}
    Then, the following statements are true:
    \begin{enumerate}
    \item If $G$ is balanced, the induced subgraph $G[S]$ always has an even number of \mlab edges;
    \item If $G$ is not balanced, the induced subgraph $G[S]$ has an odd number of \mlab edges  with probability at least $\frac{1}{4}$, over the randomness of sampling $S$. 
    \end{enumerate}
\end{lemma}

% \vik{Remove if we decide to not include this proof in the app.} \chengyuan{I wonder why we want an alternative proof? Does it really help with clarity or just because we wrote it already.} \vik{Chen is the best person to take this one.}
% It is possible to prove the correctness of \Cref{lem:neg-sample} with a combinatorial argument, albeit with a rather involved cases analysis. We include such a proof in \Cref{app:alt-proof}. 
The main idea to prove \Cref{lem:neg-sample} is by observing that the parity of $\mlab$ edges counted by our $S$-sampler is in fact captured by a \emph{degree-2} polynomial over $\IF_{2}$. The observation can be formalized as the following lemma.
% More formally, we present the following lemma.
% \vspace{-15pt}
\begin{lemma}
\label{lem:S-sampler-poly}
Let $X_{i}\in \{0,1\}$ be the indicator random variable for whether vertex $v_{i}$ is sampled by $S$-sampler.
Define the following polynomial with $(n-1)$ variables over $\IF_{2}$:
\begin{equation}
\label{equ:S-sampler-poly}
\Pstar(X) = \sum_{\substack{i,j<n\\ (v_{i},v_{j})\in \Em}} X_{i}X_{j} + \sum_{\substack{i<n\\ (v_{i},v_{n})\in \Em}} X_{i}\paren{1+\sum_{j=1}^{n-1}X_{j}}.
\end{equation}
Then, the polynomial $\Pstar(X)=\card{\Em(G[S])}$ over $\IF_2$, which is the parity of number of negative edges induced by the sample set $S$. 
% \vik{Should we say what $\card{\Em(G[S])}$ over $\IF_2$ means, or is it clear enough this way?} \chengyuan{I think it is clear, but it seems to appear the first time, so we can add it.}
% in \Cref{equ:S-sampler-poly} is 1 if and only if the parity of $\card{\Em(G[S])}$ sampled by the \underline{$S$ sampler} (as in \Cref{lem:neg-sample}) is odd.
\end{lemma}
% \vspace{-15pt}
\begin{proof}
Note that \Cref{equ:S-sampler-poly} is essentially obtained by substituting $X_{n}=1+\sum_{j=1}^{n-1}X_{j}$, and counting a \mlab edge $(v_{i},v_{j})$ if both $v_{i}$ and $v_{j}$ are sampled.
% As such, we let $v_{n}$ be the vertex $v^*$ in the $S$-sampler, and all other vertices are sampled with probability $\frac{1}{2}$.
If the number of sampled vertices other than $v_{n}$ is odd, we have $\sum_{j=1}^{n-1}X_{j}=1$, which implies $X_{n}=0$ over $\IF_{2}$; otherwise, if the number of sample vertices other than $v_{n}$ is even, we have $\sum_{j=1}^{n-1}X_{j}=0$, which implies $X_{n}=1$ over $\IF_{2}$. Therefore, the polynomial $\Pstar$ exactly captures the sampling process of the $S$-sampler. Finally, if the total number of \mlab edges is odd, the polynomial $\Pstar$ evaluates to $1$, and vice versa.
\end{proof}

As we will see shortly, \Cref{lem:S-sampler-poly} is also crucial to run our \emph{streaming} algorithm for the $S$-sampler by storing limited number of random bits.
To show that the $S$-sampler gives a useful signal for whether the graph is balanced, we show that the polynomial $\Pstar$ in \Cref{equ:S-sampler-poly} gives a useful signal.
We first show that $\Pstar$ is identically $0$ if and only if $G$ is balanced.

\begin{lemma}
\label{lem:zero-iff-balanced}
The polynomial $\Pstar$ in \Cref{equ:S-sampler-poly} is identically 0 if and only if $G$ is balanced.
\end{lemma}
\begin{proof}
Suppose first that $G$ is balanced.
Then either $\Gm$ is empty, or it is a complete bipartite graph.
Let $S$ be the sample from $S$-sampler.
In the first case, the number of $\mlab$ edges in $G[S]$ is always $0$.
In the second case, let $L$ and $R$ be the bi-partition of $\Gm$, and let $A=L\cap S$ and $B=R\cap S$.
The number of $\mlab$ edges in $G[S]$ is $\card{A} \cdot \card{B}$, which is an even number since one of $\card{A}$ and $\card{B}$ is even (recall that $\card{S}$ is odd).
By \Cref{lem:S-sampler-poly}, $\Pstar$ is thus identically $0$ when $G$ is balanced.

Suppose, on the other hand, that $G$ is not balanced.
Then there is a triangle $(v_i, v_j, v_k)$ with an odd number of $\mlab$ edges.
If $i,j,k \neq n$, setting $X_i = X_j = X_k = 1$ and $X_\ell = 0$ for $\ell \neq i,j,k$ yields $\Pstar(X) = 1$.
If on the other hand, say, $k = n$, then setting $X_i = X_j = 1$ and $X_\ell = 0$ for $\ell \neq i,j$ yields $\Pstar(X) = 1$.
Hence $\Pstar$ is not identically $0$ when $G$ is not balanced.
\end{proof}

Next, we give a lemma on the \emph{fraction} of assignments that are non-zero on the polynomial $\Pstar$, which gives us a lower bound for the success probability to capture an odd number of \mlab edges when $G$ is imbalanced. We note that the lemma is standard in the coding theory community (e.g.~\cite{Roth06Coding}), but we state a proof here for completeness.

\begin{lemma}
\label{lem:sampling-poly-arg}
    Let $P(x_1, x_2, \ldots, x_n)$ be a polynomial over $\IF_2$ with degree at most $2$.
    If $P$ is not identically $0$, then there is a set $X$, with cardinality at least $2^{n-2}$, such that $x \in X$ implies $P(x) = 1$.
\end{lemma}
\begin{proof}
    We may assume that $P$ is multilinear since $\bar{P}(x) = P(x)$ for all $x \in \set{0,1}^n$, where $\bar{P}$ is the linearization of $P$ (i.e. all monomials of the form $x_i^2$ are replaced with $x_i$).
    
    If $P$ is of degree $1$ or less, the claim holds by using the Schwartz-Zippel lemma (see~\Cref{prop:sz-lemma}).
    Suppose then that $P$ is of degree $2$.
    Up to a reordering of variables, $x_{n-1} x_n$ is a term of $P$ with coefficient $1$.
    Let $a$ be a fixed assignment of $x_k$ to $\set{0,1}$ for all $k \in [n-2]$.
    Then $P(a, x_{n-1}, x_n)$ is a multilinear degree $2$ polynomial $Q(x_{n-1}, x_n)$ with the term $x_{n-1} x_n$ having coefficient $1$.
    Checking over all $8$ possible polynomials that $Q$ may take, we see that each of them has at least one assignment $a'$ of $x_{n-1}, x_n$ to $\set{0,1}$ such that $Q(a') = 1$.
    This completes the proof since there are $2^{n-2}$ possible assignments $a$.
\end{proof}

% , which is the amount of \mlab edges in the subgraph induced by the sample.
% The parity of \negc is a signal as to whether the input graph is balanced or not. If it is even, the graph is always balanced, otherwise imbalanced with high probability. In the rest of this section, we first prove that the above claim is true (i.e. structural balance hinges on \negc), then show it takes only $O(\log^2 n)$ bits to compute \negc by a construction of 2-degree polynomial over $\mathbb{F}_2$ with pseudorandom generators (PRG).

\begin{proof}[Finalizing the proof of \Cref{lem:neg-sample}]
By \Cref{lem:S-sampler-poly} and \Cref{lem:zero-iff-balanced}, if $G$ is balanced, the polynomial $\Pstar$ is identically $0$, which means the $S$-sampler always finds an even number of \mlab edges. Otherwise, if $G$ is not balanced, $\Pstar$ is not $0$. Therefore, by \Cref{lem:sampling-poly-arg}, the $S$-sampler finds an odd number of \mlab edges with probability at least $\frac{1}{4}$. 
\end{proof}

By \Cref{lem:neg-sample}, to test the structural balance of a complete signed graph, it suffices to implement the $S$-sampler and count the number of negative edges.
However, while the counter takes $O(\log{n})$ space, it is not immediately clear how the $S$-sampler can be implemented with a small space in the streaming setting.
In particular, since we need to sample each vertex \emph{independently}, the trivial solution requires $\Omega(n)$ memory to store the random bits (for each vertex).
We address this issue in the next step. % in as long as we can design an algorithm \negc to count the number of \mlab edges induced by a sampled odd number of vertices

\subsection{Simulating the $S$-sampler in Streaming with $O(\log n)$ Space}
\label{subsec:sample-simulate}

We tackle this issue by using the fact that our $S$-sampler is a degree-2 polynomial (\Cref{lem:S-sampler-poly}).
We may consequently use a PRG that fools degree-2 polynomials (see \cite{NaorN93,Lovett09,BogdanovV10} for constructions).
More concretely, there exists a PRG that takes $O(\log n)$ truly random bits, and generates $n$ pseudorandom bits with $O(\log n)$ extra space per bit (\Cref{prop:prg-poly}) so that a degree-2 polynomial cannot distinguish a truly random input from a pseudorandom input to within a small constant error.
% We shall use this machinery to derandomize the $S$-sampler and, thereafter, implement it in streaming.
% \begin{remark}
%     In \Cref{app:S-sampler-alt}, we propose an alternative algorithm that is more combinatorial (up to the use of limited-independence and PRGs).
%     Instead of directly operating on the \mlab edges, we look into the `flip graph' -- roughly the edges that are responsible for the graph not being balanced.
%     The solution leads to slightly worse space complexity of $\polylog(n)$ bits.
%     However, this alternative approach is \emph{not} restricted to algorithms that can be represented as $O(1)$-degree polynomials.
%     More importantly, to the best of our knowledge, it is a novel application of Nisan's PRG and the first vertex-sampling approach with $o(n)$ space in the streaming setting. % Please check \Cref{app:S-sampler-alt} for more details.
% \end{remark}
% we cannot directly use the generic PRGs whose pseudorandom bits are \emph{read-once}. To address the challenge, we observe that \Cref{lem:S-sampler-poly} allows our $S$-sampler to use a PRG that takes $O(\log {n})$ truly random bits and $O(\log {n})$ extra space to construct a pseudorandom bit (\cite{NaorN93,Lovett09,BogdanovV10}, see~\Cref{prop:prg-poly}). As such,
Using this, we design an $O(\log(n))$-memory algorithm as follows. 

\begin{tbox}
\textbf{A streaming algorithm to test structural balance.}
\begin{itemize}
\item Run $100$ independent copies of the following \underline{$S$-sampler simulation}:
\begin{enumerate}
\item Maintain $O(\log n)$ truly random bits and a counter. 
\item For each arriving \mlab edge $(u,v)\in \Em$:
\begin{enumerate}
\item Generate the pseudorandom bits for vertices $u$ and $v$ (name the bits $X_{u}$ and $X_{v}$) with the PRG prescribed in \Cref{prop:prg-poly} with $\eps=\frac{1}{20}$.
\item If both $X_{u}=1$ and $X_{v}=1$, increase the counter by 1.
\end{enumerate}
\item By the end of the stream, if the parity of the counter is odd, return 1; otherwise, return 0.
\end{enumerate}
\item If any of the copies of \underline{$S$-sampler simulation} returns 1, report \textsc{Not Balanced}; otherwise, report \textsc{Balanced}.
\end{itemize}
\end{tbox}

\paragraph{Analysis of the space complexity.}
The truly random bits and the counter take $O(\log n)$ bits of space.
Furthermore, for each \mlab edge, the PRG generates the pseudorandom bits for $X_{u}$ and $X_{v}$ with $O(\log n)$ bits of extra space since $\eps=\frac{1}{20}$.
The space for this purpose can be re-used across different edges.
Each copy of the $S$-sampler simulation therefore takes $O(\log n)$ space to implement.
Finally, since we run $100$ independent copies in parallel, the final algorithm take $O(\log n)$ space.

\paragraph{Analysis of correctness.}
By \Cref{lem:neg-sample}, if the graph is balanced, the $S$-sampler simulation always returns 0.
On the other hand, if the graph is imbalanced and we sample with truly random bits $U_{n-1}$, it holds from \Cref{lem:neg-sample} that $\Pr(\Pstar(U_{n-1})=1)\geq \frac{1}{4}$ since $\Pstar$ is a valid implementation of the $S$-sampler.
By the guarantees of the PRG $g:\{0,1\}^{O(\log n)\rightarrow (n-1)}$, we have
\begin{align*}
|\Pr(\Pstar(g(U_{\ell}))=1) - \Pr(\Pstar(U_{n-1})=1)| \le \frac{1}{20},
\end{align*}
where $U_\ell$ are $O(\log n)$ truly random bits, the seed for the PRG.
We hence have $\Pr(\Pstar(g(U_{l}))=1)\geq \frac{1}{4}-\frac{1}{20} = \frac{1}{5}$.
As such, the probability for no copy to return 1 is at most $(\frac{4}{5})^{100}<\frac{1}{100}$, as desired.

\subsection{Further Results}
For lower bounds on \Cref{rst:structural-balance}, one may skip ahead to \Cref{sec:lbs}.
For a slightly weaker combinatorial algorithm which, while more complicated, is of technical interest, see \Cref{app:S-sampler-alt}.

% !TeX root = main.tex 
%!TEX root = main.tex
\section{Semi-streaming Frustration-minimizing Partition in Poly-Time}
\label{sec:frustration-index}
%\todo[inline]{Task: Add `\GG algorithm' to all the proper citations of \cite{giotis2005correlation}}\jie{I will fix this while I am making a pass}
In this section, we consider \Cref{prob:min-frust-partition} of finding, in the semi-streaming model, a partition of a complete signed graph that is as close to structurally balanced as possible.
More specifically, we want to divide the set of vertices into two parts and minimize the number of \mlab edges contained in either part and \plab edges connecting the two parts.

Giotis and Guruswami~\cite{giotis2005correlation} showed that \Cref{prob:min-frust-partition} is NP-hard.
In the streaming setting, we prove that any algorithm that gives the exact optimal frustration-minimizing partition (or computes the optimal frustration index) requires $\Omega(n^2)$ memory (see \Cref{prop:frust-min-lb}).
In view of this, we settle for finding a partition that approximates one which is as close to structurally balanced as possible.
This section shows how to adapt the correlation clustering algorithm with two clusters in~\cite{giotis2005correlation} to the semi-streaming model.
\Cref{sec:good-time-frus} then improves the running time of the \GG algorithm and completes the following theorem.

\begin{theorem}[Formalization of \Cref{rst:min-frust-partition}]
\label{thm:frus-algo-main}
    There is an algorithm that, given a complete signed graph $\grph{G}{V}{\dedge}$ and $\eps \in \paren{0,1}$, with high probability finds a partition $\paren{L,R}$ of $V$ where
    \begin{align*}
        \frust{G}{L} \le (1 + \eps) \cdot \frustind{G}.
    \end{align*}
    Moreover, the algorithm uses $O\paren{n\cdot \frac{\log^4(n/\eps)}{\eps^6}}$ space, a single pass over the stream and has running time of $O\paren{(\frac{n}{\eps})^2\cdot \log^3{n}+ n \log{n} \cdot \paren{\frac{1}{\eps}}^{O\paren{\eps^{-4}}}}$.
\end{theorem}

The remaining part of this section is organized into two parts:
\begin{enumerate}
    \item 
        \Cref{subsec:eval-frus} shows a relatively simple and standard way to efficiently evaluate an approximation of the frustration of a partition using cut sparsifiers.
        This is a key step in the algorithms to follow; they consider a number of candidate partitions and choose the one which has the (approximately) minimum frustration.
    \item
        \Cref{subsec:poly-algo}, which is the main technical ingredient of this section, goes over the correlation clustering algorithm in \cite{giotis2005correlation} in the case of two clusters.
        Additionally, we show how the algorithm can be adapted to fit in the semi-streaming model.
\end{enumerate}

\subsection{Evaluating Frustration}
\label{subsec:eval-frus}
Let $\grph{G}{V}{\dedge}$ be a complete signed graph, and let $(L,R)$ be a partition of $V$.
Then it is elementary to determine $\frust{G}{L}$; just look at every edge and tally the \mlab edges contained in $L$ (and $R$) and the \plab edges connecting $L$ and $R$.
Looking at every edge after constructing a partition, however, may not always be feasible in the semi-streaming setting; for example, an algorithm may have to make a pass over all edges to get enough information to determine a candidate partition, at which point it is too late to do the tallying (having passed through the stream).
We thus look towards non-trivial techniques to evaluate a $\frust{G}{L}$, or, more specifically, an approximation thereof.

We turn to the idea of cut sparsifiers (see \Cref{def:cutsparse}), which store a sparse representation of $G$ which approximately answers the cut queries for all $S \subseteq V$: what is the size of $E\paren{S,V \setminus S}$?

\begin{lemma}
\label{lem:eval-frus}
    Let $\grph{G}{V}{\dedge}$ be a complete signed graph.
    There is a single-pass streaming algorithm using $O\paren{n \log^3{n} / \eps^2}$ space and $O\paren{n^2 \log^2{n}}$ time that, for any $\eps \in \paren{0,1}$, with high probability produces an oracle that, when given $L\subseteq V$, returns $\frest{G}{L}{\eps}$ in $O\paren{n \log{n}/\eps^2}$ time where
    \begin{align*}
        \paren{1 - \eps}\frust{G}{L} \le \frest{G}{L}{\eps} \le \paren{1 + \eps}\frust{G}{L}.
    \end{align*}
\end{lemma}
% vikasdf
Our construction of the `frustration sparsifier' in \Cref{lem:eval-frus} follows from a standard application of the cut sparsifier, which we elaborate as follows.

\paragraph{From Cut Sparsifiers to Frustration Sparsifiers.}
Recall that
\begin{align*}
    \frust{G}{L,R} = \card{\Ep\paren{L,R}} + \card{\Em\paren{L}} + \card{\Em\paren{R}}.
\end{align*}
With a cut sparsifier, it is straightforward to estimate the $\card{\Ep\paren{L,R}}$ term; just use a cut sparsifier $H$ of $\grph{G'}{V}{\Ep}$.
Less straightforward is the problem of estimating the $\card{\Em\paren{L}}$ and $\card{\Em\paren{R}}$ terms.
The idea is as follows: the number of \mlab edges contributing to the two terms are
\begin{align*}
    \card{\Em\paren{L}} + \card{\Em\paren{R}} = \card{\Em} - \card{\Em\paren{L,R}}.
\end{align*}
One may first try to similarly use a cut sparsifier of $\grph{G'}{V}{\Em}$ to compute this quantity.
The trouble here is that the additive $\eps \card{\Em\paren{L,R}}$ error from such an approach may outsize our target error of $\eps \cdot \frust{G}{L,R}$.
This can be worked around by recalling that $G$ is a complete graph,
\begin{align*}
    \card{\Em\paren{L,R}} = \card{L} \cdot \card{R} - \card{\Ep\paren{L,R}}.
\end{align*}
Thus, we may use $H$ to not only estimate $\card{\Ep\paren{L,R}}$, but also $\card{\Em\paren{L}} + \card{\Em\paren{R}}$.
The $\eps \card{\Ep\paren{L,R}}$ error used to estimate the latter two terms is suitably small; we get a total additive error of $2 \eps \card{\Ep\paren{L,R}}$ which is at most $2 \eps \cdot \frustind{G, L}$.

\begin{tbox}
	\textbf{Frustration Sparsifier} for $\grph{G}{V}{\dedge}$ and $\eps \in \paren{0,1}$:
	\begin{itemize}
	    \item Let $\grph{H}{V}{E_H, w_H}$ be an $\frac{\eps}{2}$ cut sparsifier for $\grph{G'}{V}{\Ep}$.
    	\item Return $\frest{G}{\cdot}{\eps}$ which, when given $L \subseteq V$, computes
    	    \begin{align*}
    	        \underbrace{w_H\paren{L,V \setminus L}}_{\approx \card{\Ep\paren{L,V \setminus L}}} + \underbrace{\card{\Em} - \paren{\card{L} \cdot \card{V \setminus L} - w_H\paren{L,V \setminus L}}}_{\approx \card{\Em\paren{L}} + \card{\Em\paren{V \setminus L}}}.
    	    \end{align*}

                % Vik: What was the suggestion? Asking because there are these random underbraces which don't make sense.
    	    % \begin{align*}
    	    %     \underbrace{2w_H\paren{L,V \setminus L}}_{} + \underbrace{\card{\Em} - \card{L} \cdot \card{V \setminus L}}_{}. % \text{\chengyuan{Per Sepehr's suggestion.}}
    	    % \end{align*}
	\end{itemize}
\end{tbox}
\noindent

We defer the standard proof of \Cref{lem:eval-frus} to \Cref{app:frust-spar}, which follows from guarantees that our frustration sparsifier construction achieves.

Observe that these sparsifiers alone give us a single-pass algorithm that uses $\Ot\paren{n}$ space which, however, runs in $\Omega\paren{2^n}$ time:
construct a frustration sparsifier with parameter $\eps$, enumerate over all partitions of $V$, and return the one which gives the smallest estimated frustration.
This algorithm would yield a partition that is within a $\paren{1 + \eps}/\paren{1 - \eps}$ factor of $\frustind{G}$. %\jie{minor issue, here do we mean within $1+\eps$ and $1-\eps$ factor?}
%\todo[inline]{vik: Because the frustration can be underapproximated or overapproximated, the approximation factor from going over all partitions and choosing the best is over/under i.e. $\paren{1 + \eps}/\paren{1 - \eps}$}

In the next subsection, we show how to use these frustration sparsifiers to design a \emph{polynomial time} semi-streaming algorithm for \Cref{prob:min-frust-partition}. Henceforth, when we informally mention choosing a candidate partition that achieves the minimum frustration, there is implicitly an underlying frustration sparsifier with a small enough parameter (for example $\eps/10$).

\subsection{A Polynomial Time Algorithm: Streamification of \GG Algorithm}
\label{subsec:poly-algo}

For the remainder of this section, let $\frustind{G} = \gamma n^2$ where $\gamma$ is not necessarily a constant.
The algorithm by Giotis and Guruswami~\cite{giotis2005correlation} examines two cases, which we state below.
%informally and, perhaps, inaccurately (but we believe morally correctly): 
\begin{description}
    \item[High frustration index case ($\gamma \gtrapprox \eps$)]
        \ \\
        When $\gamma$ is large compared to $\eps$.
        In our adaptation, $\gamma > \eps/100^4$.
        This is well-approximated with an algorithm that guarantees an additive error up to $O\paren{\eps n^2}$.
    \item[Low frustration index case ($\gamma \lessapprox \eps$)]
        \ \\
        When $\gamma$ is small compared to $\eps$.
        In our adaptation, $\gamma \leq \eps/100^4$.
        This is well-approximated with an algorithm that guarantees an additive error up to $O\paren{\gamma^2 n^2}$.
\end{description}
Of course, $\gamma$ is unknown to the algorithm, which is why both algorithms are run, and the best among the two partitions is returned.

% In our adaptation, we formally specify each case like so:
% \begin{center}
%     \begin{tabular}{ |c|c| }
%         \hline
%         Informal & Formal \\
%         \hline
%         $\gamma \gtrapprox \eps$ & $\gamma > \eps/100^4$ \\ 
%         $\gamma \lessapprox \eps$ & $\gamma \le \eps/100^4$ \\ 
%         \hline
%     \end{tabular}
% \end{center} \jie{this table can be removed now}

It is straightforward to adapt the algorithm with guarantees when $\gamma \gtrapprox \eps$;
only $O\paren{n}$ edges, which we can afford to store in memory, are needed in order to perfectly simulate the algorithm in \cite{giotis2005correlation}.
We thus leave the details for $\gamma \gtrapprox \eps$ to \Cref{app:max-agree} (one can also see $\textsc{MaxAg}\paren{k,\eps}$ in Section~3 of \cite{giotis2005correlation}).
The result is encapsulated by the following proposition.
\begin{proposition}
\label{prop:maxag}
    Let $\frustind{G} = \gamma n^2$.
    If $\gamma > \eps/100^4$, then for any $\eps, \delta \in \paren{0,1}$ there is a single-pass algorithm using $\Ot\paren{n}$ space and with running time $(n/\eps)\cdot 2^{\eps^{-2}\log{(1/\eps\delta)}}$ that returns a partition $\paren{L,R}$ with $\frust{G}{L,R} \le \gamma n^2 + \eps n^2/2$ with success probability $1 - \delta$.
\end{proposition}

It remains to describe the algorithm when $\gamma \lessapprox \eps$.
When there is no constraint on space usage, the \GG algorithm works by sampling $\Theta\paren{\log{n}}$ vertices (called the sample $S$) and trying every partition $\paren{S_L, S_R}$ of $S$.
$S_L$ and $S_R$ are taken to be subsets of $L$ and $R$ respectively, where $(L,R)$ is a partition of $V$.
For each such partition $\paren{S_L, S_R}$, there should be a clear local choice for whether a vertex $v$ should be assigned to $L$ or to $R$ by choosing the smaller of $\disag{v}{S_L+v,S_R}$\footnote{In this section, we use the following notation: For any set $S$ and any vertex $v$, $S+v$ denotes $S \cup \set{v}$ and $S-v$ denotes $S \setminus \set{v}$.} and $\disag{v}{S_L,S_R+v}$ respectively; we do this for all unsampled vertices, and refer to this step of the algorithm as \merging.
The algorithm then tries to further refine the solution quality by marking each vertex that improves the frustration when moved from $L$ to $R$ or vice versa, keeping every other vertex fixed.
All marked vertices are then moved to the other part.
The process of marking vertices and moving them after all have been marked is referred to as the \switching step.

% In the semi-streaming model, we respect space constraints by executing the algorithm on sparse subgraphs of the input; as we will shortly see, the correctness of the algorithm is tolerant to small errors begotten from graph sparsification.\chengyuan{To do}
As we will see shortly, the correctness of the \GG algorithm still holds with a small error introduced by graph sparsification. This is crucial as it allows us to design space efficient streaming algorithms by simply maintaining a cut sparsifier during the stream, and performing the \GG procedure in the end. More concretely, 
only once the sparse representations of the input stream have been obtained (i.e. a single pass over the stream has been completed), do we then start iterating over every partition $\paren{S_L, S_R}$ of $S$, and in each iteration \merging, \switching, and storing the best resulting $\paren{L,R}$ seen so far.
\Merging is easily ported to the streaming setting; we just need to store the $\Theta\paren{n\log{n}}$ edges incident to $S$.
The way \switching is adapted, on the other hand, is less trivial; we store a sparse representation of the input graph by sampling $\Theta\paren{(1/\eps^2)\log{n}}$ neighbors for each vertex.
Finally, determining the (approximate) best partition seen through all iterations can be done by using a frustration sparsifier. 

% \chen{In fact, I think we can talk about the intuition why switching works even with sampling $O(\log(n)/\eps^2)$ edges for each vertex. The catch here is that on the first glance, switching seems to be hard to be adapted to the streaming setting. Note that by sampling $\log(n)/\eps^2$ edges, we can only estimate the min-cost with an additive error of $\eps\cdot n$. If \emph{both} the min-frustration and the max-frustration of $v$ are low, there's no way for such an idea to work. However, here we crucially use the property for 2 partitions where the cost must sum up to $n$. This is also the reason why our idea works only for the $2$-partition case. Something among this line will help readers understand why our algorithm works.} \todo{I think this is too much detail before even explaining the algo}

\begin{tbox}
	\textbf{Streamification of \cite{giotis2005correlation}} for $\grph{G}{V}{\dedge}$ and $\eps \in \paren{0,1}$, when $\gamma \lessapprox \eps$:
	\begin{enumerate}
	    \item Sample $100 \log{n}$ vertices uniformly at random. Call this set $S$.
	    \item Sample $100^3 \cdot \frac{\log{n}}{\eps^2}$ vertices uniformly at random for each vertex $v$. Call these sets $N_v$.
    	\item Read through the stream of (signed) edges and
    	    \begin{itemize}
    	        \item Store edges incident to $S$. Call this graph $G_S$.
    	        \item Store edges joining $v$ and $N_v$ for all $v$. Call this graph $G_N$.
    	        \item Store $\frest{G}{\cdot}{\eps/10}$, a frustration sparsifier of $G$.
    	    \end{itemize}
    	\item For every partition $\paren{S_L, S_R}$ of $S$
    	    \begin{description}
    	        \item[(\Merging)]
    	        \ \\
    	        For every $v \in V \setminus S$, assign $v$ to $L$ if $\sdisag{v}{S_L+v,S_R}{G_S} < \sdisag{v}{S_L,S_R+v}{G_S}$, and $R$ otherwise.
    	        \\
    	        $S_L$ is assigned to $L$ and $S_R$ is assigned to $R$.
    	        \item[(\Switching)]
    	        \ \\
    	        Mark $v \in L$ if $\sdisag{v}{L+v,R-v}{G_N} > \sdisag{v}{L-v,R+v}{G_N}$ and mark $v \in R$ similarly (flip the inequality).
    	        \\
    	        Switch assignments of all marked vertices from $L$ to $R$ or vice versa.
    	        \item Keep $\paren{L,R}$ if $\frest{G}{L}{\eps/10}$ is the smallest seen so far.
    	    \end{description}
    	   \item Return $\paren{L,R}$.
	\end{enumerate}
\end{tbox}
\noindent

The roles of \merging and \switching are roughly explained as follows.
When $\paren{S_L, S_R}$ is congruent with how $S$ would be partitioned in an optimal solution, \merging assigns most vertices to $L$ and $R$ in the same way they would have been assigned in said optimal solution.
We will show that these vertices do not change assignment when \switching.
The remaining vertices may have lots of disagreement with their current assignment in spite of having lots of agreement with $S_L$ or $S_R$.
\Switching rectifies this, minimizing the disagreement of these vertices with a large number of vertices (the ones that stay put, mentioned earlier).

% \chen{Add an example for why switching phase is necessary. }
\begin{remark}
\label{rmk:switch-neccessary}
It is worth mentioning that the \switching phase is essential: a vertex with high disagreement in the optimal solution may have $O(1)$-multiplicative disagreement overhead without the \switching phase (albeit the constant is small). Concretely, for the case of $\gamma<\eps=o(1)$, consider the following instance where $\card{L^*}=\card{R^*}=\frac{n}{2}$, and a high optimal-disagreement vertex $v$ has \plab edges to all vertices in $L^*$ and has \plab edges all but $\frac{n}{100}$ vertices in $R^*$. Clearly, there is roughly $\frac{n}{50}$ extra disagreement cost for $v$ is `shifted' to $R^{*}$ from $L^*$. During the \merging phase, since the algorithm may never see any \mlab neighbor for $v$, the vertex could end up in the partition induced by $R^*$. Since we can have $\Theta(\gamma n)$ such vertices (see \Cref{obs:vg-size}), if all of them are `shifted' to $R^*$, an extra cost of $O(\gamma n^2)=O(1)\cdot \frustind{G}$ breaks the $(1+\eps)$ guarantee.
\end{remark}

To continue, we state some new definitions and observations.
Then, we will look at the analysis of \merging followed by \switching, before wrapping this section up.

\begin{definition}
\label{def:v-good}
    Let $\grph{G}{V}{\dedge}$ be a complete signed graph, and let $\paren{L^*, R^*}$ be a frustration-minimizing partition of $V$ (that is, $\frust{G}{L^*,R^*} = \frustind{G}$).
    We define
    \begin{align*}
        \Vg = \set{v \in V \mid \disag{v}{L^*} \le \frac{n}{50}}.
    \end{align*}
    That is, $\Vg$ are vertices with low disagreement in an optimal solution.
    We also define the set of vertices with high disagreement in an optimal solution as $\Vb = V \setminus \Vg$.
\end{definition}

We first observe that the size of $\Vb$ can be at most $O(\gamma n)$, which means $\Vg$ has to be large.
\begin{observation}
\label{obs:vg-size}
    $\card{\Vg} \ge \paren{1 - 100\gamma}n$ and $\card{\Vb} < 100\gamma n$.
\end{observation}
\begin{proof}
    Suppose for a contradiction that $\card{\Vb} \ge 100 \gamma n$.
    Then 
    \begin{align*}
        \frustind{G}
        \ge
        &
        \frac{1}{2} \sum_{v \in \Vb} \disag{v}{L^*}
        \tag{\Cref{def:frustind}}
        \\
        >
        &
        \frac{1}{2} \cdot \card{\Vb} \cdot \frac{n}{50}
        \\
        \ge
        &
        \gamma n^2,
    \end{align*}
    which is a contradiction since $\frustind{G} = \gamma n^2$.
    The lower bound for $\card{\Vg}$ follows from $\Vb$ and $\Vg$ being complementary.
\end{proof}

Since we are in the case of $\gamma \lessapprox \eps$, $\Vg$ is a large set, and $\Vb$ is a small set.
We mentioned earlier that many vertices will have their assignments to $L$ and $R$ be congruent with that of an optimal solution after \merging, when the partition $\paren{S_L, S_R}$ of $S$ is congruent with the optimal solution.
We shortly show that $\Vg$ is indeed the set of such vertices, but first let us formalize congruence.
% \chengyuan{Does it make sense to define congruence before $V_g$?}
% vik: I think no. Deferring the definition to a point where precision is required is better. Having just an intuitive feeling for congruence up to this point is fine.

\begin{definition}
\label{def:congruent}
    Let $S \subseteq V$ and $\paren{A, B}$ be a partition of $S$.
    We say that $\paren{A, B}$ is \emph{congruent} with $\frustind{G}$ if $A = L^* \cap S$ and $B = R^* \cap S$.
    
    Furthermore, let $\algline$ be a (possibly intermediate) subroutine for an algorithm $\alg$ that outputs a partition $\paren{A, B}$ of $S$. We say $S$ is \emph{congruent} with $\frustind{G}$ after $\algline$ if we run $\alg$ till the end of $\algline$ and obtain a partition $\paren{A, B}$ of $S$ that is congruent with $\frustind{G}$.
    % We equivalently say $\paren{A, B}$ is \emph{congruent} with $\paren{L^*, R^*}$.
\end{definition}

We are now ready to delve into the guarantees of \merging and \switching.
Let us focus on the partition $\paren{S_L, S_R}$ of $S$ that is congruent with $\frustind{G}$.

\paragraph{Merging. }

Define the events
\begin{align*}
    \evemerge =& \set{\Vg \text{ is congruent with } \frustind{G} \text{ after \merging}},
    \\
    \evecon =& \set{\paren{S_L, S_R} \text{ is congruent with } \frustind{G}}.
\end{align*}

\begin{lemma}
\label{lem:good-in-place}
    %Let $S$ be a set of $100 \log{n}$ vertices sampled uniformly at random, and $\paren{S_L, S_R}$ be a partition of $S$ that is congruent with $\frustind{G}$.
    %Then, with high probability, $\Vg$ will be congruent with $\frustind{G}$ after \merging.
    With high probability, $\Vg$ will be congruent with $\frustind{G}$ after \merging.
    More specifically,
    \begin{align*}
        \prob{\evemerge \mid \evecon}
        \ge
        1 - n^{-9}.
    \end{align*}
\end{lemma}
\begin{proof}
    Let $\paren{L^*, R^*}$ be a partition of $V$ that is congruent with $\frustind{G}$.
    Note that the congruence of $\paren{S_L, S_R}$ means that $S_L \subseteq L^*$ and $S_R \subseteq R^*$.
    Suppose $v \in \Vg \setminus S$ is assigned to $L^*$ in the optimal partition.
    Then
    \begin{align*}
        \expect{\disag{v}{S_L+v, S_R} \mid \evecon}
        =
        &
        \sum_{s \in S}\frac{\disag{v}{L^*, R^*}}{n}
        \tag{Linearity of expectation}
        \\
        \le
        &
        \frac{1}{50}\card{S}.
        \tag{$v \in \Vg$ (\Cref{def:v-good})}
    \end{align*}
    Using a one-sided Chernoff bound \Cref{prop:additive-chernoff},
    \begin{align*}
        & \prob{v \text{ assigned to wrong part} \mid \evecon}\\
        =
        &
        \prob{\disag{v}{S_L+v, S_R} > \disag{v}{S_L, S_R+v} \mid \evecon}
        \\
        =
        &
        \prob{\disag{v}{S_L+v, S_R} > \card{S}/2 \mid \evecon}
        \tag{$\disag{v}{S_L+v, S_R} + \disag{v}{S_L, S_R+v} = \card{S}$}
        \\
        \le
        &
        \eexp{-\frac{144}{324}\cdot \card{S}}
        \\
        \le
        &
        n^{-10}.
    \end{align*}
    
    A similar argument holds for $v \in \Vg \setminus S$ assigned to $R^*$ in the optimal partition.
    Finally, $v \in \Vg \cap S$ are assigned to the correct part by $\evecon$.
    Thus, 
    \begin{align*}
        \prob{\evemerge \mid \evecon}
        =
        &
        1 - \prob{\cup_{v \in \Vg} v \text{ assigned to wrong part} \mid \evecon}
        \\
        \ge
        &
        1 - \sum_{v \in \Vg} \prob{v \text{ assigned to wrong part} \mid \evecon}
        \tag{Union bound}
        \\
        \ge
        &
        1 - n^{-9}.
    \end{align*}
\end{proof}

\paragraph{Switching. }
Define the event
\begin{align*}
    \eveswitch = \set{\Vg \text{ is congruent with } \frustind{G} \text{ after \switching}}.
\end{align*}

\begin{lemma}
\label{lem:good-still-in-place}
    %Let $S$ be a set of $100 \log{n}$ vertices sampled uniformly at random, and $\paren{S_L, S_R}$ be a partition of $S$ that is congruent with $\frustind{G}$.
    %Let $N_v$ be a set of $100^3 \cdot \frac{\log{n}}{\eps^2}$ neighbors of $v$ for all $v \in V$, and let $\grph{G_N}{V}{\cup_v N_v}$ be a subgraph of $G$.
    %Then, with high probability, $\Vg$ will be congruent with $\frustind{G}$ after \switching using $G_N$.
    With high probability, $\Vg$ will be congruent with $\frustind{G}$ after \switching.
    That is,
    \begin{align*}
        \prob{\eveswitch \mid \evemerge, \evecon}
        \ge
        1 - n^{-7}.
    \end{align*}
\end{lemma}
\begin{proof}
    Let $\paren{L^*, R^*}$ be a partition of $V$ that is congruent with $\frustind{G}$.
    Note that the congruence of $\paren{S_L, S_R}$ means that $S_L \subseteq L^*$ and $S_R \subseteq R^*$.
    Further, let $\paren{L, R}$ be the partition of $V$ after \merging and before \switching, conditioned on $\evemerge, \evecon$.
    
    Suppose $v \in \Vg$ is assigned to $L$ after \merging and before \switching.
    $v$ will switch if $\sdisag{v}{L,R}{G_N} > \sdisag{v}{L-v,R+v}{G_N}$.
    The expected disagreement for $v$ towards its current assignment is
    \begin{align*}
        &
        \expect{\sdisag{v}{L,R}{G_N} \mid \evemerge, \evecon}
        \\
        =
        &
        \expect{\sdisag{v}{L \cap \Vg, R \cap \Vg}{G_N} + \sdisag{v}{L \cap \Vb + v, R \cap \Vb}{G_N} \mid \evemerge, \evecon}
        \tag{$\Vg \cup \Vb = V$}
        \\
        \le
        &
        \paren{\frac{1}{50} + \frac{\card{\Vb}}{n}}\card{N_v}
        \tag{Maximizing $\sdisag{v}{L \cap \Vb + v, R \cap \Vb}{G_N}$}
        \\
        \le
        &
        \paren{\frac{1}{50} + 100 \gamma} \card{N_v}
        \tag{\Cref{obs:vg-size}}
        \\
        \le
        &
        \frac{\card{N_v}}{4}.
        \tag{$\gamma \le \frac{\eps}{100^4}$}
    \end{align*}
    By a one-sided Chernoff bound \Cref{prop:additive-chernoff},
    \begin{align*}
        &\prob{v \text{ switches} \mid \evemerge, \evecon}\\
        =
        &
        \prob{\sdisag{v}{L,R}{G_N} > \sdisag{v}{L-v,R+v}{G_N} \mid \evemerge, \evecon}
        \\
        =
        &
        \prob{\sdisag{v}{L,R}{G_N} > \card{N_v}/2 \mid \evemerge, \evecon}
        \tag{$\sdisag{v}{L,R}{G_N} + \sdisag{v}{L-v,R+v}{G_N} = \card{N_v}$}
        \\
        \le
        &
        \eexp{-\frac{\card{N_v}}{12}}
        \\
        \le
        &
        n^{-8}.
    \end{align*}
    A similar argument holds for $v \in \Vg$ assigned to $R$ after \merging and before \switching.
    Thus,
    \begin{align*}
        \prob{\eveswitch \mid \evemerge, \evecon}
        =
        &
        1 - \prob{\cup_{v \in \Vg} v \text{ switches} \mid \evemerge, \evecon}
        \\
        \ge
        &
        1 - n^{-7}.
        \tag{Union bound}
    \end{align*}
\end{proof}

We have shown so far that the assignment $\Vg$ is congruent with $\frustind{G}$ with high probability, but we have not yet shown what we had alluded to earlier: \switching will minimize disagreement between $\Vg$ and $\Vb$.
To see this, let us make two more definitions.

\begin{definition}[\Nsens, \Nres]
\label{def:nsens-nres}
    Let $\paren{L, R}$ be the partition of $V$ after \merging and before \switching, conditioned on $\evemerge, \evecon$.
    Let $v \in \Vb$.
    %Let $L_v = L \cap N_v$ and $R_v = R \cap N_v$ where $v \in \Vb$.
    \begin{itemize}
        \item $v$ is \nsens if
            \begin{align*}
                \card{\disag{v}{L \cap \Vg + v, R \cap \Vg} - \disag{v}{L \cap \Vg, R \cap \Vg + v}} < \frac{\eps}{100^2}n.
            \end{align*}
        \item $v$ is \nres otherwise.
    \end{itemize}
\end{definition}

The idea behind \nsens vertices is that there is no clear choice for whether they should belong in $L$ or $R$ after \switching.
However, the lack of there being a clear choice suggests that it does not matter which side they end up on; we can absorb the extra cost of suboptimal assignments.

\begin{claim}
\label{claim:noise-sens}
    Let $\paren{L, R}$ be the partition of $V$ after \switching, conditioned on $\eveswitch$, $\evemerge$, $\evecon$.
    Let $\paren{L^*, R^*}$ be a partition of $V$ that is congruent with $\frustind{G}$.
    Conditioned on $\eveswitch$, $\evemerge$, $\evecon$, after \switching,
    \begin{align*}
        &
        \sum_{\substack{v \in \Vb: \\v \text{ is \nsens}\\v \in L}} \disag{v}{L \cap \Vg + v, R \cap \Vg}
        +
        \sum_{\substack{v \in \Vb: \\v \text{ is \nsens}\\v \in R}} \disag{v}{L \cap \Vg, R \cap \Vg + v}
        \\
        \le
        &
        \sum_{\substack{v \in \Vb: \\v \text{ is \nsens}\\v \in L^*}} \disag{v}{L^* \cap \Vg + v, R^* \cap \Vg}
        +
        \sum_{\substack{v \in \Vb: \\v \text{ is \nsens}\\v \in R^*}} \disag{v}{L^* \cap \Vg, R^* \cap \Vg + v}
        +
        \Eebg,
    \end{align*}
    where $\Eebg = \frac{\eps}{100}\gamma n^2$. Put otherwise, the total error from assigning \nsens vertices to the wrong side is no more than $\Eebg$.
\end{claim}
\begin{proof}
    Firstly, by \Cref{def:nsens-nres}, $\disag{v}{L \cap \Vg + v, R \cap \Vg}$ and $\disag{v}{L \cap \Vg, R \cap \Vg + v}$ differ by no more than $\frac{\eps}{100^2}n$.
    %Therefore, by \Cref{obs:vg-size}, $\disag{v}{L \cap \Vg + v, R \cap \Vg}$ and $\disag{v}{L \cap \Vg, R \cap \Vg + v}$ differ by no more than $\frac{\eps}{100^2}n$.
    
    Secondly, there are at most $100 \gamma n$ vertices in $\Vb$ by \Cref{obs:vg-size}.
    
    Multiplying the two quantities, $\frac{\eps}{100^2}n$ with $100 \gamma n$, gives us $\Eebg$, an upper bound on the total error in assigning \nsens vertices to the wrong side.
\end{proof}

\Nres vertices, on the other hand, have a clearer preference for being on one side or the other after conditioning on $\evemerge, \evecon$.
The roadblock is that such a preference only becomes clear when we have total information of all incident edges.
\Cref{lem:noise-resist} shows that a sufficiently large sample of incident edges will suffice.
Define the event
% \begin{align*}
%     \everes = \set{\text{All \nres vertices are congruent with } \frustind{G} \text{ after \switching}}.
% \end{align*}
\begin{align*}
    & \everes \\
    = & \set{\text{All \nres\ } v\in\Vb \text{ minimize } \disag{v}{L \cap \paren{\Vg +v},R \cap \paren{\Vg +v}} \text{ after \switching}}.
\end{align*}
% \chengyuan{Is it possible to fit in one line?}

\begin{lemma}
\label{lem:noise-resist}
    %Let $S$ be a set of $100 \log{n}$ vertices sampled uniformly at random, and $\paren{S_L, S_R}$ be a partition of $S$ that is congruent with $\frustind{G}$.
    %Let $N_v$ be a set of $100^3 \frac{\log{n}}{\eps^2}$ neighbors of $v$ for all $v \in V$, and let $\grph{G_N}{V}{\cup_v N_v}$ be a subgraph of $G$.
    %Then, with high probability, \nres vertices will be congruent with $\frustind{G}$ after \switching using $G_N$.
    With high probability, \nres vertices will be congruent with $\frustind{G}$ after \switching.
    More specifically,
    \begin{align*}
        \prob{\everes \mid \evemerge, \evecon}
        \ge
        1 - n^{-9}.
    \end{align*}
\end{lemma}
\begin{proof}
    Let $\paren{L^*, R^*}$ be a partition of $V$ that is congruent with $\frustind{G}$.
    Note that the congruence of $\paren{S_L, S_R}$ means that $S_L \subseteq L^*$ and $S_R \subseteq R^*$.
    Further, let $\paren{L, R}$ be the partition of $V$ after \merging and before \switching, conditioned on $\evemerge, \evecon$.
    Let $v \in \Vb$ be a \nres vertex.
    
    Suppose that $\disag{v}{L \cap \Vg + v, R \cap \Vg} \ge \disag{v}{L \cap \Vg, R \cap \Vg + v}$.
    That is, with respect to $\Vg$, $v$ creates more disagreements when assigned to $L$ (we want $v$ to end up assigned to $R$).
    Then
    \begin{align*}
        &
        \expect{\sdisag{v}{L+v, R-v}{G_N} \mid \evemerge, \evecon}
        \\
        =
        &
        \sum_{u \in N_v} \prob{u \in \Vg} \cdot \prob{\paren{u,v} \text{ disagree} \lmid u \in \Vg} + \prob{u \in \Vb} \cdot \prob{\paren{u,v} \text{ disagree} \lmid u \in \Vb}
        \tag{Linearity of expectation, Law of Total Probability}
        \\
        \ge
        &
        \sum_{u \in N_v} \prob{u \in \Vg} \cdot \prob{\paren{u,v} \text{ disagree} \lmid u \in \Vg}
        \\
        \ge
        &
        \paren{\frac{1}{2} + \frac{\eps}{2 \cdot 100^2}}\frac{\card{\Vg}}{n}\card{N_v}
        % \tag{$\disag{v}{L \cap \Vg + v, R \cap \Vg} \ge \disag{v}{L \cap \Vg, R \cap \Vg + v}$}
        \tag{$v$ is \nres}
        \\
        \ge
        &
        \paren{\frac{1}{2} + \frac{\eps}{2 \cdot 100^2}}\paren{1 - 100 \gamma}\card{N_v}
        \tag{\Cref{obs:vg-size}}
        \\
        \ge
        &
        \paren{\frac{1}{2} + \frac{\eps}{2 \cdot 100^2} - 100 \gamma}\card{N_v}
        \tag{$\frac{1}{2} + \frac{\eps}{2 \cdot 100^2} < 1$}
        \\
        \ge
        &
        \paren{\frac{1}{2} + \frac{\eps}{4 \cdot 100^2}}\card{N_v}.
        \tag{$\gamma < \frac{\eps}{100^4}$}
    \end{align*}
    
    Using an additive Chernoff bound \Cref{prop:additive-chernoff}, the probability $v$ is erroneously assigned to $L$ is bounded by
    \begin{align*}
        &
        \prob{v \text{ assigned to } L \mid \evemerge, \evecon}
        \\
        \le
        &
        \prob{\card{\sdisag{v}{L+v, R-v}{G_N} - \expect{\sdisag{v}{L+v, R-v}{G_N}}} \ge \frac{\eps}{4 \cdot 100^2}\card{N_v} \mid \evemerge, \evecon}
        \\
        <
        &
        2 \eexp{-\frac{\card{N_v}^2\eps^2}{2 \cdot 100^2 \cdot \card{N_v}}}
        % \\
        % =
        % &
        % 2 \eexp{-50 \log{n}}
        % \tag{$\card{N_v} = 100^3 \frac{\log{n}}{\eps^2}$}
        % \\
        <
        % &
        n^{-10}.
    \end{align*}
    
    A similar argument holds when $\disag{v}{L \cap \Vg + v, R \cap \Vg} \le \disag{v}{L \cap \Vg, R \cap \Vg + v}$.
    Thus,
    \begin{align*}
        \prob{\everes \mid \evemerge, \evecon}
        =
        &
        1 - \prob{\cup_{v \in \Vb} v \text{ wrongly moved} \mid \evemerge, \evecon}
        \\
        \ge
        &
        1 - n^{-9}.
        \tag{Union bound}
    \end{align*}
\end{proof}

\paragraph{Everything together. }
Finally, we wrap this section up; we show an approximation guarantee in the $\gamma \lessapprox \eps$ regime, and combine it with that of the $\gamma \gtrapprox \eps$ regime to cap off the streamification of \cite{giotis2005correlation}.
We omit the complexity analysis as a very similar argument is presented in \Cref{sec:good-time-frus}, which also presents a strictly better algorithm than the one here.
%, will go through what would be a very similar argument in order to prove \Cref{thm:frus-algo-main}.

\begin{claim}
\label{claim:whp-mindis-succ}
    Events $\evecon, \evemerge, \eveswitch, \everes$ occur simultaneously with high probability.
\end{claim}
\begin{proof}
    This is shown by the chain rule; independence of $\eveswitch$ and $\everes$; and \Cref{lem:good-in-place}, \Cref{lem:good-still-in-place}, and \Cref{lem:noise-resist}.
\end{proof}

\begin{lemma}
\label{lem:mindis}
    Suppose $\evecon, \evemerge, \eveswitch, \everes$ are all satisfied.
    Let $\paren{L,R}$ be the partition of $V$ returned by our algorithm.
    Then, $\frust{G}{L,R} < \paren{1 + \eps}\cdot \frustind{G}$.
\end{lemma}
\begin{proof}
    Let $\paren{L',R'}$ be the partition of $V$ when $\evecon, \evemerge, \eveswitch, \everes$ are all satisfied (note that $\paren{L',R'}$ may actually differ from $\paren{L,R}$).
    
    First, express
    \begin{align*}
        \frust{G}{L',R'} = \Egg + \Ebg + \Eebg + \Ebb,
    \end{align*}
    where
    \begin{itemize}
        \item $\Egg$ is the number of disagreements within $V_g$,
        \item $\Ebg$ is the number of disagreements between $V_b$ and $V_g$, and
        \item $\Ebb$ is the number of disagreements within $V_b$.
    \end{itemize}
    
    We will express these summands in terms of $\frustind{G}_{g,g}$ and $\frustind{G}_{b,g}$ where the former is the number of disagreements within $V_g$, and the latter is the number of disagreements between $V_b$ and $V_g$, both with respect to the optimal partition.
    
    \begin{itemize}
        \item Since $\eveswitch$ is satisfied, we know that $\Egg = \frustind{G}_{g,g}$.
        \item Using \Cref{claim:noise-sens}, \Cref{lem:noise-resist}, and the assumption that $\eveswitch$ is satisfied, we deduce that $\Ebg + \Eebg \le \frustind{G}_{b,g} + \frac{\eps}{100}\gamma n^2$.
        \item Finally, using \Cref{obs:vg-size}, we know that $\Ebb \le (100 \gamma n)^2$.
    \end{itemize}
    
    Thus,
    \begin{align*}
        \Egg + \Ebg + \Eebg + \Ebb
        \le
        &
        \frustind{G}_{g,g} + \frustind{G}_{b,g} + \frac{\eps}{100}\gamma n^2 + (100 \gamma n)^2
        \\
        \le
        &
        \frustind{G} + \frac{\eps}{100}\gamma n^2 + (100 \gamma n)^2
        \tag{$\frustind{G}_{g,g} + \frustind{G}_{b,g} \le \frustind{G}$}
        \\
        =
        &
        \frustind{G} + \frac{\eps}{100} \frustind{G} + 100^2 \gamma \frustind{G}
        \tag{$\gamma n^2 = \frustind{G}$}
        \\
        =
        &
        \paren{1 + \frac{\eps}{100} 100^2 \gamma} \frustind{G}.
    \end{align*}
    
    The frustration of the returned partition is hence upper bounded by
    \begin{align*}
        \frust{G}{L}{R}
        \le
        &
        \frac{1 + \eps/10}{1 - \eps/10} \frust{G}{L'}{R'}
        \tag{Frustration sparsifier}
        \\
        \le
        &
        \frac{1 + \eps/10}{1 - \eps/10} \cdot \paren{1 + \frac{\eps}{100}\cdot 100^2 \gamma} \frustind{G}.
    \end{align*}
    
    Finally, using $\gamma \le \frac{\eps}{100^4}$, we have
    \begin{align*}
        \frac{1 + \eps/10}{1 - \eps/10} \cdot \paren{1 + \frac{\eps}{100}\cdot 100^2 \gamma} < \paren{1 + \eps},
        % \tag{Computer check}
    \end{align*}
    completing the proof.
\end{proof}

Combining \Cref{prop:maxag} (repeating $O\paren{\log{n}}$ times and taking the best partition to get a success guarantee with high probability), \Cref{claim:whp-mindis-succ}, \Cref{lem:mindis}, and one more call to a frustration sparsifier (to determine the better of the outputs returned by the algorithm for $\gamma \gtrapprox \eps$ and the algorithm for $\gamma \lessapprox \eps$), we have a semi-streaming algorithm for \Cref{prob:min-frust-partition} that succeeds with high probability.

Unfortunately, the running time of this algorithm is $\Theta\paren{n^{101}}$, where the inefficiency stems from having to iterate over all partitions of a sample of size $100 \log{n}$.
Optimizing over the constant is to little avail; under our analysis, if we use a sample of size $c \cdot \log{n}$ (where $c = 100$ in our presentation), $c$ must be larger than the denominator in the exponent of the Chernoff bounds used to beat a union bound over $n$ vertices. As such, even if we optimize the constant, we should not expect efficiency better than $\Omega(n^c)$ for some $c>2$. In other words, the algorithm as it stands gives us poly-time efficiency, but might still be insufficient in large-scale applications.% \chengyuan{Sepehr has a comment here which I don't think is a formal claim and necessary to add.}

The next section is a sequel, where we present an algorithm that goes around this limitation and, moreover, we give a closer look at time and space complexity thereof in order to prove \Cref{thm:frus-algo-main} stated in beginning of this section.

% !TeX root = main.tex 
%!TEX root = main.tex
\section{Frustration-minimizing Partition in Nearly-linear Time}
\label{sec:good-time-frus}

In this section, we build upon the algorithm in \Cref{sec:frustration-index} to find a $(1+\eps)$-approximation of the frustration-minimizing partition more efficiently.
We present a semi-streaming algorithm that runs in $O\paren{\frac{n\cdot \log^3(n)}{\eps^2}}$ space and $O\paren{\frac{n^2\log^3{n}}{\eps^2} + n \log{n} \cdot \paren{1/\eps}^{O\paren{\eps^{-4}}}}$ time (nearly-linear in the size of the input graph), completing \Cref{thm:frus-algo-main}.

Our presentation of the algorithm here will not be in the semi-streaming model as we believe the algorithm when run offline, in a more classical model of computation, is in and of itself of independent interest; it is, to the best of our knowledge, the first Efficient Polynomial-time Approximation Scheme (EPTAS) for \Cref{prob:min-frust-partition} and 2-correlation clustering, not to mention one that is nearly-linear in the input.
We then point out why the algorithm is easily ported to the semi-streaming setting (namely, by observing that it reads the input once, and analyzing its space complexity).
\Cref{subsec:offline_eptas} describes this algorithm in the offline setting.
\Cref{subsec:stream_eptas} brings things back to the semi-streaming model, completing \Cref{thm:frus-algo-main}.

For the remainder of this section, suppose $\frustind{G} = \gamma n^2$.
Recall that the \GG algorithm runs two algorithms in parallel.
The $\paren{\gamma \gtrapprox \eps}$ algorithm has approximation guarantees when $\gamma > \eps/100^4$.
The $\paren{\gamma \lessapprox \eps}$ algorithm, on the other hand, has approximation guarantees when $\gamma \le \eps/100^4$.
We improve on the $\paren{\gamma \lessapprox \eps}$ algorithm which, in its current incarnation, is the obstruction to achieving a nearly-linear runtime.

\subsection{An Offline Nearly-Linear EPTAS for Low Frustration Index}
% \subsection{An Offline Nearly-Linear EPTAS when $\gamma \lessapprox \eps$}
\label{subsec:offline_eptas}

Recall that the \GG algorithm when $\gamma \lessapprox \eps$ iterates over all partitions of a set of size $O\paren{\log{n}}$, where the hidden constant is much larger than $2$.
One of the takeaways of this section is that iterating over partitions of an exponentially smaller sample (one of size $O\paren{\log\log{n}}$) would not introduce so much error that our previous analysis in \Cref{sec:frustration-index} does not work; we just have to be careful with how we construct the overall partition, starting from a partition of this smaller sample.

Now we formally state the main result pertaining to the offline algorithm:
\begin{theorem}[Formalization of \Cref{rst:EPTAS-alg}]
\label{thm:offline-eptas}
    There is a randomized algorithm that, given a complete signed graph $G = \paren{V, \Ep \cup \Em}$, and for any $\eps \in \paren{0,1}$, with high probability returns a partition $(L,R)$ of $V$ where
    \begin{align*}
        \frust{G}{L} \le (1 + \eps) \cdot \frustind{G}.
    \end{align*}
    Moreover, the algorithm has runtime $O\paren{\frac{n^2\log^3{n}}{\eps^2} + n \log{n} \cdot \paren{1/\eps}^{O\paren{\eps^{-4}}}}$.\footnote{
    % \todo[inline]{Assignee: Chen; Task: mention that we actually merged $n^2\log^{100}(n)$ and $\frac{n\log^{100}(n)}{\eps^2}$ -- bound actually better.}
    We remark that the $O(\frac{n^2\log^3{n}}{\eps^2})$ runtime actually comes from a slight overestimation of two parts: an $\eps$-cut sparsifier that runs in $O(n^2\log^3{n})$ time as in \Cref{prop:eps-cutsparse-algs} and a subroutine that runs in $O(\frac{n\log^{101}(n)}{\eps^2})$ time in our algorithm.
    We hide the latter term by multiplying the former with $O(1/\eps^2)$ for simplicity, and our true runtime is stronger.
    }
\end{theorem}

\paragraph{The $\paren{\gamma \lessapprox \eps}$ Algorithm. }
The algorithm here works very similarly to that in \Cref{sec:frustration-index}.
There, we sampled vertices $S$ where $\card{S} = O\paren{\log{n}}$ and iterated over all partitions of $S$.
Here, we still sample $S$.
However, instead of iterating over all its partitions, we do the following.
Sample vertices $S'$ where $\card{S'} = O\paren{\log\log{n}}$.
Iterate over all partitions of $S'$ (instead of $S$), and let $\paren{S'_L,S'_R}$ be any such partition.
We then use $\paren{S'_L,S'_R}$ to construct a partition $\paren{S_L, S_R}$ of $S \cup S'$, in a way not unlike the \merging step in the \GG algorithm; we aptly call this step \minimerging.
Now the algorithm goes just like before: we run \merging on $\paren{S_L, S_R}$ to construct a partition $\paren{L,R}$ of $V$, and \switching to improve the frustration of this partition.

\begin{tbox}
	\textbf{Nearly-Linear Time EPTAS solving \Cref{prob:min-frust-partition}} for $\grph{G}{V}{\dedge}$ and $\eps \in \paren{0,1}$, when $\gamma <\eps/100^4$:
	\begin{enumerate}
	    \item\label{line:sampling-offline-1} Sample $2000 \log{n}$ vertices uniformly at random from $V$. Call this set $S$ and let $\grph{G_{S}}{V}{E\paren{S} \cup E\paren{S, V \setminus S}}$.
	    \item\label{line:sampling-offline-2}  Sample $100\log\log{n}$ vertices uniformly at random from $S$. Call this set $S'$, the \emph{seed-set}, and let $\grph{G_{S'}}{V}{E\paren{S'} \cup E\paren{S', V \setminus S'}}$.
	    \item\label{line:sampling-offline-3} Sample $100^3 \cdot \frac{\log{n}}{\eps^2}$ vertices uniformly at random for each vertex $v$. Call these sets $N_v$ and let $\grph{G_{N}}{V}{\bigcup_v E\paren{v, N_v}}$.
	    \item\label{line:cut-sparsifier} Construct $\frest{G}{\cdot}{\eps/10}$, a frustration sparsifier of $G$.
    	\item\label{line:best-partition-enum} For every partition $\paren{S'_L, S'_R}$ of $S'$
    	    \begin{description}
    	        \item[(\Minimerging on $S'$)]
    	        \ \\
    	        For every $v \in S \setminus S'$, assign $v$ to $S_L$ if $\sdisag{v}{S'_L+v,S'_R}{G_{S'}} < \sdisag{v}{S'_L,S'_R+v}{G_{S'}}$, and $S_R$ otherwise.
    	        \\
    	        $S'_L$ is assigned to $S_L$ and $S'_R$ is assigned to $S_R$.
    	        \item[(\Merging on $S$)]
    	        \ \\
    	        For every $v \in V \setminus S$, assign $v$ to $L$ if $\sdisag{v}{S_L+v,S_R}{G_S} < \sdisag{v}{S_L,S_R+v}{G_S}$, and $R$ otherwise.
    	        \\
    	        $S_L$ is assigned to $L$ and $S_R$ is assigned to $R$.
    	        \item[(\Switching)]
    	        \ \\
    	        Mark $v \in L$ if $\sdisag{v}{L+v, R-v}{G_N} > \sdisag{v}{L-v, R+v}{G_N}$ and mark $v \in R$ similarly (flip the inequality).
    	        \\
    	        Switch assignments of all marked vertices from $L$ to $R$ or vice versa.
    	        \item Keep $\paren{L,R}$ if $\frest{G}{L}{\eps/10}$ is the smallest seen so far.
    	    \end{description}
    	   \item Return $\paren{L,R}$.
	\end{enumerate}
\end{tbox}
\noindent

We first aim to show an approximation guarantee for the above algorithm.
The main thing to show is that, after \merging, $\Vg$ (recall: \Cref{def:v-good}) is congruent (recall: \Cref{def:congruent}) with $\frustind{G}$ with high probability.
We first show that \minimerging constructs, in some iteration, a partition that is close to being congruent with $\frustind{G}$.
This almost-congruent partition, we then show, is sufficient for the \merging step to succeed.
Contrast this with the \GG algorithm, where we insisted on a congruent partition (as opposed to an almost-congruent one) to ensure the success of \merging.
% \paragraph{Time Complexity.} The runtime of the above algorithm is $O\big( n^2 (\log n)^{100} \big)$. To enumerate all partitions of $S'$, the algorithm runs for $(\log n)^{100}$ iterations. For each candidate partition, the merging phase takes $O(n\log n)$ time and switching phase takes $O(n^2)$ time to check whether to switch. Last, computing $\frust{G}{L}$ takes no longer than $O(n^2)$ time. Therefore, the total runtime is $O(n^2 (\log n)^{100} + \frac{n}{\eps}\cdot (\frac{1}{\eps})^{\eps^{-2}})$.

% \paragraph{Approximation Guarantee. }
% The analysis hinges on some definitions and observations (\Cref{def:congruent}, \Cref{def:v-good}, \Cref{obs:vg-size}) already used in \Cref{subsec:poly-algo}. 

% There essentially remains two steps to show: First, after \minimerging on $S'$, there exists one partition $(S_L, S_R)$ that is a good approximation to the partition of $S$ that is congruent with the optimum; Second, the \merging on $S$ and \switching will succeed constructing a good partition $(L,R)$ even with the approximate partition on $(S_L, S_R)$. We close the first gap by proving \Cref{lemma:s-log-congruent}, and show that \merging with a nearly-optimal partition of $S$ ensures that most vertices are assigned correctly in \Cref{lemma:all-vg-merge}. Given that \merging is successful, the correctness of \switching follows the same proof as in \Cref{subsec:poly-algo}.

\paragraph{\Minimerging. }
Define the events
\begin{align*}
   & \evemini = \set{\Vg \cap \paren{S \cup S'} \text{ is congruent with } \frustind{G} \text{ after \minimerging}},\\
   & \eveconl = \set{\paren{S'_L, S'_R} \text{ is congruent with } \frustind{G}}.
\end{align*}

\begin{claim}
\label{claim:minimerge-vg}
    % Let $S$ be the set of $2000\log{n}$ vertices sampled uniformly at random, $S'$ be the seed-set of $100\log{\log{n}}$ vertices sampled uniformly at random,
    % and 
    Let $(S'_L, S'_R)$ be a partition of $S'$ congruent with \frustind{G}.
    Then, with $1-o\paren{1}$ probability, the vertices in $S \cup S'$ that are also in $\Vg$ will be congruent with $\frustind{G}$ after \minimerging, i.e.
    \begin{align*}
        \prob{\evemini \lmid \eveconl}
        \ge
        1 - \paren{\frac{1}{\log{n}}}^{9}.
    \end{align*}
\end{claim}
\begin{proof}
    % The proof is essentially the same as that of \Cref{lem:good-in-place}, but because we union bound over $O\paren{\log{n}}$ vertices here, our seed-set need only be of size $\Omega\paren{\log\log{n}}$.
    % \todo[inline]{vik: Proof is written in source, but commented out. I think this is good enough to be honest.}
    
    Let $\paren{L^*, R^*}$ be a partition of $V$ that is congruent with $\frustind{G}$.
    Note that the congruence of $\paren{S'_L, S'_R}$ means that $S'_L \subset L^*$ and $S'_R \subset R^*$.
    Suppose $v \in \Vg \cap \paren{S \setminus S'}$ is assigned to $L^*$ in the optimal partition.
    Then
    \begin{align*}
        % &
        \expect{\disag{v}{S'_L, S'_R} \lmid \eveconl}
        % \\
        =
        &
        \sum_{s \in S'}\frac{\disag{v}{L^*, R^*}}{n}
        \tag{Linearity of expectation}
        \\
        \le
        &
        \frac{1}{50}\card{S'}.
        \tag{$v \in \Vg$ (\Cref{def:v-good})}
    \end{align*}
    
    Using a one-sided Chernoff bound (\Cref{prop:additive-chernoff}),
    \begin{align*}
        &
        \prob{v \text{ assigned to wrong part in }\minimerging \lmid \eveconl}
        \\
        =
        &
        \prob{\disag{v}{S'_L+v, S'_R} > \disag{v}{S'_L, S'_R+v} \lmid \eveconl}
        \\
        =
        &
        \prob{\disag{v}{S'_L+v, S'_R} > \card{S'}/2 \lmid \eveconl}
        \tag{$\disag{v}{S'_L+v, S'_R} + \disag{v}{S'_L, S'_R+v} = \card{S'}$}
        \\
        \le
        &
        \eexp{-\frac{144}{324}\card{S'}}
        \\
        \le
        &
        \paren{\frac{1}{\log{n}}}^{10}.
    \end{align*}
    
    A similar argument holds for $v \in \Vg \cap \paren{S \setminus S'}$ assigned to $R^*$ in the optimal partition.
    Finally, $v \in \Vg \cap S'$ are assigned to the correct part by $\eveconl$.
    Thus, 
    \begin{align*}
        &
        \prob{\evemini \mid \eveconl}
        \\
        =
        &
        1 - \prob{\cup_{v \in \Vg \cap \paren{S \cup S'}} v \text{ assigned to wrong part} \mid \evecon}
        \\
        \ge
        &
        1 - \sum_{v \in \Vg \cap \paren{S \cup S'}} \prob{v \text{ assigned to wrong part} \mid \evecon}
        \tag{Union bound}
        \\
        \ge
        &
        1 - \paren{\frac{1}{\log{n}}}^{9},
    \end{align*}
    as desired.
\end{proof}

Define the event
\begin{align*}
    \eveminif = \set{\text{ At least }\frac{9}{10} \text{ of } \paren{S_L, S_R} \text{ is congruent with } \frustind{G} \text{ after \minimerging}}.
\end{align*}

\begin{lemma}
\label{lemma:s-log-congruent}
    % Let $S$ be the set of $2000\log{n}$ vertices sampled uniformly at random, $S'$ be the seed-set of $100\log{\log{n}}$ vertices sampled uniformly at random, and 
    Let $(S'_L, S'_R)$ be a partition of $S'$ congruent with \frustind{G}, and suppose $\evemini$ happens.
    Then, with high probability, $\frac{9}{10}$ of the vertices in $S$ will be congruent with $\frustind{G}$ after \minimerging.
    That is,
    \begin{align*}
        \prob{\eveminif \lmid \evemini, \eveconl}
        \ge
        1 - n^{-9}.
    \end{align*}
\end{lemma}
\begin{proof}
    The idea is to show that sufficiently many of the vertices in $S$ are from $\Vg$.
    Everything then follows since $\evemini$ being satisfied is part of our premise.
    
    Let $X_{S}$ be a random variable counting the number of vertices in $S$ that are also in $\Vg$.
    The expected value of $X_{S}$ is
    \begin{align*}
        \expect{X_{S}}
        =&
        \sum_{s \in S} \prob{s \in \Vg}
        \tag{Linearity of expectation}
        \\
        \ge&
        \card{S}(1-100\gamma).
        \tag{\Cref{obs:vg-size}}
    \end{align*}
    
    Observe that, since $\gamma \le \eps/100^4 \le 1/100^4$, we have $\expect{X_S} - \frac{1}{20}\card{S} \ge \frac{9}{10}\card{S}$.
    Thus,
    \begin{align*}
        \prob{X_S < \frac{9}{10}\card{S}}
        \le
        &
        \prob{\card{X_S - \expect{X_S}} > \frac{1}{20}\card{S}}
        \\
        <
        &
        2 \cdot \eexp{-\frac{\card{S}^2}{200 \cdot \card{S}}}
        \tag{Additive Chernoff bound \Cref{prop:additive-chernoff}}
        \\
        \le
        &
        n^{-9}.
    \end{align*}
    
    We complete the proof by observing that
    \begin{align*}
        \prob{\eveminif \lmid \evemini, \eveconl}
        &\ge \prob{X_S \ge \frac{9}{10}\card{S}} \\
        &= \paren{1 - \prob{X_S < \frac{9}{10}\card{S}}} \\
        & \ge
        1 - n^{-9}.
    \end{align*}
\end{proof}

\paragraph{Merging. }
Define the event
\begin{align*}
    \evemerge = \set{\Vg \text{ is congruent with } \frustind{G} \text{ after } \merging}.
\end{align*}

\begin{lemma}
\label{lemma:all-vg-merge}
    % Let $S$ be the set of $2000\log{n}$ vertices sampled uniformly at random, $S'$ be the seed-set of $100\log{\log{n}}$ vertices sampled uniformly at random, and 
    Conditioning on $\eveminif, \evemini$ happening, with high probability, all $v \in V_g$ are congruent with $\frustind{G}$ after \minimerging and \merging.
    That is,
    \begin{align*}
        \prob{\evemerge \lmid \eveminif, \evemini, \eveconl} \ge 1 - n^{-9}
    \end{align*}
\end{lemma}
\begin{proof}
    Let $\paren{L^*, R^*}$ be a partition of $V$ that is congruent with $\frustind{G}$.
    Note that the congruence of $\paren{S'_L, S'_R}$ means that $S'_L \subset L^*$ and $S'_R \subset R^*$.
    
    For every $v \in V_g$, there are three cases to consider: (1) $v \in S'$. (2) $v \in S \setminus S'$. (3) $v$ is not sampled in the algorithm, namely $v \not\in S \cup S'$.
    
    (Case 1). Suppose $v \in S'$.
    Then, since $\eveconl$ happens, $v$ is correctly assigned to $(L,R)$.
    
    (Case 2). Suppose, on the other hand, $v \in S \setminus S'$.
    Then, since $\evemini$ happens, $v$ is correctly assigned to $(L,R)$.
    
    (Case 3). Finally, suppose that $v \not\in S$.
    Let $\eve{}$ be the event that $\eveminif, \evemini, \eveconl$ all happen.
    Suppose $v$ is assigned to $L^*$ in the optimal partition.
    Then
    \begin{align*}
        \expect{\disag{v}{S_L, S_R} \lmid \eve{}}
        =
        &
        \sum_{s \in S \cup S'} \prob{v \text{ disagrees with } s \lmid \eve{}}
        \tag{Linearity of expectation}
        \\
        =
        &
        \sum_{s \in S \cup S'} \prob{v \text{ disagrees with } s \text{ and } s \text{ assigned to correct part}  \lmid \eve{}}\\&
        + \prob{v \text{ disagrees with } s \text{ and } s \text{ assigned to wrong part}  \lmid \eve{}}
        \tag{Law of Total Probability}
        \\
        \le
        &
        \sum_{s \in S \cup S'} \frac{9}{10} \prob{v \text{ disagrees with } s \lmid \eve{}, s \text{ assigned to correct part}}\\&
        +
        \frac{1}{10} \prob{v \text{ disagrees with } s \lmid \eve{}, s \text{ assigned to wrong part}}
        \tag{$\eveminif$ satisfied}
        \\
        \le
        &
        \sum_{s \in S \cup S'} \frac{9}{10} \cdot \frac{1}{50} + \frac{1}{10}
        \tag{$v \in \Vg$, and probabilities are at most $1$}
        \\
        =
        &
        \frac{59}{500}\card{S \cup S'}.
    \end{align*}
    Using a one-sided Chernoff bound (\Cref{prop:additive-chernoff}),
    \begin{align*}
        &
        \prob{v \text{ assigned to wrong part in } \merging \lmid \eveminif, \evemini, \eveconl}
        \\
        =
        &
        \prob{\disag{v}{S_L+v, S_R} > \disag{v}{S_L, S_R+v} \lmid \eveminif, \evemini, \eveconl}
        \\
        =
        &
        \, \prob{\disag{v}{S_L+v, S_R} > \card{S \cup S'}/2 \lmid \eveminif, \evemini, \eveconl}
        \tag{$\disag{v}{S_L+v, S_R} + \disag{v}{S_L, S_R+v} = \card{S \cup S'}$}
        \\
        \le
        &
        \, \eexp{-\frac{1}{5}\card{S \cup S'}}
        \\
        \le
        &
        \, n^{-10}.
    \end{align*}
    
    A similar argument holds were $v$ to have been assigned to $R^*$ in the optimal partition.
    Thus,
    \begin{align*}
        &
        \prob{\evemerge \lmid \eveminif, \evemini, \eveconl}
        \\
        &=
        1 - \prob{\cup_{v \in \Vg} v \text{ assigned to wrong part} \lmid \eveminif, \evemini, \eveconl}
        \\
        &\ge
        1 - \sum_{v \in \Vg} \prob{v \text{ assigned to wrong part} \lmid \eveminif, \evemini, \eveconl}
        \tag{Union bound}
        \\
        &\ge
        1 - n^{-9}.
    \end{align*}
\end{proof}

\paragraph{\Switching. }
\Cref{lemma:all-vg-merge} gives $\evemerge$ a $1 - o\paren{1}$ probability of occurring, much like \Cref{lem:good-in-place} from \Cref{subsec:poly-algo}.
This is all that is needed for \switching to go through successfully with $1 - o\paren{1}$ probability (see \Cref{lem:good-still-in-place}, \Cref{claim:noise-sens}, and \Cref{lem:noise-resist}).
We thus refer the reader to \Cref{subsec:poly-algo} instead of repeating the same arguments.

\subsection*{Proof of \Cref{thm:offline-eptas}}

\paragraph{Handling the $\gamma \lessapprox \eps$ case.}
We first wrap up the analysis of approximation guarantee in the $\gamma \lessapprox \eps$ case.
Conditioning on \evemerge and its preceding events, which happens with probability at least $1-o(1)$, the events \eveswitch and \everes occur with high probability, which is sufficient to get a $(1+\eps)$ approximation.
This is formalized as \Cref{claim:eptas-whp-succ} and \Cref{claim:eptas-aprrox-guarantee}.

\begin{claim}
\label{claim:eptas-whp-succ}
    Events \eveconl, \evemerge, \eveswitch, and \everes occur simultaneously with  probability at least $1-o(1)$.
\end{claim}
\begin{proof}
    % This is shown by the chain rule; independence of $\eveswitch$ and $\everes$; and \Cref{lemma:s-log-congruent}, \Cref{lemma:all-vg-merge}, \Cref{lem:good-still-in-place}, and \Cref{lem:noise-resist}.
    Conditioning on \evemerge, \Cref{lem:good-still-in-place} and \Cref{lem:noise-resist} show that the events \eveswitch and \everes happen with high probability.
    \evemerge happens with probability at least $1-o(1)$ by \Cref{lemma:all-vg-merge}
\end{proof}

\begin{claim}
\label{claim:eptas-aprrox-guarantee}
    Suppose $\gamma \leq \frac{\varepsilon}{100^4}$, and let $\paren{L, R}$ be the partition of $V$ returned by the algorithm.
    Suppose $\eveconl, \evemini, \eveminif, \evemerge, \eveswitch, \everes$ are all satisfied.
    Then, 
    \begin{align*}
        \frust{G}{L,R} < \paren{1+\eps} \cdot \frustind{G}
    \end{align*}
\end{claim}
\begin{proof}
    The proof is exactly the same as the one of \Cref{lem:mindis}.
\end{proof}

This concludes our analysis of the $\gamma \lessapprox \eps$ case.

\paragraph{Handling the $\gamma \gtrapprox \eps$ case. }
As promised, we use the algorithm in \Cref{prop:maxag} (we call it $\mathsf{MaxAg}$ as in \Cref{app:max-agree}) to handle the $\gamma \gtrapprox \eps$ case.
Concretely, we pass $\eps'=\eps^2/100^4$ to the $\mathsf{MaxAg}$ algorithm.
As such, the additive error of $\frust{G}{L,R}$ is at most $\eps^2/(2\cdot100^4)\cdot n^2\leq \eps \cdot \gamma n^2$ since $\gamma>\eps/100^4$.
This gives us a $(1+\eps)$-multiplicative approximation.

\paragraph{Runtime Analysis. }
We first analyze the $\gamma \lessapprox \eps$ case.
The sampling process (Lines~\ref{line:sampling-offline-1} to \ref{line:sampling-offline-3}) only takes $O(\frac{n\log(n)}{\eps^2})$ time.
The frustration sparsifiers can be constructed in $O(n^2\cdot \log^2(n))$ time by the cut sparsifier in \Cref{prop:eps-cutsparse-algs} and a \mlab edge counter.
For the runtime of Line~\ref{line:best-partition-enum}, note that there are $2^{\card{S'}}=\log^{100}(n)$ many partitions $(S'_L, S'_R)$ of $S'$ to iterate over.
For each iteration, \minimerging takes $O(\log(n)\log\log(n))$ time, \merging takes $O(n\log(n))$ time, \switching takes the $O(n\log(n)/\eps^2)$ time.
As such, the total runtime of Line~\ref{line:best-partition-enum} is $O(\frac{n\log^{101}(n)}{\eps^2})$. 

For the $\gamma \gtrapprox \eps$ case, we can directly use \Cref{prop:maxag} and $\eps'=\eps^2$ to get the runtime of $n\cdot(1/\eps)^{O(\eps^{-4})}$.
Therefore, combining the above analysis gives us a runtime upper bound of $O\paren{\frac{n^2\log^2(n)}{\eps^2}+n\cdot(1/\eps)^{O(\eps^{-4})}}$.

\paragraph{From $1-o(1)$ probability to high probability.}
Finally, we boost our success probability from $1-o(1)$ to $1-\frac{1}{n}$ by using the standard technique that runs $O(\log(n))$ copies of the above algorithm, and returns the copy that yields the minimum frustration.
Note that since the frustration sparsifier $\frest{G}{\cdot}{\eps/10}$ succeed with high probability, we can apply a union bound over the frustration sparsifiers of all the copies to guarantee the evaluated frustration $\frest{G}{L}{\eps/10}$ is at most $(1+\eps)\cdot \frust{G}{L,R}$.
Therefore, the final runtime is $O\paren{\frac{n^2\log^3(n)}{\eps^2}+n\log(n)\cdot(1/\eps)^{O(\eps^{-4})}}$, and the algorithm succeeds with high probability.

\subsection{Combining with the Streaming Algorithm: Proof of \Cref{thm:frus-algo-main}}
\label{subsec:stream_eptas}

With \Cref{thm:offline-eptas}, we have completed all the technical work to arrive at the nearly-linear time (for constant $\varepsilon$) streaming algorithm in \Cref{thm:frus-algo-main}.
Our final algorithm for \Cref{thm:frus-algo-main} is simply as follows.
We run two subroutines in parallel during the stream: the first for the $\gamma \lessapprox \eps$ case and the second for the $\gamma \gtrapprox \eps$ case.
For the first subroutine, we maintain a frustration sparsifier with parameter $\eps/10$ during the stream, sample according to the $\tilde{O}(n^2)$ offline algorithm as in \Cref{subsec:offline_eptas}, and perform the \minimerging, \merging, and \switching steps at the end of the stream.
For the second subroutine, we maintain a frustration sparsifier with parameter $\eps/10$ during the stream, and run the algorithm in \Cref{prop:maxag} with parameter $\eps'=\eps^2$.
We return the result of the subroutine that yields smaller frustration. 

To prove the correctness of the algorithm, we note that the arguments for $\gamma \lessapprox \eps$ case in the proof of \Cref{thm:offline-eptas} go through with the same probability.
For the $\gamma \gtrapprox \eps$ case, we now use a frustration sparsifier that guarantees the second subroutine to return a $(1+O(\eps))$-approximation of $\frustind{G}$ with high probability.
We can further run $O(\log(n))$ copies of the above streaming algorithm to boost the success probability to at least $1-\frac{1}{n}$.

To analyze the time complexity, the $\gamma \lessapprox \eps$ uses exactly the same time complexity as in \Cref{thm:offline-eptas} -- the only difference is the streaming frustration sparsifier, which can be constructed in $O(n^2\log^2(n))$ time over a single-pass of the graph stream (by \Cref{prop:eps-cutsparse-algs} and counting $O(n^2)$ \mlab edges).
For the $\gamma \gtrapprox \eps$ case, we introduce an additive $O(n^2\log^2(n))$ runtime due to the frustration sparsifier.
However, this term is asymptotically smaller than the $O(n^2\log^2(n)/\eps^2)$ runtime from the $\gamma \lessapprox \eps$ case.
As susch, the final asymptotic runtime is the same as the offline algorithm, which is $O\paren{(n^2/\eps^2)\log^2(n)+n\cdot(1/\eps)^{O(\eps^{-4})}}$.
The extra $O(\log(n))$ term does not appear since we can run the $O(\log(n))$ copies in \emph{parallel}.

Finally, we analyze the space complexity of the algorithm.
The $\gamma \lessapprox \eps$ subroutine maintains a frustration sparsifier with a space of $O((n/\eps^2)\log^3(n))$ words, and the sampling process (Lines~\ref{line:sampling-offline-1} to \ref{line:sampling-offline-3} in the streaming) only takes $O((n/\eps^2)\log(n))$ space.
On the other hand, for the $\gamma \gtrapprox \eps$ case, running the $\mathsf{MaxAg}$ algorithm with $\eps'=\eps^2$ results in $O(\frac{n}{\eps^6}\cdot\log(\frac{1}{\eps}))$ space (\Cref{clm:space-big-gamma}), and the frustration sparsifier can be implemented in $O((n/\eps^2)\log^3(n))$ space.
Hence, a space of $O\paren{(n/\eps^6)\log^3(n/\eps)}$ words is sufficient for a single copy of the algorithm, and $O\paren{(n/\eps^6)\log^4(n/\eps)}$ words are sufficient for $O(\log(n))$ copies.

\section{Space Lower Bounds}
\label{sec:lbs}
% \chen{TODO: address the comments}
In this section, we show space lower bounds to complement our algorithmic results.
\begin{itemize}
    \item Any $p$-pass deterministic streaming algorithm solving \Cref{prob:structural-balance} requires $\Omega\paren{\frac{n}{p}}$ space.
    \item Any single-pass randomized algorithm solving \Cref{prob:structural-balance} on a complete signed multi-graph, where edges are allowed to repeat twice, requires $\Omega\paren{n}$ space.
    \item Any single-pass randomized streaming algorithm solving \Cref{prob:min-frust-partition} \emph{exactly}, or that outputs the \emph{exact} frustration index, requires $\Omega\paren{n^2}$ space.
\end{itemize}
Combined with our upper bounds in \Cref{thm:structural-balance-ub} and \Cref{thm:frus-algo-main}, our results convey the following message:
$(i)$. It is possible to test structural balance with a very high space efficiency by taking advantage of the arrive-once inputs and randomness, and both aspects are \emph{necessary}; 
and $(ii)$. It is possible to achieve a $(1+\eps)$-approximation in $\Otilde(n)$ space and polynomial time for any constant $\eps$. In contrast, the exact solution requires almost all the input to be stored.

\subsection{Structural Balance Testing by Deterministic Algorithms}
\label{subsec:lb-det-algs}
We first present space lower bound for deterministic streaming algorithms. The formal statement of the lower bound is as follows.

\begin{proposition}
\label{prop:structural-balance-dlb}
    Any deterministic $p$-pass algorithm for \Cref{prob:structural-balance} requires $\Omega\paren{\frac{n}{p}}$ bits of space.
    In particular, any deterministic single-pass algorithm requires $\Omega\paren{n}$ bits of space.
\end{proposition}
        
The proof of our deterministic lower bound uses a reduction from $\EQUALITY$, which is known to require $\Omega\paren{\frac{n}{p}}$ bits of communication over $p$ rounds.

\begin{proposition}
\label{prop:equality}
    Consider a communication game, $\EQUALITY$, between Alice and Bob, where Alice holds a string $x \in \set{0,1}^n$, and Bob holds a string $y \in \set{0,1}^n$.
    The players communicate with each other and, after sufficient communication, output $1$ if $x = y$ and $0$ if $x \neq y$.
    In such a game, any $p$-round \emph{deterministic} protocol requires $\Omega\paren{\frac{n}{p}}$ bits of communication.
\end{proposition}

An instance of $\EQUALITY$ is reduced to a graph which contains $n$ pairs of shared ``bit''-vertices $v^0_i$ and $v^1_i$, one vertex $a$ for Alice, and one vertex $b$ for Bob.
A pair of bit-vertices will belong to different parts ($L$ or $R$).
Alice is in charge of choosing the orientation of each pair of bit-vertices based on her string $x \in \set{0,1}^n$; more specifically, Alice places $v^{x_i}_i \in L$ and $v^{1 - x_i}_i \in R$.
Alice also places $a$ in $L$.
\plab and \mlab edges will be added to $L$ and $R$ by Alice in the unique way that maintains structural balance.
% In terms of the information Alice and Bob have, they both know that the signed graph induced on the bit-vertices will be structurally balanced, but Bob will not know the exact partition while Alice will.
Finally, Bob attempts to place $b$ in $L$ by adding a \plab edge joining $a$ to $b$, and then by following the same rule Alice used to place $a$ in $L$, but using the information of his string $y$ instead of $x$ (which he does not know).
Now if $x=y$, Alice and Bob agree with each other on how the vertices should be partitioned and the resulting graph is structurally balanced.
Otherwise, the graph is not balanced since edges incident to $b$ introduce unbalanced triangles.
We now formally state the reduction.

\begin{tbox}
\label{red:eq-to-sb}
	\textbf{Reduction 1: $\EQUALITY$ to Structural Balance}:
	\begin{itemize}
	    \item For $p$ rounds
    	\begin{itemize}[label=($\alph*$)]
    	    \item[\textbf{Alice}]
        	\begin{enumerate}[label=($\arabic*$)]
        	    \item Feed $(v^0_i, v^1_i)$, \mlab into stream, for all $i \in \bracket{n}$.
        		\item Feed $(a, v^{x_i}_i)$, \plab and $(a, v^{1 - x_i}_i), \mlab$ into stream, for all $i \in \bracket{n}$.
        		\item For all $x_i = x_j$:
        		\begin{enumerate}
            		\item Feed $(v^0_i, v^0_j)$, \plab and $(v^1_i, v^1_j)$, \plab into stream
            		\item Feed $(v^0_i, v^1_j)$, \mlab into stream.
        		\end{enumerate}
        		\item For all $x_i \neq x_j$:
        		\begin{enumerate}
            		\item Feed $(v^0_i, v^0_j)$, \mlab and $(v^1_i, v^1_j)$, \plab into stream, for $x_i = 0$ and $x_j = 1$; 
            		\item Feed $(v^0_i, v^1_j)$, \plab into stream.
        		\end{enumerate}
        		\item Pass the memory content of the Structural Balance algorithm to Bob.
        	\end{enumerate}
        	\item[\textbf{Bob}]
        	\begin{enumerate}[label=($\arabic*$)]
        	    \item Feed $(a,b)$, \plab  into stream.
        		\item Feed $(b, v^{y_i}_i)$, \plab  and $(b, v^{1 - y_i}_i)$, \mlab into stream, for all $i \in \bracket{n}$.
        		\item Pass the memory content of the Structural Balance algorithm to Alice.
        	\end{enumerate}
    	\end{itemize}
    	\item If the graph is balanced, output $1$. Otherwise, output $0$.
	\end{itemize}
\end{tbox}
\noindent
% \chen{a,b,v in the protocol vs. s,t, a, b in the figure}
% \todo{Yeah. a,b,v is way better I think. Alice Bob Vertex}

\begin{figure}[!htb]
\centering
\includegraphics[width=1.0\columnwidth]{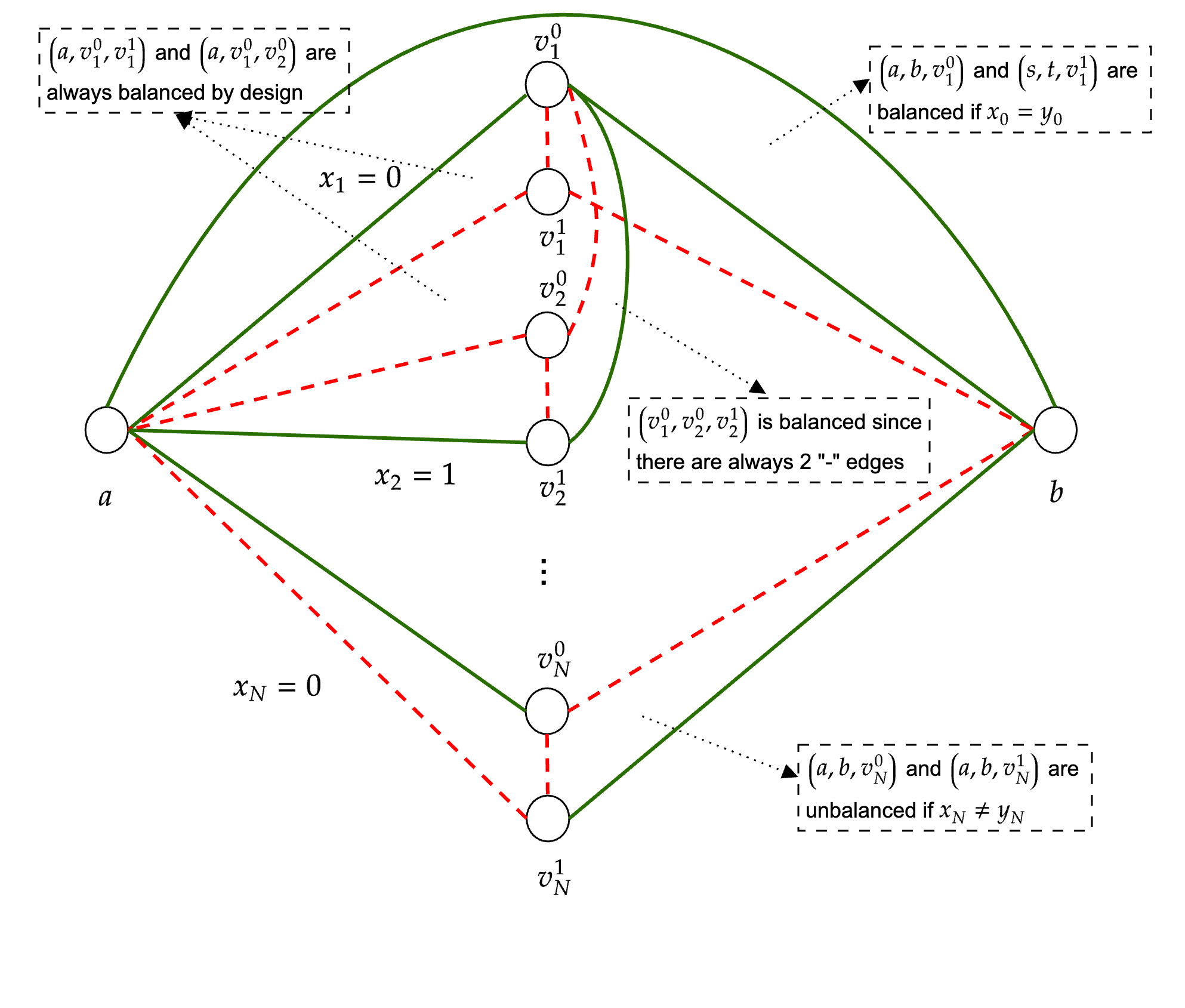}
\caption{An illustration of the correctness of \hyperref[red:eq-to-sb]{Reduction 1}.}\label{fig:equality-balance-reduction}
\end{figure}

Some bookkeeping reveals that the graph is indeed complete and well-defined (no edge signs with both \plab and \mlab).
It remains to show that the graph is balanced if and only if $x = y$.
\begin{claim}
\label{claim:det-red-correct}
    \hyperref[red:eq-to-sb]{Reduction 1} correctly outputs $1$ when $x = y$ and $0$ when $x \neq y$.
\end{claim}
\begin{proof}
    Suppose $x = y$.
    Then $L = \set{a, b, v^{x_1}_1, \dots v^{x_n}_n}$ has only \plab edges connecting its vertices, $R = \set{v^{1 - x_1}_1, \dots v^{1 - x_n}_n}$ has only \plab edges connecting its vertices, and $L$ and $R$ has only \mlab edges connecting them.
    That is, $\paren{L,R}$ is a partition of $V$ that certifies structural balance.
    Hence, the reduction outputs $1$. 
    
    Suppose, on the other hand, $x \neq y$.
    Let $x_i \neq y_i$.
    Then $(a, b, v^0_i)$ and $(a, b, v^1_i)$ are imbalanced triangles.
    The reduction thus outputs $0$.  (See \Cref{fig:equality-balance-reduction} for explanations in the picture.)
\end{proof}

\begin{proof}[Proof of \Cref{prop:structural-balance-dlb}]
    The lower bound follows from \Cref{claim:det-red-correct} and \Cref{prop:equality}.
    In particular, setting $p = 1$ shows the $\Omega\paren{n}$ lower bound for deterministic single-pass algorithms.
\end{proof}

\subsection{Structural Balance Testing by Randomized Algorithms with Multi-edges}
The second lower bound in this section shows that we need to use $\Omega\paren{n}$ space if the inputs are complete signed graphs with multi-edges.
In the regime of multi-graphs, the sign of a relation becomes meaningless if both \plab and \mlab appear between the same vertex pairs.
In view of this, we insist that the signs of the multi-edges between the same pair of vertices are \emph{consistent}: if $e_{1}$ between $(u,v)$ is \plab, a second edge $e_{2}$ between $(u,v)$ also has to be \plab.
The definition of structural balance still holds by checking one of the edges between each vertex pair (although the frustration index for imbalanced graphs may change drastically). 

The formal statement of our lower bound is as follows.
\begin{proposition}
\label{prop:structural-balance-rlb}
    Any single-pass streaming algorithm that correctly tests structural balance with probability at least $\frac{99}{100}$ on complete signed multi-graphs with sign-consistent multi-edges has to use a memory of $\Omega\paren{n}$ bits.
\end{proposition}

% is established by a variate of the well-known INDEX problem. The standard INDEX problem is described as follows: Alice holds a string $x \in \{0,1\}^{N}$, Bob holds an index $i^{*}\in [N]$, and the two players want to learn the value of $x_{i^{*}}$. It is known that in one-way communication for INDEX, $\Omega\paren{N}$ bits are required for any protocol even with randomization. 

On the high level, the plan to prove \Cref{prop:structural-balance-rlb} is a reduction from $\INDEX$.
The standard $\INDEX$ problem is described as follows: Alice holds a string $x \in \{0,1\}^{N}$, Bob holds an index $i^{*} \in [N]$, and the two players want to learn the value of $x_{i^{*}}$.

In a first attempt at a reduction, the two players are initially given $2N+2$ vertices, with each bit $x_{j}$ corresponding to two vertices $u_{j}$ and $v_{j}$ and two special vertices $s$ and $t$.
Alice creates two empty sets $A$ and $B$, and for each input bit $x_{j}$, $j\in [N ]$, Alice puts $u_{j}$ to $A$ and $v_{j}$ to $B$ if $x_{j}=0$, and $v_{j}$ to $A$ and $u_{j}$ to $B$ if $x_{j}=1$.
Then, for every vertex except $u_{i^{*}}$ and $v_{i^{*}}$\footnote{The construction is in fact not possible as we will see in the next paragraph; it is here to explain the intuition.}, Alice adds \plab edges from $s$ to the vertices in $A$ and from $t$ to vertices in $B$.
Furthermore, for each pair of vertices \emph{both inside} $A$ or $B$, Alice adds a \plab edge; and for vertex pairs such that $u \in A$ and $v \in B$, Alice adds a \mlab edge.
On the other end, Bob completes the graph by adding \plab edges to $(s,u_{i^{*}})$ and $(t,v_{i^{*}})$ and \mlab edges to $(s,v_{i^{*}})$ and $(t,u_{i^{*}})$.
Now, if $x_{i^{*}}=0$, the resulting graph is balanced; otherwise, the graph is unbalanced since $v_{i^{*}}$ is in $A$.

The plan sounds nice, but it does \emph{not} work.
Careful readers may have already found the problem: Alice does \emph{not} know which vertices correspond to $u_{i^*}$ and $v_{i^*}$; and if she somehow infers it, the problem becomes easy as she can output $x_{i^*}$ without any communication.
As such, for the above idea to work, we need to somehow `hide' the index to the holder of the string, but give them the ability to create graphs that leave everything but $x_{i^{*}}$.

To the above end, we introduce the following version of Leave-One INDEX.

\begin{problem}[Leave-One INDEX]
\label{prb:3-player-index}
Consider a communication problem between 3 players, namely Alice, Bob, and Charlie.
Both Alice and Bob hold the same string $x \in \{0,1\}^{N}$, and Charlie holds an index $i^{*} \in [N]$.
Furthermore, Alice holds $S \subset [N]$ and Bob holds $T \subset [N]$, such that $S \cup T = [N] \setminus \{i^{*}\}$.
The communication goes in the order of Alice, Bob and Charlie, and the players want to output $x_{i^{*}}$.
\end{problem}

The sets of $S$ and $T$ create some `flexibility' for Alice and Bob to infer the index.
Indeed, if $S\cap T = \emptyset$, the problem becomes easy as Bob can learn Alice's bits from a message that simply encodes which bits are missing\footnote{In particular, there is a simple protocol with 1 bit of communication when $S\cap T = \emptyset$: Alice takes XOR of all bits in $x[S]$, and send it to Bob. Bob takes XOR of all bits in $x[T]$ he will learn the XOR of $x[S\cup T]$. Bob then compares the XOR of $x$ vs. the XOR of all bits in $x[S\cup T]$ to learn $x_{i^*}$}.
Nevertheless, we prove that the worst-case communication complexity of the Leave-One INDEX problem is still $\Omega\paren{N}$.
The formal lower bound is as follows.

\begin{lemma}
\label{lem:leave-one-index-lb}
Any (possibly randomized) one-way communication protocol for the Leave-One INDEX defined in \Cref{prb:3-player-index} that outputs $x_{i^{*}}$ correctly with probability at least $\frac{99}{100}$ requires $\Omega\paren{N}$ bits of communication.
\end{lemma}

Let us first see that if \Cref{lem:leave-one-index-lb} holds, it implies a lower bound as shown in \Cref{prop:structural-balance-rlb}.

\begin{proof}[Proof of \Cref{prop:structural-balance-rlb}]
We reduce the Leave-One INDEX problem to structural balance such that the graph created by Alice and Bob is balanced if and only if $x_{i^{*}}=0$. The construction goes as follows (see also: \Cref{fig:red-2}).
\begin{tbox}
\paragraph{Reduction 2: Leave-One INDEX to Structural Balance Testing with Multi-edges}
\begin{enumerate}
\item The players create $G$ with $2N+2$ vertices and a hard-coded edge:
\begin{itemize}
\item Two special vertices $s$ and $t$; 
\item Two vertices $u_{i}$ and $v_{i}$ for every $i \in [N]$.
\item One \mlab edge between $(s,t)$.
\end{itemize}

\item Alice adds edges by the following rules:
\begin{enumerate}
\item Alice creates sets $A$ and $B$ (for book-keeping purpose, \emph{not} part of the inputs). 
\item For every $i\in [N]$, if $x_{i}=0$, Alice sets $u_{i}$ as $A_{i}$ and $v_{i}$ as $B_{i}$;
\item Otherwise, if $x_{i}=1$, Alice sets $v_{i}$ as $A_{i}$ and $u_{i}$ as $B_{i}$;
\item For every $j\in S$, Alice adds \plab edges between $(s, A_{j})$ and $(t, B_{j})$, and \mlab edges between $(s, B_{j})$ and $(t, A_{j})$.
\item For each pair of $(A_{i}, A_{j})$ and $(B_{i}, B_{j})$ such that $i \neq j$, Alice adds a \plab edge.
\item For each pair of $(A_{i}, B_{j})$, Alice adds a \mlab edge.
\end{enumerate}

\item Bob adds edges by the following rules
\begin{enumerate}
\item Bob creates sets $A$ and $B$ and assigns $A_{i}$'s and $B_{i}$'s consistent with Alice (since Bob has the same $x$).
\item For every $j\in T$, Bob adds \plab edges between $(s, A_{j})$ and $(t, B_{j})$, and \mlab edges between $(s, B_{j})$ and $(t, A_{j})$.
\end{enumerate}

\item Charlie adds edges by the following rules
\begin{enumerate}
\item Charlie adds \plab edges between $(s, u_{i^{*}})$ and $(t, v_{i^{*}})$, and \mlab edges between $(s, v_{i^{*}})$ and $(t, u_{i^{*}})$.
\end{enumerate}
\end{enumerate}
\end{tbox}

\begin{figure}[!htb]
\centering
\includegraphics[width=0.95\columnwidth]{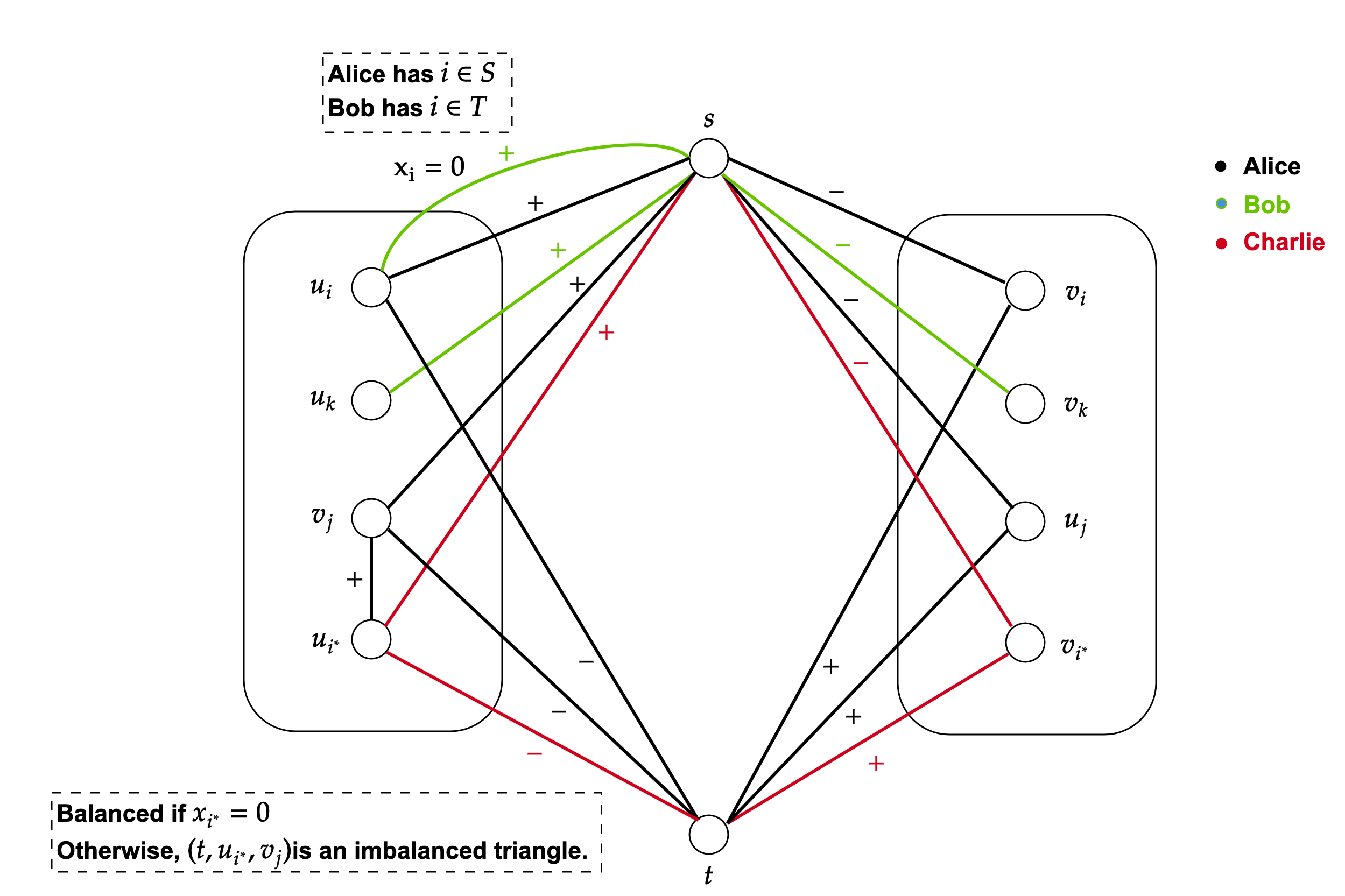}
\caption{An illustration of Reduction 2.}\label{fig:red-2}
\end{figure}

It is straightforward to verify that the graph $G$ created by the players is complete and the players can create the graph based on their inputs without communication. The resulting graph has to be multi-edge because of the indices $j \in S\cap T$, where vertex pairs of $(s, A_{j})$ (resp. $(t,B_{j})$, $(t, A_{j})$, $(s, B_{j})$) get two edges from Alice and Bob each. Their signs are always consistent since Alice and Bob hold the same $x$. 

We now show that $G$ is balanced if and only if $x_{i^{*}}=0$. If we only look at the edges created by Alice and Bob, we have a balanced partition of $A\cup \{s\}$ and $B\cup \{t\}$ with only \plab edges inside and \mlab edges between them. Now, if $x_{i^{*}}=0$, $u_{i^{*}}$ has only \plab edges to $A\cup \{s\}$ and only \mlab edges to $B\cup \{t\}$ (and vise versa for $v_{i^{*}}$). As such, the graph stays balanced. On the other hand, if $x_{i^{*}}=1$, $u_{i^{*}}$ has one \mlab edge to $s$ and \plab edges to vertices in $A$, which forms unbalanced triangles.
\end{proof}

The rest of this section is dedicated to the proof of \Cref{lem:leave-one-index-lb}.

\subsubsection*{Proof of \Cref{lem:leave-one-index-lb}}
In the proof of the communication complexity for the standard INDEX problem, the intuition is that the first player (Alice) has no knowledge of $i^{*}$ other than knowing it is uniformly distributed (in the particular hard distribution). 
As such, the problem can only be solved if a sufficiently large number of bits are communicated. However, in the Leave-One INDEX problem (\Cref{prb:3-player-index}), the first two players (Alice and Bob) might be able to learn where exactly $i^{*}$ is by comparing their sets of indices $S$ and $T$. In \Cref{prb:3-player-index}, by the promise that $i^{*}\not\in T$, Bob can learn the distribution of $i^{*}$ better than uniform even without communication. % \vik{Is the reference meant to be to equality or something else? I think it's \Cref{prb:3-player-index}.}

Our plan to prove \Cref{lem:leave-one-index-lb} is to conduct a case analysis on whether the communication between Alice and Bob somehow `solves' the problem. To be precise, we separate the cases based on whether Alice's message significantly changes the distribution of $i^{*}$ from Bob's perspective. If such a case happens, then we can use a result from \cite{AssadiR20} to show that the communication between \emph{Alice and Bob} has to be $\Omega\paren{\eps^{2} N}$. On the other hand, if the message from Alice does \emph{not} tell Bob a lot about the distribution of $i^{*}$, a large number of bits are still required for the communication between \emph{Bob and Charlie} since there are constant fraction of coordinates where $i^*$ still looks almost uniform. The latter bound cannot be obtained directly by a reduction since $i^{*}$ is already not uniform from Bob's perspective. Nevertheless, we can adapt the information-theoretic proof for INDEX to get the desired lower bound in a standard manner.

We now formalize the above intuition. Let $\Pi_{A,B}$ be the random variable for the message between Alice and Bob, and $\Pi_{B,C}$ 
% \vik{Te range of an RV is typically \IR. I'm guessing here it's precisely the messages sent. Should we say what these RVs mean? Or give examples? Can one take an expectation of these kinds of RVs?}
be the random variable for the message between Bob and Charlie. Furthermore, let $X$ be the random variable for the string being held by Alice and Bob, and let $I$ be the random variable for $i^{*}$. For the choice of the set $T$, we use $T'$ to denote the realization. Conditioning on the first message $\Pi_{A,B}=\pi_{A,B}$ and Bob's local inputs $X, T$, we let $I^{B}:=I\mid X, T, \Pi_{A,B}$ be the random variable for $i^{*}$ from \emph{Bob's internal perspective after message $\pi_{A,B}$}. 
% Finally, we let $I_{\cap S}$ be the distribution \vik{which distribution?} \emph{restricted} to a subset of supports $S$, i.e. the mass of $I(i), \forall i\in S$ is normalized by $\frac{N}{\card{S}}$ and $I(i)=0$ for all $i \not\in S$. 
We slightly abuse notation and let these random variables also represent their distributions.

As promised, the first case we are going to show is the communication lower bound between Alice and Bob if the distribution of $i^{*}$ that Bob learns is $\eps$-far from his \emph{prior} knowledge. Formally, we define the notation of $\eps$-learn as follows.
\begin{definition}
\label{def:eps-learn}
Let $\pi_{A,B}$ be the message of a randomized one-way communication protocol. % \vik{Is it a protocol, a message, or are they the same thing?} \chen{not exactly the same but usually people don't distinguish them. Add a remark above for this.}. 
We say that Bob $\eps$-learns $i^{*}$ if after the message $\pi_{A,B}$, in expectation, the distribution of $i^{*}$ from Bob's perspective is more than $\eps$-far from the distribution he knows from his own input in the total variation distance, i.e. 
\begin{align*}
\expect{\tvd{I\mid\Pi_{A,B}, X, T}{I\mid X, T}}>\eps,
\end{align*}
where the expectation is over the randomness of $X$, $T$ and $\Pi_{A,B}$.
\end{definition}

The notion of $\eps$-learn was first developed by \cite{AssadiCK19,AssadiR20}. As such, we can prove a communication lower bound for Bob to $\eps$-learn $i^{*}$.

\begin{lemma}
\label{lem:alice-bob-cc-lb}
Any such protocol $\pi_{A,B}$ for Bob to $\eps$-learn $i^{*}$ has to use $\Omega\paren{\eps^2\cdot N}$ bits of communication between Alice and Bob.
\end{lemma}
\begin{proof}
We prove the lemma by a direct reduction from \cite{AssadiR20}, which address the set intersection problem. The definition of the problem and the communication complexity is as follows.
\begin{proposition}[\!\cite{AssadiR20}]
\label{prop:set-intersect-cc}
The Set-Intersection is a two-player game between Alice and Bob. The two players are given $A\subseteq [m]$ and $B\subseteq [m]$, respectively, with the promise that there exists a unique element $e^{*}$ such that $e^{*}=A\cap B$. The goal is to find the target element $e^{*}$ after one-way communication. Let $E$ be the distribution of $e^{*}$, it is known that any communication protocol that achieves
\begin{align*}
\expect{\tvd{E\mid\Pi, B}{ E\mid B}}>\eps
\end{align*}
has to use a communication of $\Omega\paren{\eps^2 m}$ bits, where the expectation is over the randomness of $B$ and the protocol's randomness $\Pi$. In particular, the communication lower bound applies to the following distribution.
\begin{tbox}
\textbf{A hard distribution for Set-Intersection.}
\begin{itemize}
\item Sample two \emph{disjoint} sets of coordinates $A'$ and $B'$ of size $N/4-1$ each uniformly at random from $[N]$.
\item Sample an element $e^*$ uniformly at random from $[N]\setminus (A' \cup B')$, and let $A:=A'\cup \{e^*\}$, $B:=B'\cup \{e^*\}$.
\end{itemize}
\end{tbox}
\end{proposition}
The result of \Cref{prop:set-intersect-cc} can be generalized to back-and-forth communication if we allow Alice \emph{or} Bob to $\eps$-learn the distribution. However, for the purpose of our reduction, the one-way version suffices. The reduction from Set Intersection to the communication between Alice and Bob in \Cref{lem:leave-one-index-lb} is as follows.

\begin{tbox}
\paragraph{Inputs: Alice holds $A\subseteq [m]$, Bob holds $B \subseteq [m]$, $A\cap B=e^{*}$.}
\paragraph{A communication protocol \texttt{PROT} for \Cref{lem:leave-one-index-lb} that $\eps$-learns the index $i^{*}$ as in \Cref{lem:alice-bob-cc-lb}.}
\begin{enumerate}
\item Alice creates a set $S' = [m]\setminus A$.
\item Bob creates a set $T'= [m]\setminus B$.
\item Alice runs \texttt{PROT}, sends to Bob. Bob uses the distribution for $i^{*}$ as the distribution for $e^{*}$. 
\end{enumerate}
\end{tbox}
To prove the correctness of the reduction, we only need to show that a). $S'$ and $T'$ are valid inputs for \texttt{PROT} and b). $i^{*}=e^{*}$. It is straightforward to verify these conditions: we know that an element is \emph{not} covered by $S' \cup T'$ if and only if \emph{both} $A$ and $B$ cover it, which is exactly $e^{*}$. Therefore, the desired lower bound is established.
\end{proof}

\Cref{lem:alice-bob-cc-lb} implies that with $o\paren{N}$ bits of communication it is impossible for Bob to gain knowledge of the distribution of $I$ that is significantly different from what he already knows. We now observe that with the input $X$ and $T$, and conditioning on Alice's message does \emph{not} significantly change Bob's distribution, Bob's distribution of $I$ is close to uniform on the $[N]\setminus T$ coordinates.

\begin{claim}
\label{clm:bob-eps-dist}
Conditioning on the event that Bob does \emph{not} $\eps$-learn $i^*$, the distribution of $i^{*}$ that Bob learns from $T$ and Alice's message $\pi_{A,B}$ is at most $\eps$-far from uniform in expectation over the supports of $[N]\setminus T$, i.e., 
\begin{align*}
\expectr{T,\Pi_{A,B}}{\tvd{I \mid T, \Pi_{A,B}}{{U([N]\setminus T)}}}\leq \eps.
\end{align*}
\end{claim}
\begin{proof}
We first show that without Alice's message, the distribution $I$ of Bob restricted to $[N]\setminus T$ is uniform. This is a simple observation: with the promise that $i^{*}\not\in T$, Bob can safely discard all coordinates therein. Since the string $x$ is \emph{independent} of $i^{*}$, Bob does not learn anything from $x$. As such, Bob's distribution remains \emph{uniform} on the $[N]\setminus T$ coordinates.

We now condition on the fact that Bob does not $\eps$ learn $i^*$. As such, there is
\begin{align*}
 &\expectr{X,T,\Pi_{A,B}}{\tvd{I \mid X, T, \Pi_{A,B}}{U([N]\setminus T)}} \\
 \leq & \expectr{X,T,\Pi_{A,B}}{\tvd{I \mid X, T}{U([N]\setminus T)} + \tvd{I^{B}}{I\mid X, T}} \tag{triangle inequality}\\
\leq & \expectr{X,T,\Pi_{A,B}}{0+\tvd{I\mid\Pi_{A,B}, X, T}{I\mid X, T}} \leq \eps,
\end{align*}
where the last inequality directly comes from our assumption of no $\eps$-learning. 

Finally, note that the choice of $i^*$ is independent of $X$ (whether or not conditioning on $\Pi_{A,B}$). Therefore, we have 
\begin{align*}
& \expectr{X,T,\Pi_{A,B}}{\tvd{I \mid X, T, \Pi_{A,B}}{U([N]\setminus T)}} \\
& = \expectr{T,\Pi_{A,B}}{\tvd{I \mid T, \Pi_{A,B}}{{U([N]\setminus T)}}} \\
& \leq \eps,
\end{align*}
as desired.
% comes from the fact that restricting the distribution to $N-\card{T}$ coordinates can only blow up the total variation distance by a factor of $\frac{N}{\card{[N]\setminus T}}$.
\end{proof}

We now use the result of \Cref{clm:bob-eps-dist} to prove the communication complexity conditioning on Bob does not $\eps$-learn $i^*$. To continue, we insist that the distribution $T$ where $T'$ is sampled from always produces size-$t$ sets. As such, we can establish the following lemma.

% \chen{For \Cref{lem:no-eps-learn-cc}, maybe we can further divide into 2~3 claims to free people from reading the calculations.}
\begin{lemma}
\label{lem:no-eps-learn-cc}
Suppose the players sample from a distribution $T$ to ensure the size of the set given to Bob is always $t$. Conditioning on the event that Bob does \emph{not} $\eps$-learn $i^*$, any communication protocol such that Charlie correctly outputs $x_{i^*}$ with probability at least $1-\delta$ requires at least $(1-H_2(\delta)-2\eps)\cdot \paren{N-t}$ bits of communication.
\end{lemma}
\begin{proof}
We now proceed to prove the communication complexity of the problem. By Fano's inequality (\Cref{prop:fano}), for Charlie to output the $x_{i^*}$ with probability at least $1-\delta$, there should be
\begin{align*}
\HH\paren{X_{I}\mid \Pi_{A,B}, \Pi_{B,C}, T, I} \leq H_{2}(\delta).
\end{align*}
We prove a lower bound for $\HH\paren{X_{I}\mid \Pi_{A,B}, \Pi_{B,C}, T, I}$, % \vik{Is $T$ missing, or is it not supposed to be in the ineqs below. Guessing the former.}, 
which leads to a mutual information upper bound between $X_{[N]\setminus T}$ and $(\Pi_{A,B}, \Pi_{B,C})$. 
% \vik{Eq 1: Notation for the expectation looks weird (how it's over $T$, etc, and these variables are bound in the entropy also (e.g. $T = t$)). It's very likely I'm just not used to it!}
% \vik{Eq 3: The tag should use $\Pi_{B,C}, I$ instead of $\pi, i*$ no? More importantly, why is it independent?}
% \vik{Eq 6: Can someone walk me through how this step is made?}
% \vik{I am here at Eq 6.}
\begin{claim}
\label{clm:X-I-entropy-cal}
With the same setting as in \Cref{lem:no-eps-learn-cc}, there is 
\begin{align*}
\HH\paren{X_{I}\mid \Pi_{A,B}, \Pi_{B,C}, T, I} \geq \frac{1}{N-t}\cdot \HH\paren{X_{[N]\setminus T}\mid \Pi_{A,B}, \Pi_{B,C}, T} - 2\eps. 
\end{align*}
\end{claim}
\begin{proof}
The claim is established by standard manipulation of conditional entropy and the result of \Cref{clm:bob-eps-dist}. Concretely, we first write 
\begin{align*}
&\HH\paren{X_{I}\mid \Pi_{A,B}, \Pi_{B,C}, T, I} \\
&= \expectr{T' \sim T, \pi_{A,B} \sim \Pi_{A,B}}{\HH\paren{X_{I}\mid \Pi_{A,B}=\pi_{A,B}, \Pi_{B,C}, T=T', I}}\\
&= \expectr{T' \sim T, \pi_{A,B} \sim \Pi_{A,B}}{\sum_{i=1}^{N}\Pr\paren{I=i\mid \Pi_{A,B}=\pi_{A,B}, T=T'}\cdot \HH\paren{X_{I}\mid \Pi_{A,B}=\pi_{A,B}, \Pi_{B,C}, T=T', I=i}} \tag{by the definition of conditional entropy}\\
&= \expectr{T' \sim T, \pi_{A,B} \sim \Pi_{A,B}}{\sum_{i=1}^{N}\Pr\paren{I=i\mid \Pi_{A,B}=\pi_{A,B}, T=T'}\cdot \HH\paren{X_{i}\mid \Pi_{A,B}=\pi_{A,B}, \Pi_{B,C}, T=T'}} \tag{The value of $X_{i}$ is independent of $i$ conditioning on known $\pi_{A,B}$ and $t$}\\
&= \expectr{T' \sim T, \pi_{A,B} \sim \Pi_{A,B}}{\sum_{i\in [N]\setminus T'}\Pr\paren{I=i\mid \Pi_{A,B}=\pi_{A,B}, T=T'}\cdot \HH\paren{X_{i}\mid \Pi_{A,B}=\pi_{A,B}, \Pi_{B,C}, T=T'}}. \tag{$\Pr\paren{I=i\mid \Pi_{A,B}=\pi_{A,B}, T=T'}=0$ for $i \in T=T'$}
\end{align*}
For each choice of $T'$, we handle the probability for $i^*$ to be inside and outside $T'$ separately. By \Cref{clm:bob-eps-dist} and condition on the specific $T'$, we know that the probability mass for the coordinate outside $T'$ is $0$, and the distribution over the coordinates inside $T'$ is $\eps$-close to uniform. As such, we have
\begin{align*}
&\HH\paren{X_{I}\mid \Pi_{A,B}, \Pi_{B,C}, T, I} \\
&= \expectr{T' \sim T, \pi_{A,B} \sim \Pi_{A,B}}{\sum_{i\in [N]\setminus T'}  \frac{1}{N-t}\cdot \HH\paren{X_{i}\mid \Pi_{A,B}=\pi_{A,B}, \Pi_{B,C}, T=T'}} \\
& \quad + \expectr{T' \sim T, \pi_{A,B} \sim \Pi_{A,B}}{\sum_{i\in [N]\setminus T'} \paren{\Pr\paren{I=i\mid \Pi_{A,B}=\pi_{A,B}, T=T'}-\frac{1}{N-t}}}\cdot \\
& \quad \HH\paren{X_{i}\mid \Pi_{A,B}=\pi_{A,B}, \Pi_{B,C}, T=T'}\\
&\geq \expectr{T' \sim T, \pi_{A,B} \sim \Pi_{A,B}}{\sum_{i\in [N]\setminus T'} \frac{1}{N-t}\cdot \HH\paren{X_{i}\mid \Pi_{A,B}=\pi_{A,B}, \Pi_{B,C}, T=T'}} \\
& \quad - \expectr{T' \sim T, \pi_{A,B} \sim \Pi_{A,B}}{\sum_{i\in [N]\setminus T'} \card{\Pr\paren{I=i\mid \Pi_{A,B}=\pi_{A,B}, T=T'}-\frac{1}{N-t}}}\cdot \\
& \quad \HH\paren{X_{i}\mid \Pi_{A,B}=\pi_{A,B}, \Pi_{B,C}, T=T'}\\
&\geq \expectr{T' \sim T, \pi_{A,B} \sim \Pi_{A,B}}{\sum_{i\in [N]\setminus T'} \frac{1}{N-t}\cdot \HH\paren{X_{i}\mid \Pi_{A,B}=\pi_{A,B}, \Pi_{B,C}, T=T'}} \\
& \quad - \expectr{T' \sim T, \pi_{A,B} \sim \Pi_{A,B}}{\sum_{i\in [N]\setminus T'} \card{\Pr\paren{I=i\mid \Pi_{A,B}=\pi_{A,B}, T=T'}-\frac{1}{N-t}}}\tag{$\HH\paren{X_{i}\mid \pi_{A,B}, \Pi_{B,C}, T=T'}\leq \HH(X_{i})\leq 1$}\\ 
&= \expectr{T,  \Pi_{A,B}}{\sum_{i\in [N]\setminus T'}  \frac{1}{N-t}\cdot \HH\paren{X_{i}\mid \Pi_{A,B}=\pi_{A,B}, \Pi_{B,C}, T=T'}} \\
& \quad - 2\cdot \expectr{T,\Pi_{A,B}}{\tvd{I \mid T, \Pi_{A,B}}{{U([N]\setminus T)}}} \\
&\geq \expectr{T,  \Pi_{A,B}}{ \sum_{i\in [N]\setminus T'}  \frac{1}{N-t}\cdot \HH\paren{X_{i}\mid \Pi_{A,B}=\pi_{A,B}, \Pi_{B,C}, T=T'}} - 2\eps \tag{\Cref{lem:no-eps-learn-cc}}\\
&\geq \frac{1}{N-t}\cdot \HH\paren{X_{[N]\setminus T}\mid \Pi_{A,B}, \Pi_{B,C}, T} - 2\eps, \tag{subadditivity of entropy as in \Cref{fct:info-theory-facts}}
\end{align*}
as desired. \myqed{\Cref{clm:X-I-entropy-cal}}
\end{proof}

By \Cref{clm:X-I-entropy-cal}, by applying the earlier inequality of $\HH\paren{X_{I}\mid \Pi_{A,B}, \Pi_{B,C}, T, I} \leq H_{2}(\delta)$, we have 
\begin{align*}
\HH\paren{X_{[N]\setminus T}\mid \Pi_{A,B}, \Pi_{B,C}, T} \leq (H_{2}(\delta) + 2\eps)\cdot \paren{N-t}.
\end{align*}
On the other hand, we know that 
\begin{align*}
\HH\paren{X_{[N]\setminus T}\mid  T} \geq N-t,
\end{align*}
since they are uniform over the $(N-t)$ coordinates. As such, we can calculate the mutual information between $X_{[N]\setminus T}$ and $(\Pi_{A,B}, \Pi_{B,C})$ as
\begin{align*}
\II\paren{\Pi_{A,B}, \Pi_{B,C}; X_{[N]\setminus T}\mid T} &= \HH\paren{X_{[N]\setminus T}\mid  T} - \HH\paren{X_{[N]\setminus T}\mid \Pi_{A,B}, \Pi_{B,C}, T}\\
&\geq (1-H_2(\delta)-2 \eps)\cdot \paren{N-t}.
\end{align*}
By $\II\paren{\Pi_{A,B}, \Pi_{B,C}; X_{[N]\setminus T}\mid T}\leq \HH(\Pi_{A,B}, \Pi_{B,C})\leq \log(\text{supp}(\Pi_{A,B}, \Pi_{B,C}))$, we know that at least one message has to be at least as long as $(1-H_2(\delta)-2\eps)\cdot \paren{N-t}$ bits.
\end{proof}

\begin{proof}[Finalizing the proof of \Cref{lem:leave-one-index-lb}]
For completeness, we construct the hard distribution of \Cref{prb:3-player-index} as follows.
\begin{tbox}
\paragraph{A hard distribution for \Cref{prb:3-player-index}.}
\begin{enumerate}
\item Sample $A$ and $B$ according to the hard distribution for Set-Intersection prescribed in \Cref{prop:set-intersect-cc}.
\item Let $S=[N]\setminus A$, and $T=[N]\setminus B$. Sample $X$ from $\{0,1\}^{N}$ uniformly at random.
\item Give $(X, S)$ to Alice, $(X, T)$ to Bob, and $i^*$ to Charlie.
\end{enumerate}
\end{tbox}
\noindent
Observe that in the above distribution, for any choice of $T'$, there is $t=\card{T'}=\frac{3}{4}\cdot N$. Furthermore, it agrees with the hard distribution of \Cref{prop:set-intersect-cc} through the reduction in \Cref{lem:alice-bob-cc-lb}. 

If Bob $\eps$-learns $i^{*}$ for $\eps=\frac{1}{100}$, then the communication lower bound is already $\Omega\paren{N^2}$ (by \Cref{lem:alice-bob-cc-lb}). Therefore, we condition on the case when Bob does \emph{not} $(1/100)$-learns $i^{*}$. As such, let us conditioning on the event that $\expect{\tvd{I^{B}}{I\mid X, T}}\leq \frac{1}{100}$; and by \Cref{lem:no-eps-learn-cc}, to make up the success probability of $\frac{99}{100}$, the protocol needs to send a message of length at least $(1-H_2(\frac{1}{100})-2/100)\cdot (N-t) = \Omega(N)$ bits, as desired.
\end{proof}

\subsection{Frustration-minimizing Partition by Exact Algorithms}
% \todo{fix value of $\dc$}
In this section, we show that any single-pass streaming algorithm that solves the frustration index \emph{exactly} has to store almost the entire graph.

% \begin{proposition}
% \label{prop:frust-min-lb}
% Any randomized single-pass algorithm that approximates the frustration index within an additive $O\paren{n^{c/(1+c)}}$ error for any constant $c \geq 0$ requires $\Omega\paren{n^{2/(1+c)}}$ bits of space.
% % This lower bound also holds for algorithms that return an approximation of the frustration index to within the same error (without necessarily returning a partition).
% In particular, solving \Cref{prob:min-frust-partition} exactly requires $\Omega\paren{n^2}$ bits of space, and outputting the frustration index to within an additive $\sqrt{n}$ error requires $\Omega\paren{n}$ bits of space.
% \end{proposition}

\begin{proposition}
\label{prop:frust-min-lb}
Any single-pass streaming algorithm that solves \Cref{prob:min-frust-partition} exactly or returns the optimal frustration index value requires $\Omega\paren{n^2}$ bits of space.
\end{proposition}

We go through communication complexity again.
This time, our reduction is from $\INDEX$ (a matrix variant thereof) which is hard under the one-way communication regime.
The matrix $\INDEX$ is known to have an $\Omega\paren{N^2}$ unconditional communication lower bound for inputs of size $N^2$.

\begin{proposition}
\label{prop:matrix-index}
    Consider a communication game, $\INDEX$, between Alice and Bob, where Alice holds a square boolean matrix $M \in \set{0,1}^{N^2}$, and Bob holds indices $\ist$ and $\jst$.
    Alice sends a message to Bob, and Bob outputs $M_{\ist\jst}$.
    In such a game, any randomized protocol where Bob outputs $M_{\ist\jst}$ correctly with probability at least $\frac{99}{100}$ requires $\Omega\paren{N^2}$ bits of communication.
\end{proposition}

Note that the input size stated in \Cref{prop:matrix-index} is $N^2$ on Alice, and so the lower bound is linear in the input.
We choose to phrase $\INDEX$ in this way as a matter of convenience.

\paragraph{The Reduction.}
An instance of $\INDEX$ is reduced to an instance of \Cref{prob:min-frust-partition}, on the high level, in the following way.
A matrix with $N^2$ entries yields a graph with $O\paren{N}$ vertices.
First, there are the ``row'' vertices $R = \set{r_1, r_2, \dots, r_N}$.
Then, there are two sets of ``column'' vertices $C^0 = \set{c^0_1, c^0_2, \dots, c^0_N}$ and $C^1 = \set{c^1_1, c^1_2, \dots, c^1_N}$.
Finally there are the ``solution-structure forcing'' vertices $V = \set{v_1, v_2, \dots, v_{2\dc N}}$, where $\dc=O(1)$ is a large constant.
The purpose of $V$ is to force a very large partition $\paren{V_\leftarrow,V_\rightarrow}$ thereof, where all vertices in the same part are connected with \plab edges and all vertices in different parts are connected with \mlab edges.
Then, the other vertices are `assigned' to $\paren{V_\leftarrow,V_\rightarrow}$, via edges streamed by Alice and Bob, so that the part they belong to in the optimal solution is clearly determined.
In the optimal solution,
\begin{itemize}
    \item if $r_{\ist}$ and $c^0_{\jst}$ are in the same part, $M_{\ist\jst} = 0$;
    \item if $r_{\ist}$ and $c^1_{\jst}$ are in the same part, $M_{\ist\jst} = 1$.
\end{itemize}
Now, who is responsible for streaming which edges?
In the reduction to follow, Alice will stream edges that connect $R$ with $C^0$ and $C^1$.
Bob will make two very large cliques of \plab edges and connect them with \mlab edges.
Roughly, all vertices in $S \in \set{V,R,C^0,C^1}$ with indices no more than $\card{S}/2$ will belong to the first clique, and vertices with indices larger than $\card{S}/2$ will belong to the second clique.
Exceptions are made for how $r_{\ist}, c^0_{\jst}, c^1_{\jst}$ are handled; only Bob knows that these vertices are special as only he knows $\ist,\jst$.

\begin{tbox}
\label{red:mat-ind-to-exact-frust}
	\textbf{Reduction 3: $\INDEX$ to Frustration-minimizing Partition}:
	\begin{itemize}[label=($\alph*$)]
	    \item[\textbf{Alice}]
    	\begin{enumerate}[label=($\arabic*$)]
    	    \item Feed $\paren{r_i,c^{M_{ij}}_j}$, \plab into stream, for all $i,j \in \bracket{N}$.
    	    \item Feed $\paren{r_i,c^{1-M_{ij}}_j}$, \mlab into stream, for all $i,j \in \bracket{N}$.
    		\item Pass the memory of the streaming algorithm to Bob.
    	\end{enumerate}
    	\item[\textbf{Bob}]
    	\begin{enumerate}[label=($\arabic*$)]
    	    \item Let $V_\leftarrow = \set{v_1, v_2, \dots, v_{\dc N}}$ and $V_\rightarrow = \set{v_{\dc N+1}, v_{\dc N+2}, \dots, v_{2\dc N}}$, where $\dc=O(1)$ is a large constant.
    	    \item Let $R_\leftarrow$ be the $\lfloor\frac{N}{2}\rfloor$ vertices in $R \setminus \set{r_{\ist}}$ with the smallest indices, and $R_\rightarrow$ be the remaining vertices.
    	    \item Let $C^x_\leftarrow$ be the $\lfloor\frac{N}{2}\rfloor$ vertices in $C^x \setminus \set{c^x_{\jst}}$ with the smallest indices, and $C^x_\rightarrow$ be the remaining vertices, for $x \in \set{0,1}$.
    	    \item Feed \plab edges into stream to make $c^0_{\jst}, C^0_\leftarrow, C^1_\leftarrow, V_\leftarrow$ a \plab clique, $c^1_{\jst}, C^0_\rightarrow, C^1_\rightarrow, V_\rightarrow$ a \plab clique, and \mlab edges to connect the two cliques.
    	    \item Feed \plab edges into stream to make $R_\leftarrow$ and $R_\rightarrow$ \plab cliques each, and \mlab edges to connect the two cliques.
    	    \item Feed \plab edges into stream connecting $R_x$ with $V_x$, and \mlab edges into stream connecting $R_x$ with $V_y$, for $x \in \set{\leftarrow, \rightarrow}$ and $y \in \set{\leftarrow, \rightarrow} \setminus \set{x}$.
    	    \item\label{line:rist-assign} Feed \plab edges into stream to connect $r_{\ist}$ with $V_\leftarrow$, $V_\rightarrow$, $R_\leftarrow$, and $R_\rightarrow$.
    	    \item If $r_{\ist}$ and $c^x_{\jst}$ are in the same part, output $M_{\ist\jst}=x$.
    	\end{enumerate}
	\end{itemize}
\end{tbox}
\noindent

\begin{figure}[!htb]
\centering
\includegraphics[width=1.0\columnwidth]{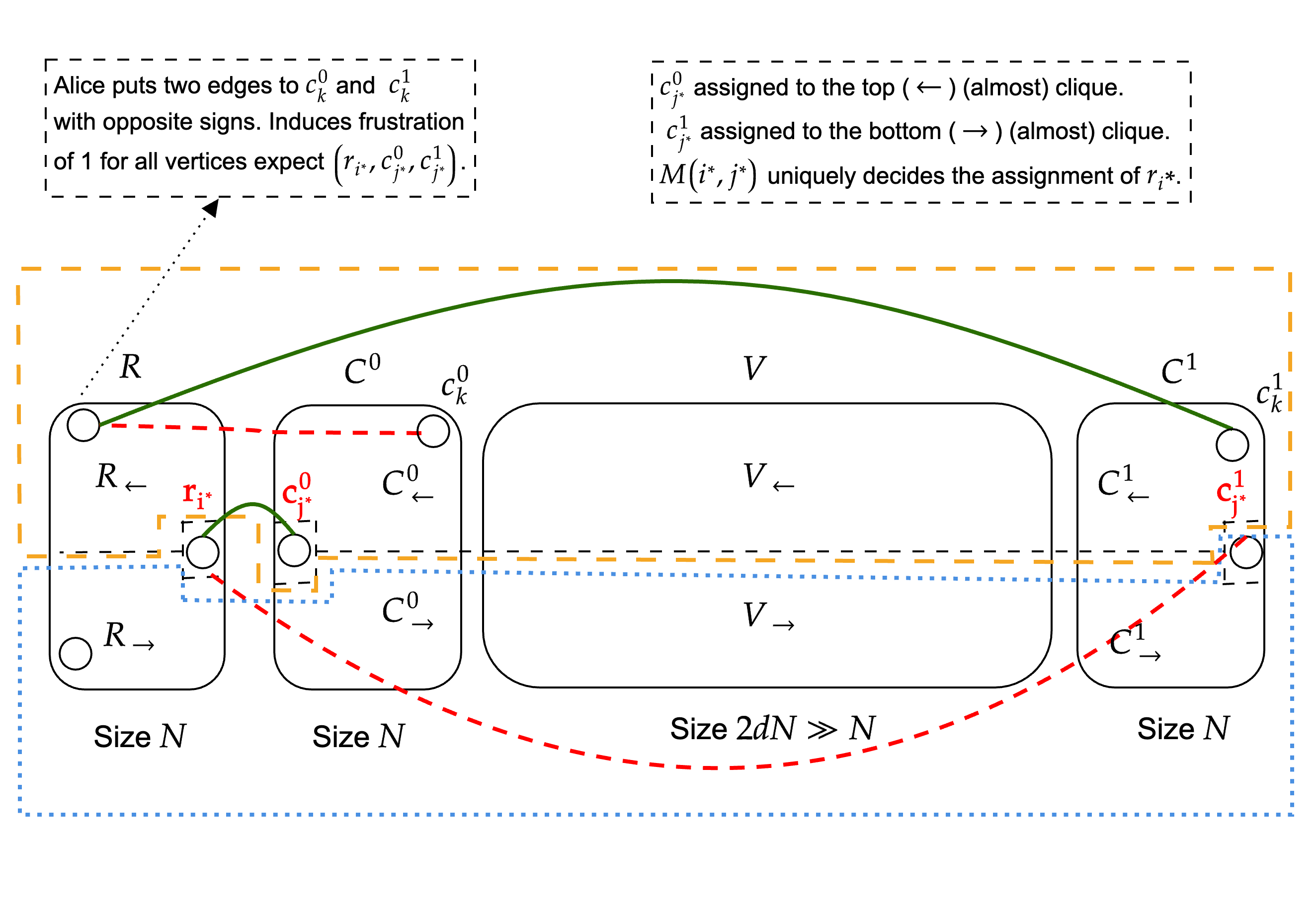}
\caption{An illustration of Reduction 3. }\label{fig:optimal-partition-index-reduction}
\end{figure}

An illustration of Reduction 3 can be found in \Cref{fig:optimal-partition-index-reduction}. We now characterize the unique optimal partition for any instance to \Cref{prob:min-frust-partition} generated by Reduction~3, and show that the next best partition has frustration that differs from the frustration of $\OPT$ by $2$. % We slightly abuse the notation to use $\OPT$ to denote both the best partition and the frustration.
% This on its own shows a $\Omega\paren{n^2}$ space lower bound for solving \Cref{prob:min-frust-partition} exactly.

\begin{claim}
\label{claim:lb-best-partn}
    The unique optimal partition (up to symmetry) for any instance to \Cref{prob:min-frust-partition} generated by Reduction~3 is
    \begin{align*}
        \OPT = \paren{V_\leftarrow \cup R_\leftarrow \cup C^0_\leftarrow \cup C^1_\leftarrow \cup \set{c^0_{\jst}} \cup S_0, V_\rightarrow \cup R_\rightarrow \cup C^0_\rightarrow \cup C^1_\rightarrow \cup \set{c^1_{\jst}} \cup S_1}
    \end{align*}
    where $S_x = \set{r_{\ist}}$ if $M_{\ist\jst} = x$ and $S_x = \set{}$ otherwise, for $x \in \set{0,1}$.
    
    The second-best partition is
    \begin{align*}
        \paren{V_\leftarrow \cup R_\leftarrow \cup C^0_\leftarrow \cup C^1_\leftarrow \cup \set{c^0_{\jst}} \cup S_1, V_\rightarrow \cup R_\rightarrow \cup C^0_\rightarrow \cup C^1_\rightarrow \cup \set{c^1_{\jst}} \cup S_0},
    \end{align*}
    whose frustration differs from that of $\OPT$ by $2$.
\end{claim}
\begin{proof}
    First, note that the frustration of $\OPT$ is exactly
    \begin{align*}
        N\paren{N-1} + \paren{\dc + 1}N + 2N = N^2 + \paren{\dc + 2} N.
    \end{align*}
    In the above, the first term comes from edges connecting $R$ with edges connecting $C^x_y$ for $x \in \set{0,1}$ and $y \in \set{\leftarrow, \rightarrow}$.
    The second term comes from edges connecting $r_{\ist}$ with $R \cup V$.
    The third term comes from edges connecting $R$ with $c^x_{\jst}$ for $x \in \set{0,1}$. % , and is in prop:equality \vik{Stopped correcting these because I feel like it's not just missing Cref. It's the wrong prop probably. I think I caught all of them.} why this bound is ``at most'' and not ``exactly''.
    
    To see that $\OPT$ is indeed the optimal partition, note first that any optimal partition needs to have at least a $2/3$ fraction of $V_\leftarrow$ in one part, and a $2/3$ fraction of $V_\rightarrow$ in the other part.
    Any partition not satisfying the above property has frustration at least $\frac{2}{9}\dc^2 N^2$, which is larger than $N^2 + \paren{\dc + 2} N$ for a sufficiently large $\dc$.
    
    Taking any partition that satisfies the above property but differs from $\OPT$, we show that there is a vertex that, when moved to the other part, strictly decreases the frustration.
    Let $\widetilde{V}_\leftarrow$ be the $2/3$ fraction of $V_\leftarrow$ in the left part, and $\widetilde{V}_\rightarrow$ be the $2/3$ fraction of $V_\rightarrow$ in the right part.
    Suppose without loss of generality that $u \neq r_{\ist}$ is in the left part in $\OPT$ and the right part in this partition.
    Then there are at least $4 \dc N / 3$ edges incident to $u$ that contribute to the frustration (these are the edges connecting $u$ with $\widetilde{V}_\leftarrow \cup \widetilde{V}_\rightarrow$.
    In other words, there are at most $2 \dc N / 3 + 3N$ edges incident to $u$ not contributing to the frustration.
    Switching $u$ to the other side thus decreases the frustration by at least $2 \dc N / 3 - 3N$, which is strictly positive for large enough $\dc$.
    Now, if only $r_{\ist}$ had been in the wrong part compared with $\OPT$, then moving it reduces the frustration by $2$.
    
    The optimal partition is thus
    \begin{align*}
        \paren{V_\leftarrow \cup R_\leftarrow \cup C^0_\leftarrow \cup C^1_\leftarrow \cup \set{c^0_{\jst}} \cup S_0, V_\rightarrow \cup R_\rightarrow \cup C^0_\rightarrow \cup C^1_\rightarrow \cup \set{c^1_{\jst}} \cup S_1},
    \end{align*}
    and the second-best partition is
    \begin{align*}
        \paren{V_\leftarrow \cup R_\leftarrow \cup C^0_\leftarrow \cup C^1_\leftarrow \cup \set{c^0_{\jst}} \cup S_1, V_\rightarrow \cup R_\rightarrow \cup C^0_\rightarrow \cup C^1_\rightarrow \cup \set{c^1_{\jst}} \cup S_0}
    \end{align*}
    with frustration $2$ higher than that of $\OPT$.
\end{proof}

\begin{proof}[Proof of \Cref{prop:frust-min-lb}]
By \Cref{claim:lb-best-partn}, an $\Omega\paren{N^2}$ communication lower bound implies a $(\frac{1}{2\dc + 3})^2\cdot n^2$ streaming memory lower bound to find the optimal partition.
Since the graph we constructed has $n=(2\dc + 3)\cdot N$ vertices, and $\dc$ is a constant (setting $\dc=100$ suffices for the proof), we get the desired $\Omega(n^2)$ lower bound.
This settles the lower bound for \emph{finding} the exact frustration-minimizing partition in \Cref{prop:frust-min-lb}.

The lower bound for streaming algorithms that return the optimal frustration value is slightly more involved.
Since there is \emph{no value gap} between the $M_{\ist\jst}=0$ and $M_{\ist\jst}=1$ cases (and Bob \emph{cannot} compute the `expected' exact frustration from the distribution), we need to apply the following trick.

\begin{tbox}
\label{red:mat-ind-to-exact-frust}
\textbf{A Trick to Adapt Reduction 3 to the Frustration Value Algorithm}:
\begin{enumerate}
    \item Alice runs the frustration value streaming algorithm in the same way as Reduction 3, and passes the memory to Bob.
    \item Bob runs two copies of the algorithm:
    \begin{enumerate}
        \item In the first copy, Bob constructs a graph in exactly the same way as Reduction 3, and runs the streaming algorithm to obtain frustration value $\textnormal{frust}_1$.
        \item In the second copy, Bob constructs a graph by changing Line~\ref{line:rist-assign} of Reduction 3: Bob adds a single \mlab edge to $V_\rightarrow$, and keeps everything else the same.
            Bob then runs the streaming algorithm to obtain frustration value $\textnormal{frust}_2$.
    \end{enumerate}
    \item Output $M_{\ist\jst}=0$ if $\textnormal{frust}_2<\textnormal{frust}_1$; output $M_{\ist\jst}=1$ otherwise. 
\end{enumerate}
\end{tbox}
\noindent
With a similar analysis as we used in \Cref{claim:lb-best-partn}, it is not hard to see that $\textnormal{frust}_2<\textnormal{frust}_1$ when $r_{\ist}$ is in the same almost-clique of $V_\leftarrow$ and $c^0_{\jst}$, which completes the proof.
\end{proof}

\subsection*{Acknowledgement} 
We thank Karthik C.S. for some preliminary discussions. 

\clearpage

\bibliographystyle{alpha}
\bibliography{general,structural-balance}

\newcommand{\etalchar}[1]{$^{#1}$}
\begin{thebibliography}{KMM{\etalchar{+}}20}

\bibitem[ACG{\etalchar{+}}15]{AhnCGMW15}
Kook~Jin Ahn, Graham Cormode, Sudipto Guha, Andrew McGregor, and Anthony Wirth.
\newblock Correlation clustering in data streams.
\newblock In Francis~R. Bach and David~M. Blei, editors, {\em Proceedings of
  the 32nd International Conference on Machine Learning, {ICML} 2015, Lille,
  France, 6-11 July 2015}, volume~37 of {\em {JMLR} Workshop and Conference
  Proceedings}, pages 2237--2246. JMLR.org, 2015.

\bibitem[ACK19]{AssadiCK19}
Sepehr Assadi, Yu~Chen, and Sanjeev Khanna.
\newblock Polynomial pass lower bounds for graph streaming algorithms.
\newblock In {\em Proceedings of the 51st Annual {ACM} {SIGACT} Symposium on
  Theory of Computing, {STOC} 2019, Phoenix, AZ, USA, June 23-26, 2019.}, pages
  265--276, 2019.

\bibitem[ACMM05]{Agarwal2005-zr}
Amit Agarwal, Moses Charikar, Konstantin Makarychev, and Yury Makarychev.
\newblock {O($\surd$log} n) approximation algorithms for min {UnCut}, min
  {2CNF} deletion, and directed cut problems.
\newblock In {\em Proceedings of the thirty-seventh annual {ACM} symposium on
  Theory of computing - {STOC} '05}, New York, New York, USA, 2005. ACM Press.

\bibitem[ACN05]{AilonCN05}
Nir Ailon, Moses Charikar, and Alantha Newman.
\newblock Aggregating inconsistent information: ranking and clustering.
\newblock In Harold~N. Gabow and Ronald Fagin, editors, {\em Proceedings of the
  37th Annual {ACM} Symposium on Theory of Computing, Baltimore, MD, USA, May
  22-24, 2005}, pages 684--693. {ACM}, 2005.

\bibitem[AG09]{AhnG09}
Kook~Jin Ahn and Sudipto Guha.
\newblock Graph sparsification in the semi-streaming model.
\newblock In Susanne Albers, Alberto Marchetti{-}Spaccamela, Yossi Matias,
  Sotiris~E. Nikoletseas, and Wolfgang Thomas, editors, {\em Automata,
  Languages and Programming, 36th Internatilonal Colloquium, {ICALP} 2009,
  Rhodes, Greece, July 5-12, 2009, Proceedings, Part {II}}, volume 5556 of {\em
  Lecture Notes in Computer Science}, pages 328--338. Springer, 2009.

\bibitem[AGM12]{AhnGM12b}
Kook~Jin Ahn, Sudipto Guha, and Andrew McGregor.
\newblock Graph sketches: sparsification, spanners, and subgraphs.
\newblock In {\em Proceedings of the 31st {ACM} {SIGMOD-SIGACT-SIGART}
  Symposium on Principles of Database Systems, {PODS} 2012, Scottsdale, AZ,
  USA, May 20-24, 2012}, pages 5--14, 2012.

\bibitem[AKR06]{Antal2006-jy}
T~Antal, P~L Krapivsky, and S~Redner.
\newblock Social balance on networks: The dynamics of friendship and enmity,
  2006.

\bibitem[AL07]{Avidor2007-do}
Adi Avidor and Michael Langberg.
\newblock The multi-multiway cut problem.
\newblock {\em Theor. Comput. Sci.}, 377(1):35--42, May 2007.

\bibitem[Alt12]{altafini2012dynamics}
Claudio Altafini.
\newblock Dynamics of opinion forming in structurally balanced social networks.
\newblock {\em PloS one}, 7(6):e38135, 2012.

\bibitem[AMW20]{Aref2020-vf}
Samin Aref, Andrew~J Mason, and Mark~C Wilson.
\newblock A modeling and computational study of the frustration index in signed
  networks.
\newblock {\em Networks}, 75(1):95--110, January 2020.

\bibitem[AR58]{alberson58symbolic}
Robert~P. Abelson and Milton~J. Rosenberg.
\newblock Symbolic psycho-logic: A model of attitudinal cognition.
\newblock {\em Behavioral Science}, 3(1):1--13, 1958.

\bibitem[AR20]{AssadiR20}
Sepehr Assadi and Ran Raz.
\newblock Near-quadratic lower bounds for two-pass graph streaming algorithms.
\newblock In Sandy Irani, editor, {\em 61st {IEEE} Annual Symposium on
  Foundations of Computer Science, {FOCS} 2020, Durham, NC, USA, November
  16-19, 2020}, pages 342--353. {IEEE}, 2020.

\bibitem[AW22]{AssadiW22}
Sepehr Assadi and Chen Wang.
\newblock {Sublinear Time and Space Algorithms for Correlation Clustering via
  Sparse-Dense Decompositions}.
\newblock In Mark Braverman, editor, {\em 13th Innovations in Theoretical
  Computer Science Conference (ITCS 2022)}, volume 215 of {\em Leibniz
  International Proceedings in Informatics (LIPIcs)}, pages 10:1--10:20,
  Dagstuhl, Germany, 2022. Schloss Dagstuhl -- Leibniz-Zentrum f{\"u}r
  Informatik.

\bibitem[Bar82]{Barahona1982-ma}
Francisco Barahona.
\newblock On the computational complexity of {Ising} spin glass models.
\newblock {\em Journal of Physics A: Mathematical and General},
  15(10):3241--3253, 1982.

\bibitem[BBC04]{BansalBC04}
Nikhil Bansal, Avrim Blum, and Shuchi Chawla.
\newblock Correlation clustering.
\newblock {\em Mach. Learn.}, 56(1-3):89--113, 2004.

\bibitem[BCMT22]{BehnezhadCMT22}
Soheil Behnezhad, Moses Charikar, Weiyun Ma, and Li{-}Yang Tan.
\newblock Almost 3-approximate correlation clustering in constant rounds.
\newblock In {\em 63rd {IEEE} Annual Symposium on Foundations of Computer
  Science, {FOCS} 2022, Denver, CO, USA, October 31 - November 3, 2022}, pages
  720--731. {IEEE}, 2022.

\bibitem[BCMT23]{BehnezhadCMT23}
Soheil Behnezhad, Moses Charikar, Weiyun Ma, and Li{-}Yang Tan.
\newblock Single-pass streaming algorithms for correlation clustering.
\newblock In Nikhil Bansal and Viswanath Nagarajan, editors, {\em Proceedings
  of the 2023 {ACM-SIAM} Symposium on Discrete Algorithms, {SODA} 2023,
  Florence, Italy, January 22-25, 2023}, pages 819--849. {SIAM}, 2023.

\bibitem[BEM17]{BateniEM17}
MohammadHossein Bateni, Hossein Esfandiari, and Vahab~S. Mirrokni.
\newblock Almost optimal streaming algorithms for coverage problems.
\newblock In {\em Proceedings of the 29th {ACM} Symposium on Parallelism in
  Algorithms and Architectures, {SPAA} 2017, Washington DC, USA, July 24-26,
  2017}, pages 13--23, 2017.

\bibitem[BJW22]{BawejaJW22}
Anubhav Baweja, Justin Jia, and David~P. Woodruff.
\newblock An efficient semi-streaming {PTAS} for tournament feedback arc set
  with few passes.
\newblock In Mark Braverman, editor, {\em 13th Innovations in Theoretical
  Computer Science Conference, {ITCS} 2022, January 31 - February 3, 2022,
  Berkeley, CA, {USA}}, volume 215 of {\em LIPIcs}, pages 16:1--16:23. Schloss
  Dagstuhl - Leibniz-Zentrum f{\"{u}}r Informatik, 2022.

\bibitem[BK96]{BenczurK96}
Andr{\'{a}}s~A. Bencz{\'{u}}r and David~R. Karger.
\newblock Approximating \emph{s-t} minimum cuts in
  \emph{{\~{O}}}(\emph{n}\({}^2\)) time.
\newblock In Gary~L. Miller, editor, {\em Proceedings of the Twenty-Eighth
  Annual {ACM} Symposium on the Theory of Computing, Philadelphia,
  Pennsylvania, USA, May 22-24, 1996}, pages 47--55. {ACM}, 1996.

\bibitem[BV10]{BogdanovV10}
Andrej Bogdanov and Emanuele Viola.
\newblock Pseudorandom bits for polynomials.
\newblock {\em {SIAM} J. Comput.}, 39(6):2464--2486, 2010.

\bibitem[CDK14]{ChierichettiDK14}
Flavio Chierichetti, Nilesh~N. Dalvi, and Ravi Kumar.
\newblock Correlation clustering in mapreduce.
\newblock In Sofus~A. Macskassy, Claudia Perlich, Jure Leskovec, Wei Wang, and
  Rayid Ghani, editors, {\em The 20th {ACM} {SIGKDD} International Conference
  on Knowledge Discovery and Data Mining, {KDD} '14, New York, NY, {USA} -
  August 24 - 27, 2014}, pages 641--650. {ACM}, 2014.

\bibitem[CGMV20]{ChakrabartiG0V20}
Amit Chakrabarti, Prantar Ghosh, Andrew McGregor, and Sofya Vorotnikova.
\newblock Vertex ordering problems in directed graph streams.
\newblock In {\em Proceedings of the 2020 {ACM-SIAM} Symposium on Discrete
  Algorithms, {SODA} 2020, Salt Lake City, UT, USA, January 5-8, 2020}, pages
  1786--1802, 2020.

\bibitem[CGW03]{CharikarGW03}
Moses Charikar, Venkatesan Guruswami, and Anthony Wirth.
\newblock Clustering with qualitative information.
\newblock In {\em 44th Symposium on Foundations of Computer Science {(FOCS}
  2003), 11-14 October 2003, Cambridge, MA, USA, Proceedings}, pages 524--533.
  {IEEE} Computer Society, 2003.

\bibitem[CH56]{Cartwright1956-fc}
Dorwin Cartwright and Frank Harary.
\newblock Structural balance: a generalization of {Heider's} theory.
\newblock {\em Psychol. Rev.}, 63(5):277--293, September 1956.

\bibitem[Che52]{chernoff1952measure}
Herman Chernoff.
\newblock A measure of asymptotic efficiency for tests of a hypothesis based on
  the sum of observations.
\newblock {\em The Annals of Mathematical Statistics}, pages 493--507, 1952.

\bibitem[CLM{\etalchar{+}}21]{Cohen-AddadLMNP21}
Vincent Cohen{-}Addad, Silvio Lattanzi, Slobodan Mitrovic, Ashkan
  Norouzi{-}Fard, Nikos Parotsidis, and Jakub Tarnawski.
\newblock Correlation clustering in constant many parallel rounds.
\newblock In Marina Meila and Tong Zhang, editors, {\em Proceedings of the 38th
  International Conference on Machine Learning, {ICML} 2021, 18-24 July 2021,
  Virtual Event}, volume 139 of {\em Proceedings of Machine Learning Research},
  pages 2069--2078. {PMLR}, 2021.

\bibitem[CLN22]{CohenAddadLN22}
Vincent Cohen{-}Addad, Euiwoong Lee, and Alantha Newman.
\newblock Correlation clustering with sherali-adams.
\newblock {\em CoRR}, abs/2207.10889, 2022.

\bibitem[CM23]{chakrabarty2023single}
Sayak Chakrabarty and Konstantin Makarychev.
\newblock Single-pass pivot algorithm for correlation clustering. keep it
  simple!
\newblock {\em arXiv preprint arXiv:2305.13560}, 2023.

\bibitem[CMSY15]{ChawlaMSY15}
Shuchi Chawla, Konstantin Makarychev, Tselil Schramm, and Grigory Yaroslavtsev.
\newblock Near optimal {LP} rounding algorithm for correlation clustering on
  complete and complete $k$-partite graphs.
\newblock In Rocco~A. Servedio and Ronitt Rubinfeld, editors, {\em Proceedings
  of the Forty-Seventh Annual {ACM} on Symposium on Theory of Computing, {STOC}
  2015, Portland, OR, USA, June 14-17, 2015}, pages 219--228. {ACM}, 2015.

\bibitem[CT06]{CoverT06}
Thomas~M. Cover and Joy~A. Thomas.
\newblock {\em Elements of information theory {(2.} ed.)}.
\newblock Wiley, 2006.

\bibitem[Dav67]{davis1967clustering}
James~A Davis.
\newblock Clustering and structural balance in graphs.
\newblock {\em Human relations}, 20(2):181--187, 1967.

\bibitem[DESZ07]{DasGupta2007-ig}
Bhaskar DasGupta, German~Andres Enciso, Eduardo Sontag, and Yi~Zhang.
\newblock Algorithmic and complexity results for decompositions of biological
  networks into monotone subsystems.
\newblock {\em Biosystems.}, 90(1):161--178, July 2007.

\bibitem[GG06]{giotis2005correlation}
Ioannis Giotis and Venkatesan Guruswami.
\newblock Correlation clustering with a fixed number of clusters.
\newblock In {\em Proceedings of the Seventeenth Annual ACM-SIAM Symposium on
  Discrete Algorithm}, SODA '06, page 1167–1176, USA, 2006. Society for
  Industrial and Applied Mathematics.

\bibitem[Har53]{Harary1953-na}
Frank Harary.
\newblock On the notion of balance of a signed graph.
\newblock {\em Michigan Math. J.}, 2(2), January 1953.

\bibitem[Har02]{Harary2002-vf}
Frank Harary.
\newblock Signed graphs for portfolio analysis in risk management.
\newblock {\em IMA Journal of Management Mathematics}, 13(3):201--210, 2002.

\bibitem[Har07]{Harary2007-yy}
Frank Harary.
\newblock On the measurement of structural balance.
\newblock {\em Behavioral Science}, 4(4):316--323, 2007.

\bibitem[HBN10]{Huffner2010-wl}
Falk H{\"u}ffner, Nadja Betzler, and Rolf Niedermeier.
\newblock Separator-based data reduction for signed graph balancing.
\newblock {\em J. Comb. Optim.}, 20(4):335--360, November 2010.

\bibitem[Hei46]{Heider1946-bt}
Fritz Heider.
\newblock Attitudes and cognitive organization.
\newblock {\em J. Psychol.}, 21:107--112, January 1946.

\bibitem[Hei82]{Heider1982-rg}
Fritz Heider.
\newblock {\em The Psychology of Interpersonal Relations}.
\newblock Psychology Press, 1982.

\bibitem[Hoe94]{hoeffding1994probability}
Wassily Hoeffding.
\newblock Probability inequalities for sums of bounded random variables.
\newblock In {\em The collected works of Wassily Hoeffding}, pages 409--426.
  Springer, 1994.

\bibitem[Ind06]{indyk2006stable}
Piotr Indyk.
\newblock Stable distributions, pseudorandom generators, embeddings, and data
  stream computation.
\newblock {\em Journal of the ACM (JACM)}, 53(3):307--323, 2006.

\bibitem[IRSA10]{Iacono2010-iz}
Giovanni Iacono, Fahimeh Ramezani, Nicola Soranzo, and Claudio Altafini.
\newblock Determining the distance to monotonicity of a biological network: a
  graph-theoretical approach.
\newblock {\em IET Syst. Biol.}, 4(3):223--235, May 2010.

\bibitem[Kho02]{Khot2002-ue}
Subhash Khot.
\newblock On the power of unique 2-prover 1-round games.
\newblock In {\em Proceedings of the thiry-fourth annual {ACM} symposium on
  Theory of computing}, STOC '02, pages 767--775, New York, NY, USA, May 2002.
  Association for Computing Machinery.

\bibitem[KLM{\etalchar{+}}14]{KapralovLMMS14}
Michael Kapralov, Yin~Tat Lee, Cameron Musco, Christopher Musco, and Aaron
  Sidford.
\newblock Single pass spectral sparsification in dynamic streams.
\newblock In {\em 55th {IEEE} Annual Symposium on Foundations of Computer
  Science, {FOCS} 2014, Philadelphia, PA, USA, October 18-21, 2014}, pages
  561--570, 2014.

\bibitem[KLM{\etalchar{+}}17]{kapralov2017single}
Michael Kapralov, Yin~Tat Lee, CN~Musco, Christopher~Paul Musco, and Aaron
  Sidford.
\newblock Single pass spectral sparsification in dynamic streams.
\newblock {\em SIAM Journal on Computing}, 46(1):456--477, 2017.

\bibitem[KMM{\etalchar{+}}20]{KapralovMMMNST20}
Michael Kapralov, Aida Mousavifar, Cameron Musco, Christopher Musco, Navid
  Nouri, Aaron Sidford, and Jakab Tardos.
\newblock Fast and space efficient spectral sparsification in dynamic streams.
\newblock In Shuchi Chawla, editor, {\em Proceedings of the 2020 {ACM-SIAM}
  Symposium on Discrete Algorithms, {SODA} 2020, Salt Lake City, UT, USA,
  January 5-8, 2020}, pages 1814--1833. {SIAM}, 2020.

\bibitem[KSGN11]{kivran2011impact}
Funda Kivran-Swaine, Priya Govindan, and Mor Naaman.
\newblock The impact of network structure on breaking ties in online social
  networks: unfollowing on twitter.
\newblock In {\em Proceedings of the SIGCHI conference on human factors in
  computing systems}, pages 1101--1104. ACM, 2011.

\bibitem[Lov09]{Lovett09}
Shachar Lovett.
\newblock Unconditional pseudorandom generators for low degree polynomials.
\newblock {\em Theory Comput.}, 5(1):69--82, 2009.

\bibitem[McG14]{McGregor14}
Andrew McGregor.
\newblock Graph stream algorithms: a survey.
\newblock {\em {SIGMOD} Rec.}, 43(1):9--20, 2014.

\bibitem[Moo78]{Moore1978-by}
Michael Moore.
\newblock An international application of {Heider's} balance theory.
\newblock {\em Eur. J. Soc. Psychol.}, 8(3):401--405, July 1978.

\bibitem[MR95]{MotwaniR95}
Rajeev Motwani and Prabhakar Raghavan.
\newblock {\em Randomized Algorithms}.
\newblock Cambridge University Press, 1995.

\bibitem[Nis92]{nisan1992pseudorandom}
Noam Nisan.
\newblock Pseudorandom generators for space-bounded computation.
\newblock {\em Combinatorica}, 12(4):449--461, 1992.

\bibitem[NN93]{NaorN93}
Joseph Naor and Moni Naor.
\newblock Small-bias probability spaces: Efficient constructions and
  applications.
\newblock {\em {SIAM} J. Comput.}, 22(4):838--856, 1993.

\bibitem[PY88]{Papadimitriou1988-bh}
Christos Papadimitriou and Mihalis Yannakakis.
\newblock Optimization, approximation, and complexity classes.
\newblock In {\em Proceedings of the twentieth annual {ACM} symposium on Theory
  of computing}, STOC '88, pages 229--234, New York, NY, USA, January 1988.
  Association for Computing Machinery.

\bibitem[Rot06]{Roth06Coding}
Ron~M. Roth.
\newblock {\em Introduction to coding theory}.
\newblock Cambridge University Press, 2006.

\bibitem[Sib14]{Sibona2014-ni}
Christopher Sibona.
\newblock Unfriending on facebook: Context collapse and unfriending behaviors.
\newblock In {\em 2014 47th Hawaii International Conference on System
  Sciences}, pages 1676--1685. ieeexplore.ieee.org, January 2014.

\bibitem[SW15]{SunW15}
Xiaoming Sun and David~P. Woodruff.
\newblock Tight bounds for graph problems in insertion streams.
\newblock In Naveen Garg, Klaus Jansen, Anup Rao, and Jos{\'{e}} D.~P. Rolim,
  editors, {\em Approximation, Randomization, and Combinatorial Optimization.
  Algorithms and Techniques, {APPROX/RANDOM} 2015, August 24-26, 2015,
  Princeton, NJ, {USA}}, volume~40 of {\em LIPIcs}, pages 435--448. Schloss
  Dagstuhl - Leibniz-Zentrum f{\"{u}}r Informatik, 2015.

\bibitem[TCAL16]{tang2016survey}
Jiliang Tang, Yi~Chang, Charu Aggarwal, and Huan Liu.
\newblock A survey of signed network mining in social media.
\newblock {\em ACM Computing Surveys (CSUR)}, 49(3):1--37, 2016.

\bibitem[TSF17]{Teixeira2017-vk}
Andreia~Sofia Teixeira, Francisco~C Santos, and Alexandre~P Francisco.
\newblock Emergence of social balance in signed networks.
\newblock In {\em Complex Networks {VIII}}, pages 185--192. Springer
  International Publishing, 2017.

\bibitem[WC81]{wegman1981new}
Mark~N Wegman and J~Lawrence Carter.
\newblock New hash functions and their use in authentication and set equality.
\newblock {\em Journal of computer and system sciences}, 22(3):265--279, 1981.

\bibitem[WLG22]{wang22coevolution}
Haotian Wang, Feng Luo, and Jie Gao.
\newblock Co-evolution of opinion and social tie dynamics towards structural
  balance.
\newblock In {\em Proceedings of the ACM-SIAM Symposium on Discrete Algorithms
  (SODA’22)}, pages 3362--3388, January 2022.

\bibitem[XHKC13]{Xu:2013:SBT:2441776.2441875}
Bo~Xu, Yun Huang, Haewoon Kwak, and Noshir Contractor.
\newblock Structures of broken ties: Exploring unfollow behavior on twitter.
\newblock In {\em Proceedings of the 2013 Conference on Computer Supported
  Cooperative Work}, CSCW '13, pages 871--876, New York, NY, USA, 2013. ACM.

\bibitem[Yan81]{Yannakakis1981-cz}
Mihalis Yannakakis.
\newblock {Edge-Deletion} problems.
\newblock {\em SIAM J. Comput.}, 10(2):297--309, May 1981.

\bibitem[Zas87]{Zaslavsky1987-gi}
Thomas Zas{\'l}avsky.
\newblock Balanced decompositions of a signed graph.
\newblock {\em J. Combin. Theory Ser. B}, 43(1):1--13, August 1987.

\end{thebibliography}

\clearpage

\appendix
\part*{Appendix}

\section{Standard Technical Tools}
\label{app:info-theoretic-facts}

\subsection{Concentration Inequalities}
\label{sub-app:concentration-inequ}
We now present the standard concentration inequalities used in our proofs. We start from the following standard variant of the Chernoff-Hoeffding bound. 

\begin{proposition}[Chernoff bound~\cite{chernoff1952measure}]\label{prop:chernoff}
	Let $X_1,\ldots,X_n$ be $n$ independent random variables with support on $\{0,1\}$. Define $X := \sum_{i=1}^{n} X_i$. Then, for every $\delta >0$, there is 
	\begin{align*}
	\Pr\paren{X\geq (1+\delta)\cdot \expect{X}} \leq \exp\paren{-\frac{\delta^2}{\delta+2}\cdot \expect{X}}.
	\end{align*}
	In particular, when $\delta\in (0,1]$, there is
	\begin{align*}
		\Pr\paren{\card{X - \expect{X}} > \delta\cdot \expect{X}} \leq 2 \cdot \exp\paren{-\frac{\delta^{2} \expect{X}}{3}}. 
	\end{align*}
\end{proposition}

\begin{proposition} [Additive Chernoff bound~\cite{chernoff1952measure, hoeffding1994probability}]\label{prop:additive-chernoff}
	Let $X_1,\ldots,X_n$ be $n$ independent random variables with support in $[0,1]$. Define $X := \sum_{i=1}^{n} X_i$. Then, for every $t>0$,
	\begin{align*}
	    \Pr\paren{\card{X - \expect{X}}>t} \leq 2\cdot 
	    \exp\paren{-\frac{2t^2}{n}}.
	\end{align*}
\end{proposition}

\subsection{Basic Background on Information Theory}

We review basic definitions from information theory that we used in our lower bound proof. We start with the basic definition of total variation distance. 

%Similar to the KL divergence we defined above, the total variation distance (TVD) is another statistical distance between two distributions. 
\begin{definition}
\label{def:tvd}
Let $X$ and $Y$ be two discrete random variables supported over the domain $\Omega$ with distributions $\mu_{X}$ and $\mu_{Y}$. The \textnormal{\textbf{total variation distance (TVD)}} between $X$ and $Y$ is defined as 
\begin{align*}
\tvd{X}{Y} := \frac{1}{2}\sum_{\omega\in \Omega}\card{\mu_X(\omega)-\mu_Y(\omega)}. 
\end{align*}
\end{definition}

The total variation distance to \emph{uniform distributions} is closely related to the success probability of algorithms. A useful property for the TVD is the triangle inequality for metrics.

\begin{fact}
\label{fct:tvd-triangle}
Let $X$, $Y$ and $Z$ be 3 discrete random variables supported over the domain $\Omega$, there is
\begin{align*}
\tvd{X}{Z} \leq \tvd{X}{Y} + \tvd{Y}{Z}.
\end{align*}
\end{fact}

For a random variable $X$, we let $\HH(X)$ be the \emph{Shannon entropy} of $X$, defined as follows
\begin{definition}
\label{def:entropy}
Let $X$ be a discrete random variable with distributions $\mu_{X}$, the \emph{Shannon entropy} of $X$ is defined as
\begin{align*}
\HH(X) := \expect{\log(1/\mu(X))} =\sum_{x \in \text{supp}(X)} \mu(x)\cdot \log(\frac{1}{\mu(x)}),
\end{align*}
where $\text{supp}(X)$ is the support of $X$. If $X$ is a Bernoulli random variable, we use $H_2(p)$ to denote its Shannon entropy, where $P$ is the probability for $X=1$.
\end{definition}

Let $X$, $Y$ be two random variables, we define the \emph{conditional entropy} as $\HH(X|Y)=\mathbb{E}_{y \sim Y}[\HH(X\mid Y=y)]$. Subsequently, we can define the \emph{mutual information} between $X$ and $Y$ as $\II\paren{X;Y}:=\HH(X)-\HH(X\mid Y) = \HH(Y)-\HH(Y|X)$, as standard.

We list the following facts~\cite{CoverT06} that are useful for our lower bound proof.
\begin{fact}
\label{fct:info-theory-facts}
Let $X$, $Y$, $Z$ be three discrete random variables:
\begin{itemize}
\item $0\leq \HH(X)\leq \log(\card{\text{supp}(X)})$. In particular, if $X$ is a Bernoulli random variable, there is $H_2(p)\leq 1$.
\item $0\leq \II(X;Y)\leq \min\{\HH(X), \HH(Y)\}$.
\item Conditioning on independent random variable: let $X$ be independent of $Z$, then $\II(X;Y)\leq \II(X;Y\mid Z)$.
\item Chain rule of mutual information: $\II(X, Y; Z)=\II(X;Z)+\II(Y;Z\mid X)$.
\item Sub-additivity of entropy: $\HH(X,Y) \leq \HH(X) + \HH(Y)$, where $\HH(X, Y)$ is the joint entropy of variables $X, Y$.
\end{itemize}
\end{fact}
Shannon entropy and mutual information are closely related to the necessary number of bits to be revealed in the communication problems to get a desired `success' probability. In particular, we use the celebrated Fano's inequality in our proof.
\begin{proposition}[Fano's inequality]
\label{prop:fano}
Let $X$, $Y$ be two discrete random variables and suppose there is a function $g: \text{supp}(Y)\rightarrow \text{supp}(X)$ such that $\Pr\paren{X\neq g(b)}\leq \delta$. Then, $\HH(X\mid Y)\leq H_2(\delta)$.
\end{proposition}

\subsection{Communication Games and Streaming Lower Bounds}\label{sec:communication}
In a multi-party communication game, each party is given certain inputs and aims to solve the problem together. The communication complexity is defined as the minimum number of bits required to send between the parties to solve the game. It is standard machinery to prove lower bounds of streaming algorithms via reduction from multi-player communication games, that is, communication complexity lower bounds imply streaming lower bounds.\par 
Suppose there are $n$ players $p_1, \dots, p_n $, $p_1$ send a message $m_1$ to $p_2$ and so on, we want the message $m_{n}$ to have enough information of the solution. This process can be simplified as a two-player game: Alice sends a message to Bob so that the problem can be solved. We briefly introduce a protocol showing how the reduction works:
\begin{itemize}
    \item Suppose Alice and Bob are trying to solve a function $f$. Alice is given the first half of the stream and Bob is given the second half.
    \item Alice internally computes the streaming algorithm with her input and sends the state of the streaming algorithm to Bob.
    \item Bob continues with his input and calls the streaming algorithm to output $f$.
\end{itemize}

If the communication complexity of this game is $t$, then sending the state of the streaming algorithms requires $t$ bits, which indicates that the streaming algorithm must use space at least $t$, therefore, we can show streaming lower bounds by reduction from a  communication game.

% \subsection{Technical Tools from Prior Work}\label{sec:tools}

% Now we include several technical tools needed to come to our results, which may be skipped and returned to on an as-needed basis. We briefly introduce cut sparsifiers, limited-independence hash functions and pseudorandom generators. The cut sparsifier is a constructive element of frustration evaluation for \Cref{prob:min-frust-partition} and the latter two are used in efficient sampling for \Cref{prob:structural-balance}.

\subsection{Cut Sparsifiers}\label{sec:cut-sparsifiers}
A cut sparsifier aims to `compress' the original graph to its subgraph with a substantially smaller number of edges to preserve the cut size in a graph. Concretely, an $\eps$-cut sparsifier $H$ of $G$ preserves the size of every cut in $G$ within a $(1\pm \eps)$ factor, formalized as follows. 

\begin{definition}[\textbf{Cut Sparsifier}]
\label{def:cutsparse}
	Given a graph $G=(V,E,w_G)$, we say that a weighted subgraph $H :=(V,E_H,w_H)$ is an $\eps$-cut sparsifier of $G$ if for all non-empty $S \subset V$, the following holds:
	\[
% 	(1-\eps)\cdot \sum_{(u,v)\in E_G}w_G(u,v)\cdot \mathbb{I}_{\{u\in S, v\notin S\}}
% 	\leq \sum_{(u,v)\in E_H}w_H(u,v) \cdot \mathbb{I}_{\{u\in S, v\notin S\}}
% 	\leq (1+\eps)\cdot  \sum_{(u,v)\in E_G}w_G(u,v)\cdot \mathbb{I}_{\{u\in S, v\notin S\}},
\sum_{(u,v)\in E_H}w_H(u,v) \cdot \mathbb{I}_{\{u\in S, v\notin S\}}
	\leq (1\pm\eps)\cdot \sum_{(u,v)\in E_G}w_G(u,v)\cdot \mathbb{I}_{\{u\in S, v\notin S\}},
	\]
	where $w_G(u,v)$ (resp. $w_H(u,v)$) denotes the  weight of edge $(u,v)$ in $G$ (resp. in $H$). 
\end{definition}

The work by Bencz{\'{u}}r and Karger \cite{BenczurK96} first shows that an $\eps$-cut sparsifier exists for any graph with $O(n\log(n)/\eps^2)$ edges, and it can be constructed in near-linear time ($\Otilde(n^2)$). Furthermore, in the graph streaming model, it is known that with $\tilde{O}(n/\eps^2)$ space, one can deterministically construct an $\eps$-sparsifier in a \emph{single} pass. We present the results for $\eps$-cut sparsifiers as follows.

\begin{proposition}[\!\cite{BenczurK96,AhnG09,McGregor14}]
\label{prop:eps-cutsparse-algs}
The following algorithms for $\eps$-cut sparsifiers are known.
\begin{enumerate}
\item There exists a randomized algorithm that given a unweighted graph $G=(V,E)$, with high probability computes an $\eps$-cut sparsifier with $O(\frac{n\log(n)}{\eps^2})$ edges in $O(n^2\cdot \log^2(n))$ time.
\item There exists a randomized streaming algorithm that given a graph $G=(V,E, w_G)$ in a single-pass graph stream, with high probability computes an $\eps$-cut sparsifier in $O(n^2\cdot \log^2(n))$ time with a memory of $O(\frac{n\cdot \log^3(n)}{\eps^2})$ words.
\end{enumerate}
\end{proposition}

The algorithms in \Cref{prop:eps-cutsparse-algs} are used in a blackbox way in our results.

\subsection{Limited-independent Hash functions}
\label{app:k-independent-hash}
Given $n$ random variables, we say they are $k$\textit{-wise independent} $(n\geq k)$ if any subset of size no more than $k$ of the variables are independent of each other. Then a $k$-wise independent hashing function~\cite{wegman1981new} ensures that hash codes of any $k$ inputs are $k$-wise independent random variables. We give the formal definition as follows:
\begin{definition}
\label{def:k-hash}
A family of hash functions $\mathcal{H} = \{h: U\rightarrow R \}$ is $k$-wise independent if for any distinct $x_1, \dots, x_k \in U$ and any $y_1, \dots, y_k \in R$, we have:
\begin{align*}
    \prob{h(x_1)=y_1\land \dots \land h(x_k)=y_k} = \frac{1}{|R|^k}.
\end{align*}
\end{definition}

The following result states that such a family of $k$-wise independent hash functions exists and only consumes roughly logarithm space when $k=O(1)$.
\begin{proposition}~\cite{MotwaniR95}
\label{prop:k-ind-hash}
For integers $n,m,k \geq 2$, there exists a family of $k$-wise independent hash functions $\mathcal{H} = \{h: [n]\rightarrow [m] \}$ such that sampling and storing a function $h\in \mathcal{H}$ takes $O(k\cdot(\log n + \log m))$ bits of space.
\end{proposition}

\subsection{Low-degree polynomials and the Schwartz–Zippel lemma}
We give the Schwartz–Zippel lemma for low-degree polynomial that is useful in identity testing. On the special case of polynomials over $\IF_{2}$, the lemma statement is as follows.
\begin{proposition}
\label{prop:sz-lemma}
Let $P$ be a degree-$C$ polynomial over $\IF_{2}$ with $n$ variables, and let $X=(X_{1},X_{2},\cdots,X_{n})$ be random variables chosen independently and uniformly at random from $\IF_{2}$. Then, there is
\begin{align*}
\Pr\paren{X=0}\leq \frac{C}{2}.
\end{align*}
\end{proposition}

\subsection{Pseudorandom Generators (PRGs)}
A pseudorandom generator is a deterministic procedure that converts \textit{true randomness} into \textit{pseudo-randomness}, therefore is a widely-used tool in derandomization. The notion of randomness here stems from \textit{computational indistinguishability}, which basically claims that a statistical test on two random variables returns the same output with probability $1-\varepsilon$. Formally, random variables $X, Y \in \{0,1\}^m$ are $\varepsilon$-\textit{indistinguishable}, then for any $f: \{0,1\}^m \rightarrow \{0,1\}$, $|\prob{f(X)=1}-\prob{f(Y)}=1|\leq \varepsilon$. Based on this, a pseudorandom generator stretches truly random seeds to longer strings, which are \textit{indistinguishable} with the seeds for any $f$. The definition follows:

\begin{definition}
\label{def:prg}
Given a family of functions $\mathcal{F}$, an $\varepsilon$-pseudorandom generator (PRG) is a deterministic mapping $g: \{0,1\}^l \rightarrow \{0,1\}^n$, where $n\geq l$, such that for any $f \in \mathcal{F}: \{0,1\}^n \rightarrow \{0,1\}$, $f(U_n)$ and $f(g(U_l))$ are $\varepsilon$-indistinguishable, where $U_n$ denotes a uniformly random string.
\end{definition}
% \jie{of length m? should we use n instead of m? here f operates on strings of length m but then $U_n$ has length n.}

Nisan \cite{nisan1992pseudorandom} gives an efficient construction of PRG for algorithms with bounded space, using a significantly smaller number of random bits. We formally state its properties below.

\begin{proposition}[Corollary~1 in \cite{nisan1992pseudorandom}]
\label{prop:nis}
    Any randomized algorithm running in $\textrm{space}\paren{S}$ and using $R$ (read-once) random bits may be converted to one that uses only $O\paren{S \log{R}}$ random bits (and runs in $\textrm{space}\paren{S \log{R}}$).
\end{proposition}

\paragraph{PRGs for Polynomials.} Besides the general notion in \Cref{def:prg}, pseudorandom generators are also studied under testers of special families of functions. In particular, if we restrict our attention to the family of \emph{polynomial functions} with constant (total) degree, there exist stronger PRGs that circumvent the read-once property of the psuedorandom bits. Formally, we define $\eps$-pseudorandom generators for degree-$C$ polynomials as follows.
\begin{definition}
\label{def:prg-poly}
Let $ \mathcal{P}_{C}$ be the family of degree-$C$ polynomial functions over $\IF_{2}$ with $n$ variables. An $\varepsilon$-pseudorandom generator (PRG) for degree-$C$ polynomials is a deterministic mapping $g: \{0,1\}^l \rightarrow \{0,1\}^n$, where $n\geq l$, such that for any $P\in \mathcal{P}_{C}$, there is 
\begin{align*}
\card{\Pr\paren{P\paren{g(U_{l})}=1}-\Pr\paren{P\paren{U_{n}}=1}}\leq \eps,
\end{align*}
where $U_n$ denotes a uniformly random string.
\end{definition}

It is known that one can construct a PRG for constant degree polynomials with logarithm number of truly random bits and logarithm extra space. In particular, for $C=2$ (degree-2) polynomials, we know the following results.

\begin{proposition}[\cite{NaorN93,Lovett09,BogdanovV10}]
\label{prop:prg-poly}
There exists an $\eps$-pseudorandom generator $g: \{0,1\}^l \rightarrow \{0,1\}^n$ for degree-2 polynomials that takes $O(\log{\frac{n}{\eps}})$ truly random bits and uses $O(\log{\frac{n}{\eps}})$ space to construct each pseudorandom bit. The pseudorandom bits are \emph{not} necessarily read-once.
\end{proposition}

The pseudorandom generator in \Cref{prop:prg-poly} avoids the read-once restriction of the pseudorandom bits in classical PRGs -- a property that is extremely useful in the streaming setting.

\subsection{$2$-lifts of Graphs}
Here we describe $2$-lifts, which is a graph transformation used in, for example, \Cref{prop:semi-streaming-test}.
\begin{definition}[$2$-lift]
\label{def:two-lift}
    Given a signed graph $G$, its $2$-lift is a graph $\hat{G}$.
    The vertices of $\hat{G}$ are defined in the following way: for each vertex $v \in V(G)$, there are two distinct vertices $v_0, v_1 \in V(\hat{G})$; thus $|V(\hat{G})| = 2|V(G)|$.
    The edges of $\hat{G}$ are defined in the following way: for each edge $(u,v) \in E(G)$:
    \begin{itemize}
        \item If the sign of $(u,v)$ is `$+$', then $(u_0,v_0), (u_1,v_1) \in E(\hat{G})$.
        \item If the sign of $(u,v)$ is `$-$', then $(u_0,v_1), (u_1,v_0) \in E(\hat{G})$.
    \end{itemize}
\end{definition}

\section{Structural Balance Testing in the Semi-streaming Setting}
\label{app:semi-streaming-testing}

% \chen{TO-DO: give a lower bound for general graph (lower bound for bipartiteness). }

We present the bounds of testing the structural balance in the semi-streaming memory scenario, i.e., algorithms with memory $\Ot(n)$. Our results for this type of algorithms include
\begin{enumerate}
\item A simple deterministic algorithm that uses a memory of $O(n)$ words to test structural balance for complete graphs.
\item For general graphs, we show that a memory of $\Omega(n\log{n})$ bits is necessary, which establishes a separation between the testing of complete vs. general graphs.
\end{enumerate}

\subsection{Upper bound: a deterministic semi-streaming algorithm}
We first present our algorithm that uses $O(n)$-words memory as follows.

\begin{proposition}
\label{prop:semi-streaming-test}
There exists a deterministic algorithm that given a signed complete graph $G=(V, \Ep\cup \Em)$, returns correctly whether $G$ is structurally balanced with a memory of $O\paren{n}$ words.
\end{proposition}

The high-level idea of the algorithm is as follows. Consider the following graph transformation that creates two copies $v_L$ and $v_R$ for each vertex $v$. For each arriving edge $(u,v)$, we connect $(u_L, v_L)$ and $(u_R, v_R)$ if the label is \plab, and connect $(u_L, v_R)$ and $(u_R, v_L)$ if the label is \mlab. As such, if we look at any triplet of vertices $(u,v,w)$, the graph forms disconnected components when there are zero or twice of `$L$-$R$ crosses', and the graph becomes connected whenever there is an imbalanced triangle. 

We now formalize the above idea as the following algorithm.
\begin{tbox}
\textbf{A Semi-streaming algorithm for structural balance testing}
\begin{enumerate}
\item Construct a graph $\tilde{G} = (\tilde{V}, \emptyset)$ with $\card{\tilde{V}}=2n$ vertices. For each vertex $v \in V$, align two vertices in $\tilde{V}$, and name them $v_L$ and $v_R$.
\item For each arriving edge $(u,v)$:
\begin{enumerate}
\item If $(u,v)$ is a \plab edge, add edges $(u_L, v_L)$ and $(u_R, v_R)$ to $\tilde{G}$.
\item If $(u,v)$ is a \mlab edge, add edges $(u_L, v_R)$ and $(u_L, v_R)$ to $\tilde{G}$.
\end{enumerate}
\item Run the standard connectivity algorithm on $\tilde{G}$ (i.e. add an edge if including the edge does not no create any cycle). 
\item If $\tilde{G}$ is connected, output `imbalanced'. Otherwise, output `balanced'.
\end{enumerate}
\end{tbox}

It is obvious that the algorithm uses $O(n)$ words since the connectivity test algorithm only maintain trees. We now give a simple analysis of the correctness of the algorithm.

\begin{lemma}
\label{lem:semi-streaming-test-analysis}
$\tilde{G}$ is connected if and only if $G$ is imbalanced.
\end{lemma}
\begin{proof}
For a triplet of vertices $(u,v,w)$ in $G$, if it forms an imbalanced triangle, the cycle induced by $(u_L, v_L, w_L, u_R, v_R, w_R)$ is connected. As such, for a pair of vertices $(x,z)$, suppose $x$ is connected with the component induced by $u_L$ and $z$ is connected with the component induced by $u_R$, $x$ and $y$ become connected in $\tilde{G}$ through $u_L$ and $u_R$. Note that since $G$ is connected, every vertex should be connected to the component induced by either $u_L$ or $u_R$.

Conversely, if \emph{every} triplet of vertices $(u,v,w)$ in $G$ forms balanced triangle, it induces two cycles among the vertices. As such, for any $u_L$ and $u_R$, they are always in different components.
\end{proof}

\subsection{An $\Omega(n\log{n})$ lower bound for testing structural balance on general graphs}

We now give a lower bound for testing structural balance on \emph{general} graphs, i.e. the graphs are not necessarily complete. The lower bound works against randomized algorithms; as such, we have shown that in the one-pass setting, testing whether the graph is balanced on general graphs is (much) harder.

To begin with, we first formally define the notion of testing whether a graph is balanced for general graphs. 
\begin{problem}
\label{prob:structural-balance-general}
    Given a signed graph $G = \paren{V, \Ep \cup \Em}$, where $\Ep$ and $\Em$ can be arbitrary sets of edges, decide if there exists a way to add $\plab$ and $\mlab$ edges to form $\tilde{\Ep}$ and $\tilde{\Em}$ such that
    \begin{itemize}
    \item The edges in $\tilde{\Ep} \cup \tilde{\Em}$ form a complete graph; and
    \item $G'=(V, \tilde{\Ep} \cup \tilde{\Em})$ is balanced as prescribed in \Cref{prob:structural-balance}.
    \end{itemize}
In other words, the answer is `NO' if and only if the graph is not balanced with $\Ep$ and $\Em$, i.e. having cycles with an odd number of negative edges.
\end{problem}

We now show a lower bound by a straightforward reduction from bipartiteness problem.
% \chen{TO-DO: include the reduction from bipartiteness here.}
\begin{proposition}
\label{prop:general-graph-test-lb}
Any single-pass streaming algorithm that correctly solves \Cref{prob:structural-balance-general} with probability at least $\frac{99}{100}$ requires a space of $\Omega(n \, \log{n})$ bits.
\end{proposition}
\begin{proof}
We use the following lower bound of \cite{SunW15} on testing whether a graph is bipartite. 

\begin{proposition}[\cite{SunW15}]
Any single-pass algorithm that given a graph $G=(V,E)$ in a stream, outputs correctly whether the graph is bipartite with probability at least $\frac{99}{100}$ requires a space of $\Omega(n \, \log{n})$ bits.
\end{proposition}
\noindent
We now establish a reduction as follows: for a given graph $G=(V,E)$, we label every edge as $\mlab$ and send it to the streaming algorithm for \Cref{prob:structural-balance-general}. As such, the only way for the graph to have cycles of an odd number of $\mlab$ edges is to have odd cycles in $G$. Therefore, the graph is balanced if and only if $G$ is bipartite, i.e. without odd cycles.
\end{proof}

\section{Approximation Guarantee of Frustration Sparsifiers (Proof of \Cref{lem:eval-frus})}
\label{app:frust-spar}

Recall that our frustration sparsifier is defined as follows.
\begin{tbox}
	\textbf{Frustration Sparsifier} for $\grph{G}{V}{\dedge}$ and $\eps \in \paren{0,1}$:
	\begin{itemize}
	    \item Let $\grph{H}{V}{E_H, w_H}$ be an $\frac{\eps}{2}$ cut sparsifier for $\grph{G'}{V}{\Ep}$.
    	\item Return $\frest{G}{\cdot}{\eps}$ which, when given $L \subseteq V$, computes
    	    \begin{align*}
    	        \underbrace{w_H\paren{L,V \setminus L}}_{\approx \card{\Ep\paren{L,V \setminus L}}} + \underbrace{\card{\Em} - \paren{\card{L} \cdot \card{V \setminus L} - w_H\paren{L,V \setminus L}}}_{\approx \card{\Em\paren{L}} + \card{\Em\paren{V \setminus L}}}.
    	    \end{align*}
	\end{itemize}
\end{tbox}
\noindent

\begin{proof}[Proof of \Cref{lem:eval-frus}]
    The time and space bounds follow directly from the algorithm and an invocation of \Cref{prop:eps-cutsparse-algs}.
    
    It remains to show
    \begin{align*}
        \paren{1 - \eps}\frust{G}{L} \le \frest{G}{L}{\eps} \le \paren{1 + \eps}\frust{G}{L}.
    \end{align*}
    for all $L \subseteq V$.
    
    Observe that
    \begin{align*}
        \frest{G}{L}{\eps}
        =
        &
        w_H\paren{L,R} + \card{\Em} - \paren{\card{L} \cdot \card{R} - w_H\paren{L,R}}
        \\
        \ge
        &
        \paren{1 - \frac{\eps}{2}} \card{\Ep\paren{L,R}} + \card{\Em} - \card{L} \cdot \card{R} + \paren{1 - \frac{\eps}{2}} \card{\Ep\paren{L,R}}
        \tag{Cut sparsifier: $\paren{1 - \frac{\eps}{2}} \Ep\paren{L,R} \le w_H\paren{L,R}$}
        \\
        =
        &
        \paren{1 - \eps} \card{\Ep\paren{L,R}} + \card{\Em} - \card{L} \cdot \card{R} + \card{\Ep\paren{L,R}}
        \\
        =
        &
        \paren{1 - \eps} \card{\Ep\paren{L,R}} + \card{\Em} - \paren{\card{L} \cdot \card{R} - \card{\Ep\paren{L,R}}}
        \\
        =
        &
        \paren{1 - \eps} \card{\Ep\paren{L,R}} + \card{\Em} - \card{\Em\paren{L, R}}
        \tag{$G$ is a complete graph}
        \\
        \ge
        &
        \paren{1 - \eps} \frustind{G},
        \tag{\Cref{def:frustind}}
    \end{align*}
    which asserts the left side of the inequality.
    
    Similarly, the right side of the inequality is asserted by
    \begin{align*}
        \frest{G}{L}{\eps}
        =
        &
        w_H\paren{L,R} + \card{\Em} - \paren{\card{L} \cdot \card{R} - w_H\paren{L,R}}
        \\
        \le
        &
        \paren{1 + \frac{\eps}{2}} \card{\Ep\paren{L,R}} + \card{\Em} - \card{L} \cdot \card{R} + \paren{1 + \frac{\eps}{2}} \card{\Ep\paren{L,R}}
        \tag{Cut sparsifier: $w_H\paren{L,R} \le \paren{1 + \frac{\eps}{2}} \Ep\paren{L,R}$}
        \\
        =
        &
        \paren{1 + \eps} \card{\Ep\paren{L,R}} + \card{\Em} - \card{L} \cdot \card{R} + \card{\Ep\paren{L,R}}
        \\
        =
        &
        \paren{1 + \eps} \card{\Ep\paren{L,R}} + \card{\Em} - \paren{\card{L} \cdot \card{R} - \card{\Ep\paren{L,R}}}
        \\
        =
        &
        \paren{1 + \eps} \card{\Ep\paren{L,R}} + \card{\Em} - \card{\Em\paren{L, R}}
        \tag{$G$ is a complete graph}
        \\
        \le
        &
        \paren{1 + \eps} \frustind{G}.
        \tag{\Cref{def:frustind}}
    \end{align*}
    
    The proof is thus complete.
\end{proof}

% \section{An Algorithm for \Cref{prob:min-frust-partition} in the $\gamma \gtrapprox \eps$ Regime}
\section{A Standard Algorithm for High Frustration Index}
\label{app:max-agree}

Details and proofs of the algorithm can be found in \cite{giotis2005correlation} (see $\textsc{MaxAg}\paren{k,\eps}$ in Section~3 there).
Here, we just state the algorithm with the aim of making its portability to the semi-streaming model salient.

\begin{tbox}
	\textbf{$\textsc{MaxAg}\paren{2,\eps}$} for $\grph{G}{V}{\dedge}$ and $\eps \in \paren{0,1}$:
	\begin{enumerate}
	    \item Set $t = \lceil \frac{4}{\eps} \rceil$.
	    \item Let $\paren{V_1, V_2, \dots, V_t}$ be a partition of $V$ where $\card{\card{V_i} - \card{V_j}} \le 1$.
	    \item Let $S_1, S_2, \dots, S_t$ be sets of $O\paren{\frac{1}{\eps^2}\log{\frac{1}{\eps}}}$ vertices, where $S_i$ samples vertices from $V \setminus V_i$ uniformly at random.
	    \item For edges $e$ in stream: Store $e$ if it is incident to any $S_i$.
	    \item For each partition $\paren{A,B}$ of $S_i$ for all $i \in \bracket{t}$:
	        \begin{enumerate}
	            \item Let $A_i$ be the vertices in $V_i$ with least disagreement towards $\paren{A,B}$ when placed in $A$.
	                This uses edges incident to $S_i$.
	            \item Let $B_i$ be $V_i \setminus A_i$.
	            \item Store $\paren{A_i, B_i}$ in $\paren{A^*_i, B^*_i}$ if it is the ``best'' partition of $V_i$ seen so far; that is, if $\card{\Em\paren{A_i,A}} + \card{\Ep\paren{A_i, B}} + \card{\Em\paren{B_i,B}} + \card{\Ep\paren{B_i, A}}$ is minimized, which can be checked using edges incident to $S_i$.
	        \end{enumerate}
	    \item Return $\bigcup_{i \in \bracket{t}} \paren{A^*_i,B^*_i}$.
	\end{enumerate}
\end{tbox}
\noindent

To get a guarantee with high probability, we repeat the above $O\paren{\log{n}}$ times and pick the best partition (using a Frustration Sparsifier to evaluate an approximate value of the frustration for each partition).

We observe the space complexity to be 
\begin{claim}
\label{clm:space-big-gamma}
    Algorithm $\textsc{MaxAg}\paren{2,\eps}$ uses space of $O(\frac{n}{\eps^3}\log(\frac{1}{\eps}))$ words.
\end{claim}
\begin{proof}
    We only store edges incident to $S_i$ sets.
    As such, the total number of edges to be stored is at most $$n\cdot t \cdot \card{\cup_i S_i} + {\card{\cup_i S_i} \choose 2}=O(\frac{n}{\eps^3}\log(\frac{1}{\eps})).$$
    The second term does not dominate since we sample at most $n$ vertices.
\end{proof}

\section{An Alternative Approach to Simulate the $S$-Sampler in Streaming}
\label{app:S-sampler-alt}

This section provides a combinatorial algorithm which solves \Cref{prob:structural-balance}, up to the use of limited-independence hash functions and PRGs.
To recap, we gave an algebraic solution in \Cref{sec:stuctural-balance-ub} by observing, firstly, that if we sample an odd sized set of vertices there is a reasonable chance that the induced graph will have an odd number of negative edges when the graph is imbalanced and in any other case there would always be an even number of negative edges; and secondly, that the parity of edges was counted by a degree-2 polynomial.
This was then derandomized using sums of epsilon biased generators.

We take the same approach here, but stop short of using a PRG for low-degree polynomials.
Instead, we aim to use Nisan's PRG to derandomize the process.
When it is the edges being sampled in a sketching-based algorithm, using Nisan's PRG to derandomize the sampling procedure was first developed by~\cite{indyk2006stable}.
This technique has since been used in other works, such as~\cite{AhnGM12b,KapralovLMMS14,KapralovMMMNST20}, and its usage requires that the random bits which we want to derandomize are read-once.
The key observation is that sketching-based algorithms are invariant under the order of inputs.
Therefore, it is sufficient to consider a fixed order of inputs, namely, one where all the updates for each edge are bundled together; here the random bits on each edge are indeed read-once.
Our situation is not immediately suited towards this approach: we are sampling vertices and not edges, and so there is no order of the input stream that works.

This section is organized as follows: We first introduce the notion of flip graphs and define the sub-flip-graph sampling problem which \Cref{prob:structural-balance} reduces to; reframing the problem this way is a matter of convenience.
We then show how to solve sub-flip-graph sampling using two vertex samplers: \lod and \hid which use $\Omega(n)$ bits of space.
For any graph, at least one of the samplers will work.
\lod is the easier case, where limited-independence hash functions alone suffice to keep the number of random bits down.
\hid is the harder case, which samples in a way that is amenable to the use of Nisan's PRG in order to get the space used down to $O(\polylog(n))$.
The key idea behind \hid stems from the following setup: What if the graph were bipartite and we were only sampling vertices from one side?
Then, there is indeed an order of edges that renders the random bits of each vertex to be read-once.
This is too much to hope for, but we constrain the problem enough so that an invocation of Nisan's PRG does indeed go through similarly.

\subsection{Flip-graphs}

We begin by formalizing the notion of flip.

\begin{definition}[Flip-edge]
\label{def:flip}
    Let $\grph{G}{V}{\dedge}$ be a complete signed graph, and $\paren{L,R}$ be an arbitrary partition of $V$.
    With respect to $\paren{L,R}$,
    \begin{itemize}
        \item $\paren{u,v} \in \Ep$ is a \flp if $u \in L, v \in R$ or $u \in R, v \in L$.
        \item $\paren{u,v} \in \Em$ is a \flp if $u,v \in L$ or $u,v \in R$.
    \end{itemize}
\end{definition}

\begin{definition}[Flip-graph]
\label{def:flip-graph}
    Let $\grph{G}{V}{\dedge}$ be a complete signed graph, and $\paren{L,R}$ be an arbitrary partition of $V$.
    The \emph{flip-graph} of $\paren{L,R}$ is the graph
    \begin{align*}
        \grph{\flpgrph{G}{L,R}}{V}{\set{e \in \dedge \lmid e \text{ is a \flp with respect to } \paren{L,R}}}.
    \end{align*}
\end{definition}

Suppose that, for some partition $\paren{L,R}$, the number of \flps in $G\bracket{S}$ is odd, but $\card{S}$ is even.
Is there an easy way to change the parity of $\card{S}$ without affecting the parity of \flps captured?
For an arbitrary \emph{flip-graph}, this is not possible.

\begin{example}
    Consider a complete bipartite graph $K$.
    Then every $S$ such that the number of edges in $K\bracket{S}$ is odd implies $\card{S}$ is even.
\end{example}

We thus look at a particular kind of \emph{flip-graph} using \Cref{def:bpart}.

\begin{definition}[Basic Partition]
\label{def:bpart}
    Let $\grph{G}{V}{\dedge}$ be a complete signed graph.
    For any vertex $v \in V$, a $v$-basic partition is the following partition of $V$:
    \begin{align*}
        \bpart{v} = \paren{\set{v} \cup \Np{v}, \Nm{v}} = \paren{\bpart{v}_L, \bpart{v}_R}.
    \end{align*}
\end{definition}

\begin{observation}
\label{obs:bpart-flip}
    The following statements hold:
    \begin{enumerate}
        \item If $\grph{G}{V}{\dedge}$ is balanced, then $\flpgrph{G}{\bpart{u}} = \flpgrph{G}{\bpart{v}}$ for all $u,v \in V$.
        \item If $\grph{G}{V}{\dedge}$ is imbalanced, then $\flpgrph{G}{\bpart{v}}$ is a graph with at least one isolated vertex and at least one edge.
    \end{enumerate}
\end{observation}
\begin{proof}
    Suppose that $\grph{G}{V}{\dedge}$ is balanced.
    Pick any vertex $v \in V$.
    Observe that $\bpart{v}$ is the partition of $V$ that certifies that $G$ is balanced; every pair of vertices in $\bpart{v}_L$ (or in $\bpart{v}_R$) is connected by a \plab edge, and every vertex in $\bpart{v}_L$ is connected to every vertex in $\bpart{v}_R$ by a \mlab edge.
    Thus, by \Cref{def:flip} $\flpgrph{G}{\bpart{v}}$ is the graph of $n$ isolated vertices.
    Since $v$ was chosen arbitrarily, the conclusion for item~1 follows.
    
    Suppose, on the other hand, that $\grph{G}{V}{\dedge}$ is imbalanced.
    Pick any vertex $v \in V$.
    Then $v$ is isolated in $\flpgrph{G}{\bpart{v}}$, and there is at least one edge in $\flpgrph{G}{\bpart{v}}$ as, otherwise, $\bpart{v}$ certifies that $G$ is balanced, which contradicts our assumption.
    Item~2 is thus proved.
\end{proof}

In view of \Cref{obs:bpart-flip}, it will therefore be very useful to use $\flpgrph{G}{\bpart{\vst}}$ in our analysis, for some arbitrary but fixed vertex $\vst \in V$.
We continue our detour from directly attacking \Cref{prob:structural-balance} and study the problem of how to sample $S$, an odd number of vertices, that induce an odd number of edges in a flip-graph defined on some $\vst$-basic partition when the input graph is imbalanced.

\begin{problem}[Sub-flip-graph Sampling]
\label{prob:flip-graph}
    Let $\vst$ be an arbitrary fixed vertex.
    Let $\grph{G}{V}{\dedge}$ be a complete signed graph, and let $\grph{\flpgrph{G}{\bpart{\vst}}}{V}{E}$ be its flip-graph.
    Denote $n = \card{V}$ and $m = \card{E}$. 
    The task is to find a subset of vertices $S \subseteq V$ such that $\card{S}$ is odd and, if $G$ is imbalanced, $\flpgrph{G}{\bpart{\vst}}\bracket{S}$ has an odd number of edges.
\end{problem}

The following claim shows that if we can solve \Cref{prob:flip-graph}, we can also solve \Cref{prob:structural-balance}.

\begin{claim}
\label{claim:flip-red}
    Let $S \subseteq V$ be a subset that solves \Cref{prob:flip-graph}.
    Then, $\card{\Em_G\paren{S}}$ is even if $G$ is balanced, and odd if $G$ is imbalanced.
\end{claim}
\begin{proof}
    Suppose $G$ is balanced.
    Using \Cref{def:flip} and the fact that $G$ is a complete graph, $\card{\Em_G\paren{S}} = \card{S \cap \bpart{\vst}_L} \cdot \card{S \cap \bpart{\vst}_R}$ is even since $\card{S}$ is odd (because at least one of $\card{S \cap \bpart{\vst}_L}$ and $\card{S \cap \bpart{\vst}_R}$ should be even).
    
    Now suppose $G$ is imbalanced.
    Let $P = E_{\flpgrph{G}{\bpart{\vst}}}\paren{S \cap \bpart{\vst}_L, S \cap \bpart{\vst}_R}$ and $M = E_{\flpgrph{G}{\bpart{\vst}}}\paren{S} \setminus P$.
    Then, using \Cref{def:flip} and the fact that $G$ is a complete graph, \[\card{\Em_G\paren{S}} = \card{S \cap \bpart{\vst}_L} \cdot \card{S \cap \bpart{\vst}_R} + \card{M} - \card{P}\] which is odd since both $\card{S}$ and $\card{M} + \card{P} = \card{E_{\flpgrph{G}{\bpart{\vst}}}\paren{S}}$ are odd.
\end{proof}

\begin{observation}
\label{obs:vst}
    Let $S \subseteq V$.
    Suppose $\flpgrph{G}{\bpart{\vst}}\bracket{S}$ has an odd number of edges, but $\card{S}$ is even.
    Then, adding or removing $\vst$ from $S$ changes the parity of $\card{S}$ without changing the parity of the number of edges in $\flpgrph{G}{\bpart{\vst}}\bracket{S}$.
\end{observation}
\begin{proof}
    This follows from \Cref{obs:bpart-flip}; $\vst$ is an isolated vertex in $\flpgrph{G}{\bpart{\vst}}$.
\end{proof}

By \Cref{obs:vst}, we may assume from here on out that any sampler for \Cref{prob:flip-graph} returns $S$ where $\card{S}$ is always odd; using just $\log{n}$ space an algorithm can check if $\card{S}$ is even and test if $\vst \in S$, then simply add or remove $\vst$ from $S$ if $\card{S}$ is even (using one extra ``override'' bit in its output to indicate membership of $\vst$).
In view of this, note that conditioned on $G$ being balanced, any sampler for \Cref{prob:flip-graph} succeeds with probability $1$.
Thus, if we can sample $S$ that solves \Cref{prob:flip-graph} with $\Omega\paren{1}$ probability when $G$ is imbalanced, we can solve \Cref{prob:structural-balance} with the exact same probability: by \Cref{claim:flip-red}, just count the parity of \mlab edges in $G\bracket{S}$.
We henceforth focus on the case when $G$ is imbalanced.

One approach to \Cref{prob:flip-graph} is to sample each vertex independently with probability $1/2$; one can then show that $S$ satisfies our desiderata with probability at least $1/4$.
This approach, however, uses $\Omega\paren{n}$ bits of space to store the $n$ bits of randomness,  as the algorithm needs to read these random bits more than once and cannot generate them on the fly (for example, if the edges arrive in the stream in lexicographic order, $n$ passes are made through the random bits).

To remedy this, we run two vertex samplers, \lod and \hid, simultaneously. Let $m$ be the number of edges in $\flpgrph{G}{\bpart{\vst}}$. Our samplers, we will show, have the following guarantees:
\begin{itemize}
    \item \lod succeeds with $\Omega\paren{1}$ probability when the maximum degree $\Delta$ of graph $\flpgrph{G}{\bpart{\vst}}$  is at most $12 \sqrt{m}$.
    \item \hid succeeds with $\Omega\paren{1}$ probability when the maximum degree $\Delta$ of graph $\flpgrph{G}{\bpart{\vst}}$  is at least $12 \sqrt{m}$.
\end{itemize}
% Notice that for any graph $\flpgrph{G}{\bpart{\vst}}$ one of the conditions must be true.

\lod takes in a parameter $\mwt$, and \hid takes in a parameter $\kwt$.
When the parameters are set correctly (taking values in $\cands$, a set of geometrically scaling values), one of \lod or \hid succeeds with $\Omega\paren{1}$ probability, depending on which of the exhaustive maximum degree conditions is satisfied by $\flpgrph{G}{\bpart{\vst}}$.

% To simplify the exposition slightly, we will describe \lod and \hid as non-deterministic algorithms (they will each non-deterministically guess a number from $\cands$, a set of geometrically scaling values).
% What is really happening, when viewed as deterministic algorithms, is that each of \lod and \hid returns $O\paren{\log{n}}$ random sets, \jie{maybe add "according to value $m$ taking each possibility of the set"} one of which succeeds with $\Omega\paren{1}$ probability when $\flpgrph{G}{\bpart{\vst}}$ meets the appropriate maximum degree condition.
% \todo[inline]{vik: after your comments, maybe the right approach is to eschew non-determinism completely. lod and hid will each take a parameter, m and k respectively. the final algorithm will make log n calls to each, one for each possible parameterization of m and k}

\subsection{\lod}

\lod is the simpler one of the two samplers.
The goal of \lod is to sample vertices $\slod \subseteq V$, and capture exactly one edge in $\flpgrph{G}{\bpart{\vst}}\bracket{\slod}$.
% \begin{itemize}
%     \item The representation of $\slod$ is a \indep{4} $\hlod$, used as an indicator function for $\slod$; this consumes $O\paren{\log{n}}$ space.
% \end{itemize}

\begin{tbox}
	\textbf{\lod}\paren{\mwt}:%\\
	    \begin{enumerate}[label=($\arabic*$)]
    	    \item Let $\hlod$ be a \indep{4} that indicates $v \in \slod$ with probability $\frac{1}{50\sqrt{\mwt}}$.
    	    \item Return $\slod$.
    	        \begin{itemize}
    	            \item $\slod$ is represented by $\hlod$: \qquad $v \in \slod \iff \hlod\paren{v} = 1$.
    	        \end{itemize}
	    \end{enumerate}
\end{tbox}
\noindent

\paragraph{Analysis. }
We now show that when $G$ is imbalanced and $\mwt$ is set correctly, \lod succeeds with $\Omega\paren{1}$ probability.

\begin{lemma}
\label{lem:lod-success}
    Suppose $m \ge 1$.
    Let $m \le \mwt \le 2m$.
    Further, suppose $\Delta \le 12 \sqrt{m}$, where $\Delta$ is the maximum degree of vertices in $\flpgrph{G}{\bpart{v}}$.
    Let $\slod$ be the set of vertices returned by \lod.
    Then,
    \begin{align*}
        \prob{\card{\edgs{\slod}} = 1} \ge \frac{1}{7500}.
    \end{align*}
\end{lemma}
\begin{proof}
    \begin{align*}
        \prob{\card{\edgs{\slod}} = 1}
        &=
        \sum_{e \in E} \prob{\edgs{\slod} = \set{e}}
        \tag{Disjoint events}
        \\
        &\ge
        m \cdot \frac{2}{3 \cdot 50^2 \mwt}
        \tag{\Cref{claim:slod-bound}}
        \\
        &\ge 
         \frac{1}{7500}.
        \tag{$\mwt \le 2m$}
    \end{align*}

    To finish off the proof, we prove \Cref{claim:slod-bound}.
    \begin{claim}
    \label{claim:slod-bound}
        $\prob{\edgs{\slod} = \set{\paren{u,v}}} \ge \frac{2}{3 \cdot 50^2 \mwt}$.
    \end{claim}
    \begin{proof}
        Let $\paren{u,v}$ be an arbitrary edge in $E$.
        Let $X$ be a random variable counting the number of edges in $\edgs{\slod}$, except for $\paren{u,v}$.
        Then
        \begin{align*}
            \prob{\edgs{\slod} = \set{\paren{u,v}}}
            & =
            \prob{u,v \in \slod} \cdot \prob{X < 1 \lmid u,v \in \slod}
            \\
            & =
            \prob{u \in \slod} \cdot \prob{v \in \slod} \cdot \paren{1 - \prob{X \ge 1 \lmid u,v \in \slod}}
            \tag{\indepce{4} of $\hlod$, and complementarity}
            \\
            & \ge
            \frac{2}{3 \cdot 50^2 \mwt}.
            \tag{Markov's inequality}
        \end{align*}
        
        The invocation of Markov's inequality can be seen to follow from
        \begin{align*}
            &
            \expect{X \lmid u,v \in \slod}
            \\
            &=
            \sum_{
                \substack{
                    \paren{x,y} \in E \setminus \set{\paren{u,v}}:
                    \\x,y \notin \nbr{u} \cup \nbr{v}
                }
            }
            \prob{x,y \in \slod \lmid u,v \in \slod}
            % \\
            % &
            +
            \sum_{
                \substack{
                    \paren{x,y} \in E \setminus \set{\paren{u,v}}:
                    \\y = u \text{ or } y = v
                }
            }
            \prob{x \in \slod \lmid u,v \in \slod}
            \tag{Linearity}
            \\
            & \le
            m \cdot \prob{x,y \in \slod \lmid u,v \in \slod}
            +
            24 \sqrt{m} \cdot \prob{x \in \slod \lmid u,v \in \slod}
            \tag{$\Delta \le 12 \sqrt{m}$}
            \\
            & =
            m \cdot \prob{x \in \slod} \cdot \prob{y \in \slod}
            +
            24 \sqrt{m} \cdot \prob{x \in \slod}
            \tag{\indepce{4} of $\hlod$}
            \\
            & =
            \frac{m}{50^2 \cdot \mwt}
            +
            \frac{24 \cdot \sqrt{m}}{50 \cdot \sqrt{\mwt}}
            % \\
            % & 
            \le
            \frac{1}{3}
            .
            \tag{$m \le \mwt$}
        \end{align*}
    \end{proof}
    The proof of \Cref{lem:lod-success} is thus complete: $\prob{\card{\edgs{\slod}} = 1} \ge \frac{1}{7500}$.
\end{proof}

\subsection{\hid}

% \hid is the more involved one of the two samplers.
% The moral behind \hid is to capture an odd number of edges in $\flpgrph{G}{\bpart{\vst}}\bracket{\shid{S}}$, where $\shid{S}$ is a random subset of vertices.
% On failure, we reduce the problem to that of sampling a vertex with an odd degree into $\shid{S}$.
% To go about this, \hid first finds a random partition $\paren{\shid{L},\shid{R}}$ of $V$; the goal here is that a high-degree vertex is separated from lots of its neighbors.
% Then, \hid samples $\shid{S} \subseteq \shid{L}$.
% If $\flpgrph{G}{\bpart{\vst}}\bracket{\shid{S}}$ has an odd number of edges, \hid has succeeded.
% Otherwise, \hid samples $\shid{T} \subseteq \shid{R}$.
% Here, we hope for $\flpgrph{G}{\bpart{\vst}}\bracket{\shid{T}}$ to contain no edges, and for there to be exactly one vertex in $\shid{T}$ with an odd degree into $\shid{S}$.

\hid is more involved, but the moral behind it is to capture an odd number of edges via two sampling procedures, where the second one is a `failsafe' of the first one.
We first sample a random set of vertices $\shid{S}$ and examine the parity of edges in $\flpgrph{G}{\bpart{\vst}}\bracket{\shid{S}}$; if it is odd, we are good.
Otherwise, we create an easier situation by reducing to the problem of sampling a vertex with an odd degree into $\shid{S}$. 

\begin{figure}[h]
    \centering
    \includegraphics[width=0.9\columnwidth]{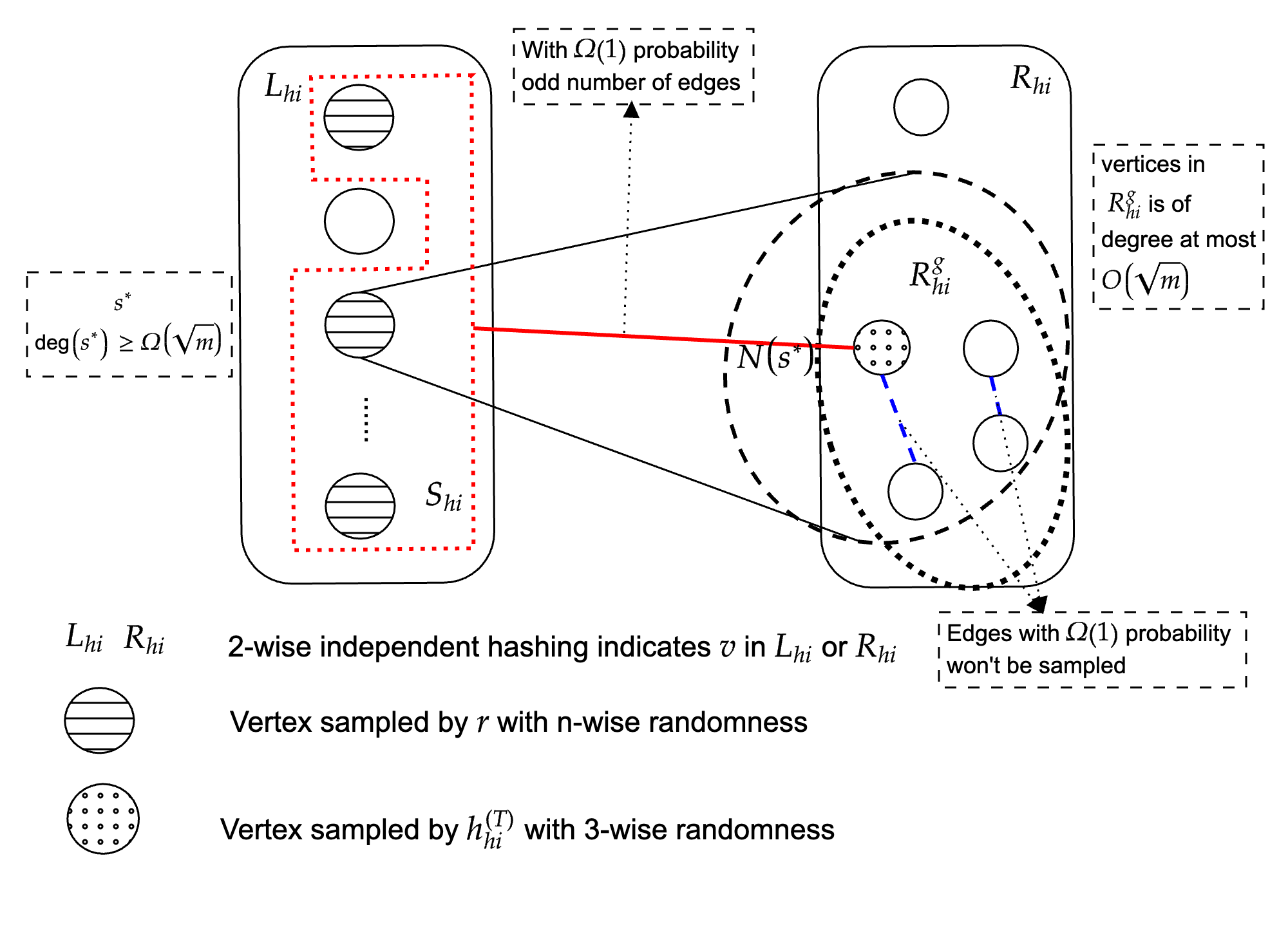}
    \caption{An illustration of the sampling strategy of \hid}
    \label{fig:high-degree-sampler}
\end{figure}

To explain the entire sampling process informally, \hid first finds a random partition $\paren{\shid{L},\shid{R}}$ of $V$; the goal here is that a high-degree vertex is separated from lots of its neighbors.
Then, \hid samples $\shid{S} \subseteq \shid{L}$.
If $\flpgrph{G}{\bpart{\vst}}\bracket{\shid{S}}$ has an odd number of edges, \hid has succeeded.
Otherwise, \hid samples $\shid{T} \subseteq \shid{R}$; this is the 'failsafe' mentioned earlier.
Here, we hope for $\flpgrph{G}{\bpart{\vst}}\bracket{\shid{T}}$ to contain no edges, and for there to be exactly one vertex in $\shid{T}$ with an odd degree into $\shid{S}$. We demonstrate  the design of \hid in \Cref{fig:high-degree-sampler}.

% \begin{itemize}
%     \item The representation of $\paren{\shid{L},\shid{R}}$ is a \indep{2} $\hhid{LR}$ which indicates $v \in \shid{L}$ or $v \in \shid{R}$; this consumes $O\paren{\log{n}}$ space.
%     \item The representation of $\shid{S}$ uses $n$ truly independent random bits; this consumes $\Omega\paren{n}$ space, but we will show later how this can be derandomized using Nisan's pseudorandom generator \cite{nisan1992pseudorandom}.
%     \item The representation of $\shid{T}$ uses a \indep{3} $\hhid{T}$, used as an indicator function; this consumes $O\paren{\log{n}}$ space.
% \end{itemize}
% Note that $v \in \shid{S}$ if and only if both $\hhid{LR}$ indicates $v \in L$ and an independent random bit for $\shid{S}$ indicates $v$, and similarly $v \in \shid{T}$ if and only if both $\hhid{LR}$ indicates $v \in R$ and $\hhid{T}$ indicates $v$.

We now formally introduce our \hid as follows.
\begin{tbox}
	\textbf{\hid}\paren{\kwt}:%\\
	    \begin{enumerate}[label=($\arabic*$)]
    		\item Let $\hhid{LR}$ be a \indep{2} that indicates $v \in \shid{L}$ and $v \in \shid{R}$ each with probability $\frac{1}{2}$.
    		\item Let $\hhid{T}$ be a \indep{3} that indicates $v$ with probability $\frac{1}{10\kwt}$.
    		\item Let $r$ be a string of $n$ truly random bits, each taking value $0$ or $1$ with probability $1/2$.
    		\item Return $\shid{S}$ and $\shid{T}$.
    		    \begin{itemize}
    		        \item $\shid{S}$ is represented with $\hhid{LR}$ and $r$: \[v \in \shid{S} \iff \hhid{LR}\paren{v} = 0 \text{ and } r_v = 1.\]
    		        \item $\shid{T}$ is represented with $\hhid{LR}$ and $\hhid{T}$: \[v \in \shid{T} \iff \hhid{LR}\paren{v} = 1 \text{ and } \hhid{T}\paren{v} = 1.\]
    		    \end{itemize}
	    \end{enumerate}
\end{tbox}
\noindent

Note that $r$ takes $\Omega\paren{n}$ bits of space, but we will show later how this can be `derandomized' using Nisan's PRG \cite{nisan1992pseudorandom} -- the resulting algorithm still uses randomness, but it uses an exponentially smaller number of truly random bits.

\paragraph{Analysis. }
To see that \hid succeeds, we will first show that $\hhid{LR}$ succeeds in separating a high degree vertex from most of its neighbors; this sets us up for the main lemma (\Cref{lem:hid-success}) of the analysis.

\begin{claim}
\label{claim:hid-sep}
    Suppose $m \ge 1$, and $\sst \in V$ has degree at least $12 \sqrt{m}$ in $\flpgrph{G}{\bpart{v}}$.
    Define the event
    \begin{align*}
        \evesep = \set{\sst \in \shid{L} \text{ and } \card{\nbr{\sst} \cap \shid{R}} \ge \card{\nbr{\sst}}/4}.
    \end{align*}
    That is, $\evesep$ is the event where $\sst \in \shid{L}$ and at least a quarter of its neighbors are in $\shid{R}$.
    Then,
    \begin{align*}
        \prob{\evesep} \ge \frac{1}{6}.
    \end{align*}
\end{claim}
\begin{proof}
    Let $X$ count the number of vertices in $\nbr{\sst} \cap \shid{L}$.
    Then,
    \begin{align*}
        \prob{\evesep}
        & =
        \prob{\sst \in \shid{L}} \cdot \prob{X < \frac{3}{4} \card{\nbr{\sst}} \lmid \sst \in \shid{L}}
        \\
        & =
        \frac{1}{2} \cdot \paren{1 - \prob{X \ge \frac{3}{4} \card{\nbr{\sst}} \lmid \sst \in \shid{L}}}
        \tag{Complementarity}
        \\
        & \ge
        \frac{1}{6}.
        \tag{Markov's inequality}
    \end{align*}
    
    The invocation of Markov's inequality can be seen to follow from
    \begin{align*}
        \expect{X \lmid \sst \in \shid{L}}
        &=
        \sum_{v \in \nbr{\sst}} \prob{v \in \shid{L} \lmid \sst \in \shid{L}}
        \tag{Linearity}
        \\
        &=
        \card{\nbr{\sst}} \cdot \prob{v \in \shid{L}}
        \tag{\indepce{2} of $\hhid{L,R}$}
        \\
        &=
        \frac{\card{\nbr{\sst}}}{2}.
    \end{align*}
    The proof is thus complete: $\prob{\evesep} \ge \frac{1}{6}$.
\end{proof}

$\evesep$ being satisfied implies that $\card{\shid{R} \cap \nbr{\sst}} \ge 3 \sqrt{m}$.
Why is this useful?
Roughly, we want $\shid{T} \subseteq \shid{R}$ to contain an odd number of edges into $\shid{S}$.
If $\sst \in \shid{S}$, then $\shid{R} \cap \nbr{\sst}$ being large suggests there are many `good' candidate vertices to be sampled into $\shid{T}$ -- the candidates who have edges towards $\shid{S}$ and whose degree inside $\shid{R}$ is low.
Let us formalize this idea.

\begin{claim}
\label{claim:many-low-in-R}
    Suppose $m \ge 1$, and $\sst \in V$ has degree at least $12 \sqrt{m}$ in $\flpgrph{G}{\bpart{v}}$.
    Suppose $\evesep$ and $\sst \in \shid{S}$ are both satisfied.
    Then, there is a set $\shid{R^g} \subseteq \shid{R}$ where
    \begin{enumerate}
        \item $\card{\shid{R^g}} \ge \card{\nbr{\shid{S}} \cap \shid{R}}/3$;
        \item For all $v \in \shid{R^g}$, there is at least one edge from $v$ to $\shid{S}$;
        \item For all $v \in \shid{R^g}$, the degree of $v$ is no more than $\sqrt{m}$.
    \end{enumerate}
\end{claim}
\begin{proof}
    Let $\shid{R^g} = \set{v \in \nbr{\shid{S}} \cap \shid{R} \lmid \card{\nbr{v}} \le \sqrt{m}}$.
    Properties (2) and (3) follow by definition.
    It remains to show property (1): $\card{\shid{R^g}} \ge \card{\nbr{\shid{S}} \cap \shid{R}}/3$.
    
    Observe that
    \begin{align*}
        \card{\nbr{\shid{S}} \cap \shid{R}}
        & \ge
        \card{\nbr{\sst} \cap \shid{R}}
        \tag{$\sst \in \shid{S}$}
        \\
        & \ge
        \card{\nbr{\sst}}/4
        \tag{$\evesep$ satisfied}
        \\
        &\ge
        3 \sqrt{m}.
        \tag{$\card{\nbr{\sst}} \ge 12 \sqrt{m}$}
    \end{align*}
    Suppose for a contradiction that $\card{\shid{R^g}} < \card{\nbr{\shid{S}} \cap \shid{R}}/3$.
    Then at least $2 \sqrt{m}$ vertices have degree more than $\sqrt{m}$.
    Summing over degrees to get twice the number of edges, this implies that $m = \card{E} \ge 2 m$, which is a contradiction since $m \ge 1$.
\end{proof}

We are now ready to show the main result for \hid.

\begin{lemma}
\label{lem:hid-success}
    Suppose $m \ge 1$, and $\sst \in V$ has degree at least $12 \sqrt{m}$ in $\flpgrph{G}{\bpart{v}}$.
    Let $\shid{L}$, $\shid{R}$, $\shid{S}$, and $\shid{T}$ be the sets of vertices returned by \hid.
    Let $\card{\nbr{\shid{S}} \cap \shid{R}} \le \kwt \le 2 \card{\nbr{\shid{S}} \cap \shid{R}}$.
    Then,
    \begin{align*}
        \prob{\card{\edgs{\shid{T}} \cup \edgs{\shid{T},\shid{S}}} \mbox{ is odd}} \ge \frac{1}{4320}.
    \end{align*}
\end{lemma}
\begin{proof}
    Start with the following:
    \begin{itemize}
        \item Let $\shid{R^g}$ be the largest set of vertices in $\shid{R}$ such that (same as in \Cref{claim:many-low-in-R})
            \begin{enumerate}
                \item For all $v \in \shid{R^g}$, there is at least one edge from $v$ to $\shid{S}$;
                \item For all $v \in \shid{R^g}$, the degree of $v$ is no more than $\sqrt{m}$,
            \end{enumerate}
            conditioned on $\evesep$ and $\sst \in \shid{S}$ being satisfied.
        \item Define 
            \begin{align*}
                \eve{u} = \set{\shid{T} \cap \nbr{\shid{S}} = \set{u} \text{ and } \card{\edgs{\shid{T}}} = 0}
            \end{align*}
            for all $u \in \shid{R^g}$.
    \end{itemize}
    Then
    \begin{align*}
        &
        \prob{\card{\edgs{\shid{T}} \cup \edgs{\shid{T},\shid{S}}} \text{ is odd}}
        % \\
        % &
        \ge
        \sum_{u \in \shid{R^g}} \prob{\eve{u}, \evesep, \sst \in \shid{S}, \card{\edgs{u,\shid{S}}} \text{ is odd}}
        \tag{Disjoint events}
        \\
        &\ge
        \frac{\card{\nbr{\shid{S}} \cap \shid{R}}}{3} \prob{\eve{u}, \evesep, \sst \in \shid{S}, \card{\edgs{u,\shid{S}}} \text{ is odd}}
        \tag{\Cref{claim:many-low-in-R}}
        \\
        &\ge
        \frac{\card{\nbr{\shid{S}} \cap \shid{R}}}{1080 \kwt} \prob{\card{\edgs{u,\shid{S}}} \text{ is odd} \lmid \eve{u}, \evesep, \sst \in \shid{S}} \tag{\Cref{claim:hid-bound}}
        \\
        &\ge
        \frac{\card{\nbr{\shid{S}} \cap \shid{R}}}{2160 \kwt}
        \tag{Independence of $r_{s^*}$ from the other bits in $r$, and of $r$ from $\hhid{L,R}, \hhid{T}$}
        \\
        &\ge
        \frac{1}{4320}.
        \tag{$\kwt \le 2 \card{\nbr{\shid{S}} \cap \shid{R}}$}
    \end{align*}
    We finish off with a proof of \Cref{claim:hid-bound}.
    
    \begin{claim}
    \label{claim:hid-bound}
        $\prob{\eve{u}, \evesep, \sst \in \shid{S}} \ge \frac{1}{360 \kwt}$.
    \end{claim}
    \begin{proof}
        Define the random variable $X_1$ that counts the number of vertices in $\shid{T} \cap \nbr{\shid{S}}$, except for $u$.
        Define the random variable $X_2$ that counts the number of vertices in $\shid{T} \cap \nbr{u}$.
        Define the random variable $X_3$ that counts the number of edges in $\edgs{\shid{T}}$ whose endpoints are not incident to $\nbr{\shid{S}} \cup \nbr{u}$.
        Then
        \begin{align*}
            & \prob{\eve{u}, \evesep, \sst \in \shid{S}} \\
            &\ge
            \frac{1}{12} \cdot \prob{\eve{u} \lmid \evesep, \sst \in \shid{S}}
            \tag{\Cref{claim:hid-sep} and $\prob{\sst \in \shid{S}} = 1/2$}
            \\
            &\ge
            \frac{1}{12} \cdot \prob{u \in \shid{T} \lmid \evesep, \sst \in \shid{S}} \cdot \prob{X_1 + X_2 + X_3 < 1 \lmid u \in \shid{T}, \evesep, \sst \in \shid{S}}
            \\
            &=
            \frac{1}{12} \cdot \prob{u \in \shid{T} \lmid \evesep, \sst \in \shid{S}} \cdot \paren{1-\prob{X_1 + X_2 + X_3 \ge 1 \lmid u \in \shid{T}, \evesep, \sst \in \shid{S}}}
            \tag{Complementarity}
            \\
            &\ge
            \frac{1}{12} \cdot \prob{u \in \shid{T}} \cdot \frac{1}{3}
            \tag{$\evesep, \sst \in \shid{S}$ are independent of $\hhid{T}$, and Markov's Inequality}
            \\
            &=
            \frac{1}{360 \kwt}.
        \end{align*}
        
        The invocation of Markov's Inequality can be seen to follow from
        \begin{align*}
            &
            \expect{X_1 + X_2 + X_3 \lmid u \in \shid{T}, \evesep, \sst \in \shid{S}}
            \\
            \le&
            \sum_{v \in \nbr{\shid{S}} \cap \shid{R} \setminus \set{u}} \prob{v \in \shid{T} \lmid u \in \shid{T}, \evesep, \sst \in \shid{S}}
            \\
            &+
            \sum_{v \in \nbr{u}} \prob{v \in \shid{T} \lmid u \in \shid{T}, \evesep, \sst \in \shid{S}}
            \\
            &
            +
            \sum_{\paren{x,y} \in E} \prob{x,y \in \shid{T} \lmid u \in \shid{T}, \evesep, \sst \in \shid{S}}
            \tag{Linearity}
            \\
            &\le
            \card{\nbr{\shid{S}} \cap \shid{R}} \cdot \prob{v \in \shid{T} \lmid u \in \shid{T}}
            \tag{$\evesep, \sst \in \shid{S}$ are independent of $\hhid{T}$}
            \\
            &
            +
            \sqrt{m} \cdot \prob{v \in \shid{T} \lmid u \in \shid{T}}
            \tag{$\evesep, \sst \in \shid{S}$ are independent of $\hhid{T}$, and $\card{\nbr{u}} \le \sqrt{m}$}
            \\
            &
            +
            m \cdot \prob{x,y \in \shid{T} \lmid u \in \shid{T}}
            \tag{$\evesep, \sst \in \shid{S}$ are independent of $\hhid{T}$}
            \\
            =
            &
            \card{\nbr{\shid{S}} \cap \shid{R}} \cdot \prob{v \in \shid{T}}
            \\
            &
            +
            \sqrt{m} \cdot \prob{v \in \shid{T}}
            \\
            &
            +
            m \cdot \prob{x \in \shid{T}} \cdot \prob{y \in \shid{T}}
            \tag{\indepce{3} of $\hhid{T}$}
            \\
            =
            &
            \,\frac{\card{\nbr{\shid{S}} \cap \shid{R}}}{10 \kwt} + \frac{\sqrt{m}}{10 \kwt} + \frac{m}{10^2 \kwt^2}
            \\
            \le
            &
            \,\frac{\card{\nbr{\shid{S}} \cap \shid{R}}}{10 \kwt} + \frac{\card{\nbr{\shid{S}} \cap \shid{R}}}{10 \kwt} + \frac{\card{\nbr{\shid{S}} \cap \shid{R}}^2}{10^2 \kwt^2}
            \tag{$\evesep, \sst \in \shid{S}$, and \Cref{claim:many-low-in-R}}
            \\
            \le
            & 
            \,\frac{1}{3}.
            \tag{$\card{\nbr{\shid{S}} \cap \shid{R}} \le \kwt$}
        \end{align*}
    \end{proof}
    
    The proof is thus complete: $\prob{\card{\edgs{\shid{T}} \cup \edgs{\shid{T},\shid{S}}} \text{ is odd}} \ge \frac{1}{4320}$.
\end{proof}

\subsubsection{Using \hid: A Derandomized Linear Sketch}

\Cref{lem:hid-success} intimates at what kind of a linear sketch, using \hid, we would like to design.
For random sets $\shid{S}$ and $\shid{T}$ given by \hid, the sketch counts $\card{\edgs{\shid{T}} \cup \edgs{\shid{T},\shid{S}}}$.
With the view of derandomizing this sketch, we will ``open'' \hid up in the following algorithm.

\begin{tbox}
	\textbf{\hidbs}:%\\
	    \begin{enumerate}[label=($\arabic*$)]
    		\item Receive $\hhid{LR}$, $\hhid{T}$ and $r$ from \hid\paren{\kwt}.
    % 		\item Receive $\hhid{T}$ from \hid\paren{\kwt}.
    % 		\item Receive $r$ from \hid\paren{\kwt}.
    		\item For edges $e =\paren{u,v}$ in stream:
        		\begin{itemize}
        		    \item If $\hhid{LR}$ indicates $u,v \in \shid{R}$, $\hhid{T}(u)=1$ and $\hhid{T}(v)=1$, increment counter -- $e$ is in $\edgs{\shid{T}}$.
        		    \item If $\hhid{LR}$ indicates $u \in \shid{L}$, $r_u=1$, and if $\hhid{LR}$ indicates $v \in \shid{R}$ and $\hhid{T}(v)=1$, increment counter -- $e$ is in $\edgs{\shid{T}, \shid{S}}$.
        		    \item Swap the roles of $u$ and $v$ in the preceding point and check with the same rule.
        		\end{itemize}
        	\item Return the counter's parity.
	    \end{enumerate}
\end{tbox}
\noindent

\paragraph{Derandomization. }
The following approach towards derandomizing a sketch originates from \cite{indyk2006stable}.
More examples of its use can be found in \cite{AhnGM12b,kapralov2017single}.
The technique uses Nisan's PRG, which we recapitulate with \Cref{prop:nis}.

% \begin{proposition}[Corollary~1 in \cite{nisan1992pseudorandom}]
% \label{prop:nis}
%     Any randomized algorithm running in $\textrm{space}\paren{S}$ and using $R$ random bits may be converted to one that uses only $O\paren{S \log{R}}$ random bits (and runs in $\textrm{space}\paren{S \log{R}}$).
%     The random bits are read-once.
% \end{proposition}

Our sketch fails to satisfy the read-once condition on random bits and so we are unable to invoke \Cref{prop:nis} directly.
In particular, the bits in $r$ may be read up to $n$ times.
Consider the following, however.
Run \hidbs until just after $\hhid{LR}$ and $\hhid{T}$ are stored in memory (each requiring $O\paren{\log{n}}$ space).
Now, for the sake of argument, assume an order on the vertices in $\shid{L}$.
When the edges in $\edgs{\shid{L},\shid{T}}$ arrive before any other edge in the stream and are moreover ordered by the endpoint in $\shid{L}$, the random bits of $r$ can be accessed in a read-once way (in the order of $\shid{L}$) while using just $O\paren{1}$ more memory.
For this order of edge-arrivals, \hidbs uses $O\paren{\log{n}}$ space and the random bits are read-once \footnote{Contrast this with the initial approach of just sampling vertices independently. Unlike with \hid, there is no way to order the edges so as to make the random bits read-once.}.
We can therefore invoke \Cref{prop:nis} and replace $r$ of length $n$ with $r'$, a truly random bit-string of length $O\paren{\log^2{n}}$ that can be stretched to $O\paren{n}$ pseudorandom bits that fool our algorithm.
Since our algorithm is a linear sketch, the order in which edges arrive in the stream does not alter its output.
We can therefore use the PRG to stretch $r'$ in a just-in-time fashion to access the pseudorandom bits that need to be read for any particular edge, and hence handle a stream of arbitrarily ordered edges.
We refer to the derandomized version of \hid with \dhid.

\subsection{The Complete Algorithm}

Recall \Cref{claim:flip-red}; we now return to attacking \Cref{prob:structural-balance} directly.
Using \lod to count \mlab edges in $\Em\paren{\slod}$ is straightforward.
We have not yet shown what to do with \dhid vis-a-vis counting \mlab edges.
Note that our sketch, used directly, counts \mlab edges in $\Em\paren{\shid{T}} \cup \Em\paren{\shid{T},\shid{S}}$, which is not necessarily an edge set for an induced subgraph.
We can, however, count edges in $\Em\paren{\shid{S}}$.
If the number of edges in $\Em\paren{\shid{S}}$ is odd, we are home-free.
Otherwise, noting that
\begin{align*}
    \card{\Em\paren{\shid{S} \cup \shid{T}}} = \card{\Em\paren{\shid{S}}} + \card{\Em\paren{\shid{T}} \cup \Em\paren{\shid{T},\shid{S}}}
\end{align*}
and appealing to \Cref{lem:hid-success} when the degree condition is satisfied, the number of edges in \Em\paren{\shid{S} \cup \shid{T}} will be odd with $\Omega\paren{1}$ probability on an imbalanced instance.

\begin{tbox}
	\textbf{\btest}:\\
	    Run $15000 \log{n}$ copies in parallel:
	    \begin{enumerate}[label=($\arabic*$)]
    		\item For $\mwt$ in $\cands$, receive $\hlod$ from \lod$\paren{\mwt}$.
    		\item For $\kwt$ in $\cands$, receive $\hhid{LR}, \hhid{T}, r'$ from \dhid\paren{\kwt}.
    		%\item Receive $r'$ from \dhid\paren{\cdot} (the value of $\kwt$ does not matter here).
    		\item For \mlab edges $\paren{u,v}$ in stream, for all pairs $\mwt, \kwt$: Use PRG on $r'$ to get indicators $r_u, r_v$ and
    		    \begin{itemize}
    		        \item If $\hlod$ indicates both $u, v \in \slod$, increment this copy's counter for $\Em\paren{\slod}$.
    		        \item If $\hhid{LR}$ indicates $u,v \in \shid{L}$, and $r_u, r_v$ indicate $u, v \in \shid{S}$, increment this copy's counter for $\Em\paren{\shid{S}}$ and $\Em\paren{\shid{S} \cup \shid{T}}$.
    		        \item If $\hhid{LR}$ indicates $u,v \in \shid{R}$ and $\hhid{T}$ indicates $u,v \in \shid{T}$, increment this copy's counter for $\Em\paren{\shid{S} \cup \shid{T}}$.
    		        \item If $\hhid{LR}$ indicates $u \in \shid{L}$ and $v \in \shid{R}$, and $r_u$ indicates $u \in \shid{S}$, and $\hhid{T}$ indicates $v \in \hhid{T}$, increment this copy's counter for $\Em\paren{\shid{S} \cup \shid{T}}$.
    		        \item Same as the preceding point, but with $u$ and $v$ swapped.
    		    \end{itemize}
	    \end{enumerate}
	    If any copy of counters for $\Em\paren{\slod}$, $\Em\paren{\shid{S}}$, or $\Em\paren{\shid{S} \cup \shid{T}}$ is odd, return \textsc{Not Balanced}.
	    Otherwise, return \textsc{Balanced}.
\end{tbox}
\noindent

\begin{proof}[Proof of \Cref{thm:structural-balance-ub}]
    Depending on the maximum degree of $\flpgrph{G}{\bpart{\vst}}$, one of the premises of \Cref{lem:lod-success} or \Cref{lem:hid-success} will be satisfied, for the right choice of $\mwt \in \cands$ or $\kwt \in \cands$ respectively.
    The correctness and probabilistic guarantee of \btest then follows from \Cref{claim:flip-red}, \Cref{obs:vst}, \Cref{lem:lod-success}, and \Cref{lem:hid-success}.
    \btest succeeds with probability at least $1 - (1 - 1/7500)^{15000 \log{n}} \ge 1 - \frac{1}{n}$.
    
    The space used by \btest for $\hlod$, $\hhid{L,R}$, and $\hhid{T}$ are $O\paren{\log^3{n}}$ each since they are \indep{k} for $k \le 4$ (meaning $O\paren{\log{n}}$ bits to store one of them, by \Cref{prop:k-ind-hash}), and there are $O\paren{\log^2{n}}$ copies of each.
    The space used for $r'$ is $O\paren{\log^4{n}}$, since there are $O\paren{\log^2{n}}$ copies, each of which are $O\paren{\log^2{n}}$ bits long.
    Finally, the space used for storing $O\paren{\log^2{n}}$ counters that count up to at most $n^2$ is $O\paren{\log^3{n}}$.
    The total space used by \btest is thus $O\paren{\log^4{n}}$.
\end{proof}

\end{document}